\documentclass[a4paper, amsfonts, amssymb, amsmath, reprint, showkeys, nofootinbib, twoside, superscriptaddress]{revtex4-2}

\usepackage{graphicx}
\usepackage{dcolumn}
\usepackage{bm}
\usepackage{xcolor}

\begin{document}

\title{Measuring the dynamical evolution of the United States lobbying network}

\author{Karol A. Bacik}
\affiliation{Department of Mathematics, Massachusetts Institute of Technology, \\ 77 Massachusetts Avenue, 
Cambridge, MA 02142, USA}

\author{Jan Ondras}
\affiliation{Department of Mathematics, Massachusetts Institute of Technology, \\ 77 Massachusetts Avenue, 
Cambridge, MA 02142, USA}

\author{Aaron Rudkin}
\affiliation{Department of Political Science, Massachusetts Institute of Technology, \\ 77 Massachusetts Avenue, 
Cambridge, MA 02142, USA}

\author{J\"orn Dunkel}
\affiliation{Department of Mathematics, Massachusetts Institute of Technology, \\ 77 Massachusetts Avenue, 
Cambridge, MA 02142, USA}

\author{In Song Kim}
\affiliation{Department of Political Science, Massachusetts Institute of Technology, \\ 77 Massachusetts Avenue, 
Cambridge, MA 02142, USA}

\date{\today}



\maketitle

\textbf{Lobbying networks constitute complex political systems that mobilize vast human and financial resources to influence governmental decision-making, often with profound national and global consequences~\cite{Bombardini2020,DeFigueiredo2014,halllobbyingapsr,baumgartner}. A comprehensive understanding of lobbying strategies and dynamics requires time-resolved, system-wide data, which are largely unavailable for most political systems~\cite{Lazer2020,You2017}. In the United States (U.S.), the Lobbying Disclosure Act (LDA) of 1995 mandates public reporting of all federal lobbying activities in detailed quarterly filings. However, extracting structured, quantitative information from these filings has remained technically challenging and labor-intensive.
Here we present and analyze \textsf{LobbyView}~\cite{LV:link}, a relational database that integrates and disambiguates data from more than 1.6 million LDA reports. \textsf{LobbyView} provides access to detailed lobbying disclosures, reconciled corporate entities, and tools for linking LDA data to external legislative and corporate databases.
We demonstrate the utility of \textsf{LobbyView} by examining both macro-level and highly granular lobbying dynamics. Specifically, we reconstruct the connectivity patterns of the U.S.~lobbying network, and we show how they evolve over time, we identify organizational principles such as the accumulation of professional contacts within a small set of firms, and reveal how lobbying activity is synchronized with electoral cycles. Moreover, we introduce a probabilistic framework for analyzing lobbying behavior at the scale of individual bills, issues, or firms.
We envision \textsf{LobbyView}~\cite{LV:link} as a resource not only for political scientists, but also for quantitative interdisciplinary research, enabling the application of methods from statistical physics, systems biology, and machine learning to the study of lobbying systems.
}
\par
Political networks~\cite{Porter2005, Fowler2006,Lazer2011,Bond2012}
are complex dynamical systems that evolve and adapt in response to
national and global events~\cite{Igan2012,Bavel2020}. A prime example
is the U.S.~lobbying system~\cite{Bombardini2020,DeFigueiredo2014}, a
self-organized multilayered network (Fig.~1A) that, fueled by billions of dollars annually~\cite{Bombardini2020}, has significantly shaped the course of policymaking in the U.S.~\cite{scha:1935,trip:etal:2002,Hall2006}. The interplay between lobbyists, firms, legislators, and agencies within the U.S.~lobbying network has profound consequences for both national~\cite{kang:2016} and
international~\cite{kim:milner:20} issues, including the economy~\cite{kim:17}, global
climate~\cite{brulle:2018,kenn:2020} and health~\cite{stein:2009}
crises, and armed~\cite{miln:ting:2015}
conflicts. Understanding these interactions is a central challenge in the social and political sciences: which political entities engage in lobbying and which
strategies do they pursue~\cite{Bertrand2014}? What determines the scale of lobbying expenditures~\cite{Ansolabehere2003}, and how do they affect downstream policy decisions~\cite{Hall2006}? How does the lobbying network reorganize in response to disruptive events such as the 2007--2008 financial crisis or the COVID-19 pandemic?

Despite the importance of a systemic understanding of these and other critical issues, and notwithstanding important methodological advances~\cite{Lazer2009,Lazer2020,You2017}, a quantitative end-to-end analysis of the network of actors in lobbying and their interactions remains elusive. This gap primarily persists due to the absence of comprehensive, high-fidelity data that captures both the monetary and informational flows, as well as the intricate dynamics among the various components of the U.S.~lobbying system.
The availability of such data holds the key to substantial future progress as it opens the possibility of utilizing recently developed methods from network theory~\cite{Barabasi_book,Newman_book,Goyal_book,Watts2004,Boccaletti2006}, applied mathematics~\cite{Strogatz2001,Bick2023} and statistical physics~\cite{Albert2002,Bardoscia2021} to achieve a predictive understanding of political decision processes.  In particular, such data will offer a chance to compare the evolution and adaptation within an influential human interaction network with the emergent dynamical behaviors and scaling laws found in other complex social~\cite{Fowler2010,Szell2010,Traud2012}, biological~\cite{Jeong2003,Rosenthal2015,Bassett2017}, physical~\cite{Papadopoulos2018,Posfai2024}, and information-processing~\cite{Albert1999,broder2000graph} network systems.

\par 
In general, lobbying can take different forms (advertisements, donations, campaigning through political action committees (PACs), intervention in primary elections, federal policy advisory committees, etc.), some more elusive than others~\cite{lapira2015lobbying,furnaspartisan,uclaparties,kalladonations,tripathipac}.
Nevertheless, critical insight into federal lobbying in the U.S.~can be gained by analyzing the data from the Lobbying Disclosure Act (LDA) reports.
Here we present and analyze  \textsf{LobbyView}~\cite{LV:link}, a comprehensive database enabling quantitative analysis of all federal lobbying activities in the United States since 1999. 
Continuously expanded over the past decade and now made publicly available in complete form, \textsf{LobbyView} integrates more than 1.6 million filings, representing over 87 billion USD in lobbying expenditures. The records provide granular detail on each reported interaction among clients, lobbying firms, lobbyists, government entities, and politicians, with half-yearly—and more recently, quarterly—temporal resolution.

\par
To showcase the potential of \textsf{LobbyView} as a resource for interdisciplinary quantitative research, we use our data to characterize the complex dynamics of the lobbying network at different time scales. We find that long-term evolution of professional connections is governed by features shared with other complex systems, such as preferential attachment; mid-term system dynamics track election cycles; and short-term perturbations in the network are often responses to critical global events, such as the global financial crisis of 2008 or the COVID-19 pandemic. Importantly, \textsf{LobbyView} and the LDA data can also be used to estimate and compare the conditional probabilities of different lobbying strategies to reveal the lobbying behaviors of individual clients or sectors at increasing levels of resolution.

\section*{The LobbyView  database}

\paragraph*{Lobbying reports.} The Lobbying Disclosure Act (LDA) of 1995, later amended by the Honest Leadership and Open Government Act (HLOGA) of 2007, mandates that federally registered lobbyists and lobbying firms (registrants) file quarterly
reports~\cite{LDA1995regs}. These reports, known as LD-2 disclosure forms, detail their
lobbying activities on behalf of their clients. The disclosures include the total lobbying expenditures and payments by the clients during the relevant period, along with a breakdown across 79 policy issues of concern, including taxation, international trade, healthcare, and more. For each issue area, registrants must provide detailed information about the specific subjects of their lobbying, including any lobbied congressional bills or resolutions. These disclosures are unstructured text and require considerable effort to extract and structure. Reports are additionally required to identify the federal agencies, departments, and chambers of Congress that were
targeted for lobbying. The names of individual lobbyists involved and any past
working relationships between these lobbyists and government officials or legislators are also disclosed. We have collected and
curated this data into the \textsf{LobbyView} database~\cite{LV:link,mm},
which is continuously updated with new records as they become available.

\par

\paragraph*{Technical aspects.} A substantial yet often underappreciated technical challenge is that the same lobbying actor (e.g., a lobbyist) may appear in LDA reports under multiple variations of their name (e.g., Joseph Smith, Joe Smith, J. Smith). Similarly, firms are frequently represented inconsistently, and when companies change names, merge, or are acquired, the raw report data typically lacks a reliable identifier to uniquely track them. To resolve these and other text disambiguation challenges and produce unique identifiers for every agent and firm in the database,  we developed and implemented matching algorithms (see~\cite{mm}), which make it possible not only to accurately infer the connectivity patterns of the lobbying network but also to interface unstructured LDA reports to major external databases, such as WRDS Compustat~\cite{compustat}, BoardEx~\cite{boardex}, Moody's Orbis~\cite{orbis}, NOMINATE/VoteView~\cite{voteview} congressional ideological scores, interest group scores of legislators, and personal political donation records. To enable future research, \textsf{LobbyView} provides access to the identifiers required to cross-link with external databases. 

\begin{figure*}
    \centering
    \includegraphics[width=\textwidth]{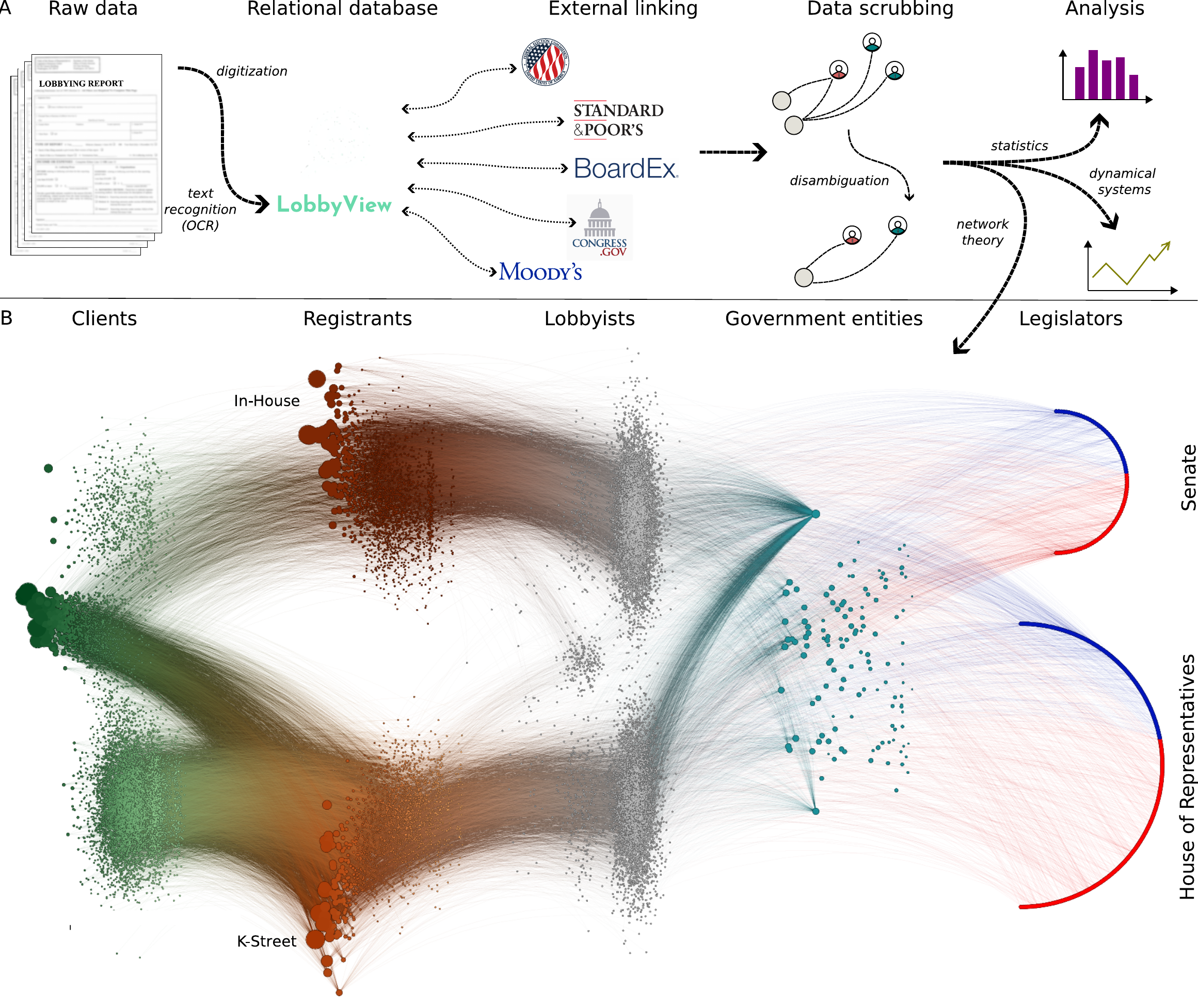}
    \caption{\textbf{Our \textsf{LobbyView} database comprises over 1.6
        million public records, covering all federal U.S.~lobbying
        since 1999, and enabling a time-resolved reconstruction of the
        multilayered architecture and scaling behaviors of the
        lobbying network.} 
        (A)~The data processing from the Lobbying Disclosure Act (LDA) reports to lobbying network starts with the digitalized text of publicly available disclosures. The atomistic data is then compiled into a relational database, which can be interfaced with other political datasets. To correctly infer the statistics of lobbying activities and infer the lobbying network structure, we clean the data and disambiguate the lobbying actors.  
        (B)~Network representation of all
      U.S.~federal lobbying activities during the year 2017.
      The first layer comprises 10,694 clients initiating lobbying (green), and the second layer comprises 4,405 registrants (brown). The node size in these layers is proportional to the cumulative USD amount
      spent/received by each individual client/registrant (c.f. Sec.~IV and V of~\citep{mm}). 
      The registrants are connected to 11,543 lobbyists they employ (third layer, gray). 
      We also reconstructed historical associations of lobbyists with 128 specific government agencies (turquoise) or current members of Congress (374 legislators; Democrats in blue and Republicans in red).
      The details of the visualization can be found in~\cite{mm}, Sec.~VII. 
      }
      \label{fig:network_graph_2017_and_degree_distributions}
\end{figure*}

\section*{Reconstructing the lobbying network}


The disambiguated \textsf{LobbyView} data allow us to discern structural features of the complex lobbying network. To illustrate this, we elucidate the network dynamics and the statistical properties of interactions between different types of actors by adopting a meso-level approach that treats classes of actors as the relevant unit of analysis. Specifically, we distinguish five primary
categories of lobbying actors: \emph{clients} (firms or interest
groups initiating lobbying efforts), \emph{registrants} (lobbying
firms responsible for filing reports), \emph{lobbyists} (individuals
directly involved in lobbying activities), \emph{government entities}
(House(s) of Congress, and federal agencies and departments targeted
by lobbyists), and \emph{legislators} (members of the U.S.~Congress).
Accordingly, we visualize the lobbying network as a multipartite
five-layered graph (Fig.~\ref{fig:network_graph_2017_and_degree_distributions}A)~\cite{Kivela2014}. The
first (most upstream) layer of the network corresponds to clients
(green), contracting registrants in the second layer (brown), who
employ lobbyists in the third layer (gray). Registrants have two primary types: either \textit{In-House} departments of the client
firm, or external lobbying firms (often metonymically called \textit{K-Street} firms, after the street in Washington, D.C.~where many are headquartered). We assign connections between the first three layers of the network by adding edges $c \to r$ and $r \to l$ if lobbyist $l$ lobbied
for client $c$ through registrant $r$.  The resulting representation is a trade-off: it is a useful first-order approximation (see Sec.~VI of~\cite{mm}), even as it obscures the richer polyadic character of lobbying interactions~\cite{Battiston2020,Bick2023}. The remaining two downstream layers represent the targets of the lobbying efforts: government entities (turquoise) and legislators
(members of Congress) (red/blue). The LDA does not require that registrants disclose direct contacts with legislators, so we infer these connections indirectly by leveraging past employment of lobbyists by legislators, which we make available through \textsf{LobbyView}. Over time, more and more lobbyists have documented these past connections. In
2023, $20\%$ of lobbyists had a known connection to the government (Sec.~III of~\cite{mm}). Moreover, in recent years $80\%$ of Senators and $60\%$ of
Representatives have at least one active lobbyist connected to them through past employment (Fig.~S9 of~\cite{mm}). This professional network is believed to play an important role in the lobbying process~\cite{Blanes2012,Shepherd2020}, though, of course, lobbyists' contacts are not limited to their prior employers~\cite{Carpenter1998}.

\begin{figure*}
    \centering
    \includegraphics[width=.75\textwidth]{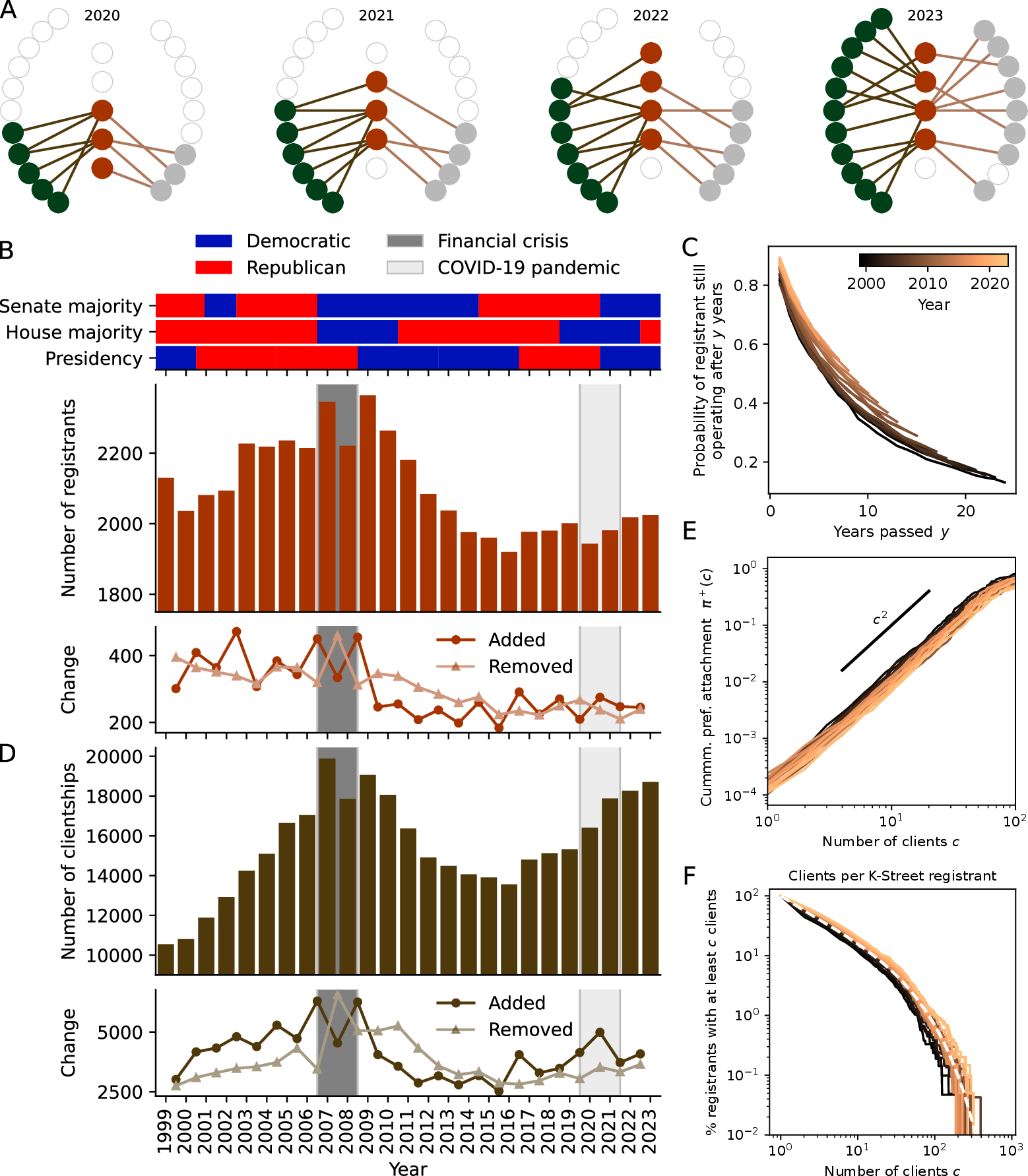}
    \caption{\textbf{The lobbying network is evolving and adapting in response to political landscape and crisis events.}  (A)~Attachment
      and detachment processes in a sample client-registrant-lobbyist
      (green-brown-gray) network component during 2020--2023 (for selection methodology
      see Sec.~VIII.D of~\cite{mm}).  (B)~Number of
      active K-Street registrants per year (bars) and their yearly
      increments (circles) and decrements (triangles).  The top
      horizontal bars show the party (Democratic in blue and
      Republican in red) of the Senate majority, House majority, and
      President by year.  The financial crisis (2007--2008) and
      COVID-19 pandemic (2019--2021) are shown by gray overlays.
      (C)~Number of active clientships (client-registrant
      connections) per year (bars) and their yearly increments
      (circles) and decrements (triangles).  (D)~K-Street
      registrant survival probability.  (E)~Cumulative attachment function $\pi^+(c)$ in terms of K-Street registrant in-degree. The quadratic functional form implies that the probability that a new link `selects' a registrant with $c$ clients is proportional to $c$. 
      (F)~Complementary cumulative distribution (CCDF) of the number of clients per K-Street registrant (K-Street registrant in-degree), 
      with a fitted cut-off power law probability mass function $p(c) \propto c^{-d}\exp(c /c_0)$, where $d = 1.46$ and $c_0 = 86$ (white dashed line). 
      Quantities
      in (D--F) are computed separately for each year and their precise mathematical definitions can be found in~\cite{mm}, Sec.~VIII.}
    \label{fig:network_evolution}
\end{figure*}

\section*{Long-term dynamical evolution and accumulated advantage}

\paragraph*{Ever-changing actors.} The lobbying network is highly dynamic, demonstrating continual evolution and adaptation to changes in the national and geopolitical environment. Because our database encompasses lobbying reports from 1999 to present, it offers a unique opportunity for longitudinal analysis of such processes~\cite{Kossinets2006,Newman2001,Jeong2003,Wang2024}. Focusing on the K-Street registrants (firms that at least once lobbied on behalf of another client), we observe an annual influx of 200--500 new registrants into the network, alongside a comparable number exiting ones (Fig.~\ref{fig:network_evolution}B). 
As the set of registrants regenerates and rejuvenates continuously, the network drifts over time. For instance,  comparing the set of registrants in
2000 and 2020, we find that only 20\% of the registrants from 2000 are
still active (Fig.~\ref{fig:network_evolution}C). 

\paragraph*{Preferential attachment and detachment.} Even registrants who remain in operation gain and lose clients (Fig.~\ref{fig:network_evolution}C). Every year, some clients
leave the lobbying network, new clients join, and the remaining
clients establish relations with new registrants. The scale of this
\textit{rewiring} process is quite substantial, as it involves up to a
third of the clientship connections each year. 
The \textsf{LobbyView} data suggest that the rewiring does not occur at random, but instead the measured probability of a registrant attracting a new client is proportional to the number of its existing clients~(Fig.~\ref{fig:network_evolution}E)~\cite{Eriksen2001,Newman2001,Wang2024}. 
In other words, we find evidence of linear \textit{preferential attachment}. 
Preferential attachment~\cite{Barabasi1999}, also known as the Matthew principle~\cite{price1976general}, describes a network growth mechanism in which newly formed links are statistically more likely to attach to nodes with higher degree.
It has been previously observed in other socio-economic networks~\cite{Barabasi1999,Stumpf2012,Schweitzer2009}, and is known to provide a mechanism for generating heavy-tailed degree distributions~\cite{Barabasi1999,Bianconi2001,Courtney2018,bassetti2009statistical}.
Consistent with these findings, the distribution of clients per K-Street registrant is also heavy-tailed, and it can be approximated by a power law with exponent $d \approx 1.5$, and a cut-off $c_0 \approx 85$, (Figure~\ref{fig:network_evolution}F). 
Practically, the heavy tail distribution implies the existence of a small group of elite registrants dominating the lobbying market.
Quantitatively, we find that about 80\% of K-Street income is generated by only 20\% of
registrants (Sec.~V of~\cite{mm}). 
\par
In summary, by analyzing the dynamical evolution of the client-registrant ties, we found evidence of preferential attachment, which offers a plausible explanation for the hierarchical structure of K-Street lobbying. 
We note, however, that preferential attachment is not the only principle governing the evolution of the lobbying network; for example,  the detachment
probability is also proportional to the number of existing clients
(Fig.~S24 of~\cite{mm}).
The preferential attachment and preferential detachment act as countervailing forces, which appear to stabilize the statistical equilibrium of the network~\cite{Saavedra2008,Wang2024}. 
Indeed, despite significant churn in the membership of core political actors involved in lobbying, the global statistics of the lobbying network (layer-to-layer degree distributions) have not changed in any substantial way in the past 25 years.


\begin{figure*}
    \centering
\includegraphics[width=.95\textwidth]{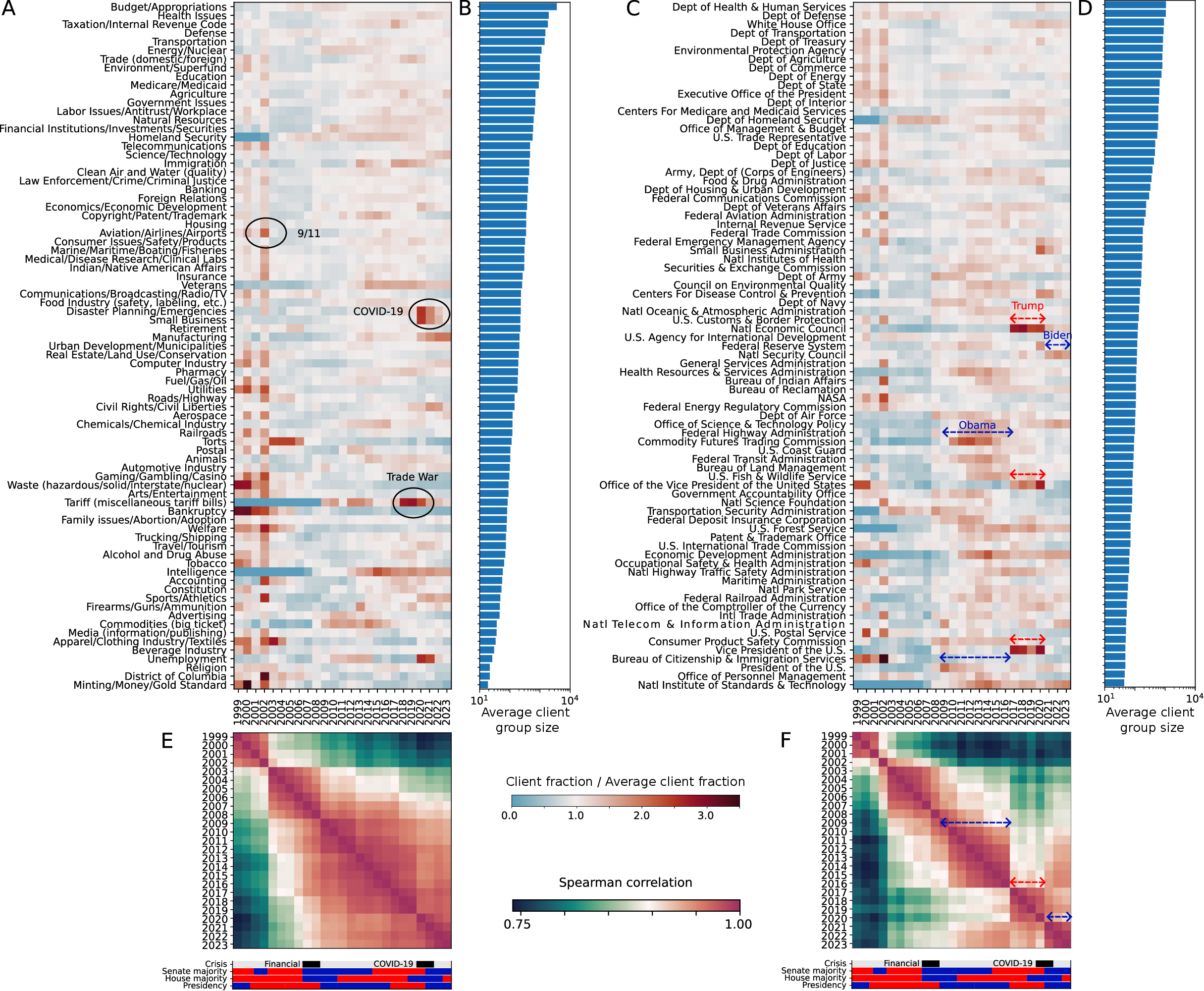}
    \caption{\textbf{Our data enables fine-grained analysis of the lobbying network adaptations.}     (A)~Fraction of clients lobbying on a given issue, scaled by its average value in all years considered.
    Some of the spikes of interest can be easily associated with crisis events, such as the terrorist attacks of 9/11 or the COVID-19 pandemic. 
    (B)~Average number of clients lobbying on a given issue. 
    (C) Fraction of clients approaching a given government entity in a given year, scaled by the average value in the period considered.
    The government entities are listed according to the average number of targeting clients.
    In this Figure, we present only the most-approached government entities; a more extensive list can be found in~\cite{mm}, Sec.~XI.
    Some of them, such as ``President of the United States'' (direct lobby) and ``Executive Office of the President'' (lobby with President's staff), are closely related. 
    Nevertheless, we retain the reported format in order to preserve these subtle distinctions.
    (D)~Average number of clients approaching different government entities. 
    (E)~Year-to-year correlation of the \textit{issue portfolio vectors} (columns of the matrix in panel (A)).
    (F)~Year-to-year correlation of the \textit{government approach portfolios} presented as columns in panel (C). The correlation matrix has a manifest block-diagonal structure, synchronous with the presidential terms. 
    }
    \label{fig:heatmaps}
\end{figure*}

\section*{Short-term adaptation driven by the political landscape }

\par
The lobbying network is not a closed ecosystem, but it responds and adapts to national and geopolitical events and changes. For instance, Fig.~\ref{fig:network_evolution}B clearly shows a decline in the number of active lobbying firms after 2009, persistent for much of the subsequent decade.
Plausible causes of the decline include the global financial crisis of 2007--2008, and the enactment of HLOGA in 2007, which restricted certain forms of \textit{revolving--door} lobbying \citep{straushloga, revolvingdoor}. 
As expected, firms open, close, and adapt their activities in response to events. 
We now explore the general characteristics of these adaptations.

\paragraph*{Lobbying subnetworks: robust vs.~volatile interests.} 
The LDA requires that lobbying reports tie each activity to one of the 79 general issue areas~\cite{LDA1995regs}.
Using this information, we can reconstruct \textit{issue subnetworks}, differing in size and connectivity. 
Issues representing core government activities, such as taxation and healthcare, are consistently lobbied by a large number of clients. 
Less popular issues fluctuate in popularity according to the vicissitudes of current controversies or policy initiatives. 
For instance, we observe substantial spikes in lobbying activity concerning disaster planning and small businesses, ordinarily fairly uncommon lobbying issues, during the early period of the COVID-19 pandemic (Fig.~\ref{fig:heatmaps}A-B).
In general, by comparing the size of interest groups across issues and years, we can quantify the evolution of the \textit{lobbying issue portfolio}. 

Complementary information about the adaptations of the lobbying network can be obtained by analyzing \textit{target subnetworks} and \textit{lobbying target portfolios}. 
The targets in this case correspond to the various government agencies approached by lobbyists~(Fig.~\ref{fig:heatmaps}C-D). 
Large federal executive departments, such as the Department of Health and Human Services, are consistently lobbied by many clients every year, so the size of the corresponding subnetwork remains relatively stable. 
In contrast, smaller target subnetworks experience substantial fluctuations in size.
Some entities, including the National Economic Council, display a periodic pattern, synchronized with the presidential election cycle, which economists call the \textit{political business cycle}~\citep{nordhaus, drazen, schultz}.

Quantitatively, the lobbying issue portfolio and lobbying target portfolio can be represented as feature vectors: columns of the matrix in
Fig.~\ref{fig:heatmaps}A and Fig.~\ref{fig:heatmaps}C, respectively. 
To characterize the issue/target adaptations in a systematic manner, we compute year-to-year Spearman correlation matrices. 
This analysis reveals that the lobbying issue portfolio is responsive to crisis events (Fig.~\ref{fig:heatmaps}E), and the lobbying target portfolio reflects changes in government (Fig.~\ref{fig:heatmaps}F).
The issue portfolio underwent a major rearrangement following the 9/11 terrorist attacks and during the COVID-19 pandemic. 
On the other hand, the lobbying target correlation matrix exhibits a distinct block-diagonal structure, which synchronizes with the election cycle.
Thus, it appears that lobbyists tend to approach the same entities with consistency during the course of a single presidential administration, but adapt their lobbying efforts quickly when changes in leadership occur. 
This adaptation underscores the lobbying network's sensitivity to the differing views of the two major political parties in the U.S~regarding the scope and responsibilities of government departments and agencies.

In sum, while the global structure of the lobbying network remains relatively stable, the content and routing of information flow can quickly adapt to new issues and political contexts. 
In the next section, we will show how the \textsf{LobbyView} data can be used to reveal and examine lobbying pathways at high resolution.


\begin{figure*}
    \centering
    \includegraphics[width=.75\textwidth]{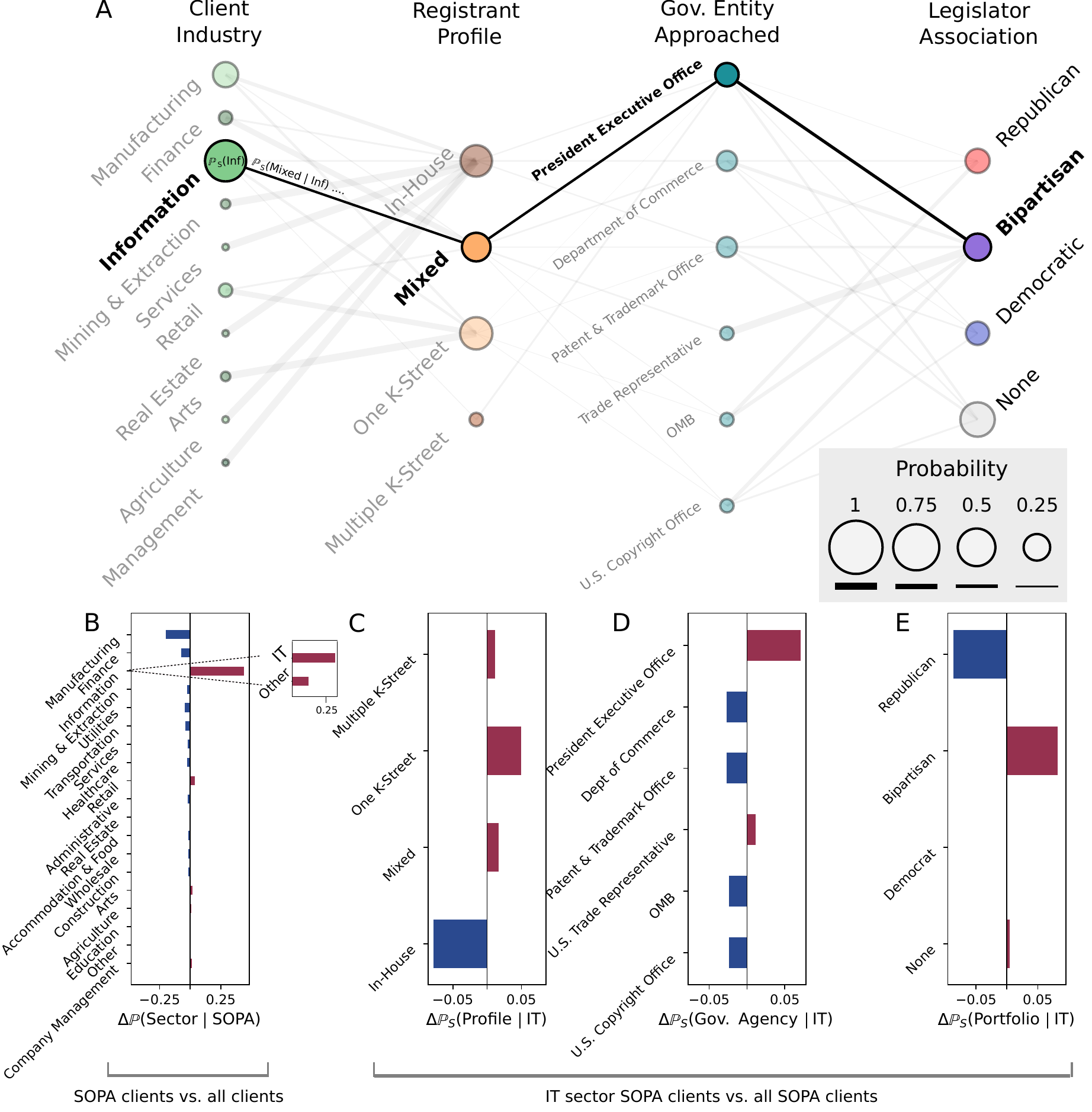}
    \caption{\textbf{The \textsf{LobbyView} database enables granular analysis of the lobbying dynamics.} 
    As an example, in this figure, we analyze lobbying surrounding the Stop Online Piracy Act (SOPA). 
    We restrict our attention to the lobbying strategies of the 67 publicly-traded U.S.~firms that lobbied on SOPA in the last quarter of 2011 when the bill was introduced.
    (A)~Technically, our analysis amounts to evaluating the probability of different lobbying pathways.
    The pathway highlighted in this diagram corresponds to an information sector company, lobbying through both In-House and K-Street firms (mixed profile), which approached the Executive Office of the President (EOP), and whose lobbyists had connections to both of the major parties (bipartisan portfolio). 
    More generally, this diagram can be understood as a coarse-grained version of the lobbying network in Fig.~\ref{fig:network_graph_2017_and_degree_distributions}B, with nodes corresponding to different compound random variables. 
    The node sizes represent different marginal probabilities, and the edge thickness represents conditional probabilities.
    (B) SOPA-specific shifts in client sector probability. 
    The sectors correspond to 2-digit NAICS codes. 
    Positive probability shift (most prominent for the information sector, containing IT companies, as well as publishers, broadcasters and motion picture industry) indicates elevated interest in SOPA.
    (C) IT-specific probability shifts in \textit{registrant profile} (in-house lobbying department only, both in-house and a K-Street firm (mixed), one K-Street firm, or multiple K-Street firms). 
    (D) IT-specific shifts in the probability of approaching different government entities. 
    Note that multiple government entities can be approached at the same time, so the events in this layer are not mutually exclusive.  
    For brevity, in this figure, we only show the 6 most often mentioned government entities.    
    (E) IT-specific shifts in the probability of having an association with the current legislator of a particular party (Democratic, Republican, both, or none). In panels C-E, we restrict our analysis to the SOPA lobby probability distribution $\mathbb{P}_S$. Full details of the probabilistic analysis are presented in the SI, Section XII.
 }
    \label{fig:bill_lobby}
\end{figure*}

\section*{Uncovering probable lobbying pathways}

Fundamentally, \textsf{LobbyView} defines a joint probability distribution over a high-dimensional space, with each lobbying report corresponding to one elementary event.
Graphically, each such elementary event can be represented as a pathway through the lobbying network linking all the agents involved in the report: client, registrant, lobbyists, government entities, and legislators (c.f. Fig.~\ref{fig:network_graph_2017_and_degree_distributions}B), with additional information such as issue code or lobbied bills, provided in an annotation. 
Under a random null model, any pathway and any annotation would be equally likely, but the empirical lobbying network is markedly more structured. 
\par
By comparing the empirical probability of different elementary pathways, we can conduct detailed case studies, e.g., tracking lobbying activities of a specific company, or the career trajectory of a particular lobbyist.
We can also aggregate elementary events into compound pathways and investigate increasingly general aspects of the lobbying dynamics (Fig.~\ref{fig:bill_lobby}A). 
We will now illustrate this approach through a case study on industry lobbying related to the Stop Online Piracy Act (SOPA) of 2011.
\par
SOPA was a bipartisan bill, introduced in the House of Representatives in October 2011, aimed at combating copyright infringement. 
It was broadly supported by entertainment and media companies, but equally opposed by information technology (IT) firms, who viewed the bill as exposing them to liability and chilling user expressions of free speech. 
The bill's eventual legislative failure was deemed by contemporary media analyses as surprising, and attributed to concerted efforts by information technology companies to lobby and generate public backlash~\citep{nytsopa}. 
Data from \textsf{LobbyView} confirms that IT companies indeed lobbied SOPA in a major way, and they allow us to investigate their lobbying strategies in depth.
\par 
To quantify the involvement of different industry sectors in SOPA lobby, we analyze the probability space of lobbying pathways pursued by publicly-traded firms in the last quarter of 2011. 
We are able to assign clients to industry sectors because \textsf{LobbyView} provides bindings to S\&P Compustat~\cite{compustat} -- an external corporate database that contains information about the North American Industry Classification System (NAICS) code of corporate clients.
If we randomly select a lobbying client $C$, the probability that $C$ belongs to the IT sector  $\mathbb{P}(C \in \text{Inf})$ is about 0.1, reflecting the general structure of the lobbying market. 
However, the conditional probability that $C$ belongs to the information sector, given that this client mentioned SOPA in one of their reports $\mathbb{P}(C \in \text{IT}\,|\, \text{SOPA})$ is more than 0.4.
The substantial \textit{probability shift} 
\[ \Delta \mathbb{P}(\text{IT}\,|\, \text{SOPA}) = \mathbb{P}(C \in \text{IT}\,|\, \text{SOPA}) - \mathbb{P}(C \in \text{IT}) \]confirms that the information industry indeed took particular interest in SOPA when it was first introduced~(inset of Fig.~\ref{fig:bill_lobby}B).\\

This basic computation is a starting point for answering more detailed questions about lobbying strategies. 
Which registrants did the information companies hire? 
Which government entities or agencies did they approach? 
Do they rely on connections between lobbyists and legislators? 
All of these questions can be answered by computing appropriate information-specific probability shifts.  
Thus, we discover that IT firms were more likely to hire multiple external K-street lobbying firms (Fig.~\ref{fig:bill_lobby}C), more likely to contact the Executive Office of the President (Fig.~\ref{fig:bill_lobby}D), and more likely to assemble lobbying teams with pre-existing connections to both the Democratic and Republican parties (Fig.~\ref{fig:bill_lobby}E). 
These findings reveal that the IT industry rolled out a broader lobbying campaign compared to other companies interested in SOPA, supporting the hypothesis that a diverse lobbying portfolio can be conducive to legislative success.

\par
In SI Section XII, we apply the same Bayesian analysis to additional bill-lobbying case studies, demonstrating its potential to elucidate connections between lobbying and legislation.
As the relational structure of the \textsf{LobbyView} database enables rapid calculation of lobbying pathway likelihood,  similar probabilistic methodologies~\cite{pearl1988probabilistic,kindermann1980markov} can be deployed to investigate a wide range of topics, from partisan polarization~\citep{Fiorina2017} (c.f. Sec XIII of~ \cite{mm}) to revolving door effects~\citep{straushloga, revolvingdoor} in lobbying, quantitatively. \\

\section*{Discussion}

We introduced \textsf{LobbyView}, a high-dimensional dataset that uses Lobbying Disclosure Act filings to construct a dynamical network enabling quantitative end-to-end analysis of the U.S.~lobbying system. To understand the interactions between the various political stakeholders, we represented the lobbying system as a layered network that transforms competing interests (left-most layer) into lobbying pressure (subsequent layers). We illustrated the potential of \textsf{LobbyView} as a research resource that allows us to investigate lobbying dynamics at different levels of granularity.
At the global level, we computed key characteristics of the lobbying network, including degree distributions, dynamic evolution patterns, and adaptations triggered by political events.
Our analysis revealed that preferential attachment is a significant mechanism in the long-term evolution of connections, the dominant role of the election cycle for the selection of lobbying targets, and the adaptability of information flow in response to critical events.
We also introduced a template for fine-scale probabilistic analysis of lobbying dynamics, using lobbying strategies for a specific bill as an illustrative example.
The probabilistic formulation provides a unified framework that can be used to explore different aspects of the lobbying dynamics.
Interesting avenues for future research include also the role of higher-order (Sec.~VI of the SI) and multilayer interactions~\cite{Battiston2020,Bick2023,Kivela2014}, as well as the multiscale community structure of the lobbying network~\cite{Delvenne2010,Mucha2010}. \textsf{LobbyView}~\cite{LV:link} is continuously updated, and we hope it will drive a comprehensive evidence-based reflection on interest politics that can increase public awareness and help improve the democratic process. 
A collective interdisciplinary effort across and beyond the academic community will be essential for achieving this goal. 


\section*{Acknowledgments}

This research received support through Schmidt Sciences, LLC (to J.D.), the MathWorks Professorship Fund (to J.D.), the National Science Foundation (SES-1725235, SES-2017315) (to I.K.), and the Russell Sage Foundation (\# 1908-17912) (to I.K.).  I.K.~and J.D.~are grateful for seed grant support from the  Social and Ethical Responsibilities of Computing (SERC) Initiative of the MIT Schwarzman College of Computing, and from the ICSR Seed Fund of the MIT Institute for Data, Systems, and Society.

\bibliography{scibib}

\end{document}



\title{Supplementary Information for\\
\lq Measuring the dynamical evolution of the United States lobbying network\rq}
\author{Karol A. Bacik}

\author{Jan Ondras}

\author{Aaron Rudkin}

\author{J\"orn Dunkel}

\author{In Song Kim}

\maketitle
\tableofcontents
\normalsize
\clearpage

\setcounter{page}{1}

\section{Data sources and processing}
\label{sec:sm:sources}

In this Section, we describe the data processing procedure, from a lobbying disclosure report to the lobbying network. 
We start by describing the nature of the data and the general features of the \textsf{LobbyView} database, which contains plenty of additional data that we do not analyze here. 
We then specialize to the subset of data that we use in this article and describe its structure more formally.

\subsection{Lobbying report structure}
\label{sec:sm:report_structure}

The origin of the lobbying disclosure data included in \textsf{LobbyView}~\cite{LV:link} and referred to in this article are disclosures made by registrants, and legally required under the Lobbying Disclosure Act of 1995~\cite{LDA1995}.  Under current federal regulations, registrants file three standard disclosure forms to maintain compliance with the act, labeled LD-1, LD-2, and LD-203~\cite{LDA1995regs}. Form LD-1 contains the initial registration of a registrant or the initial registration of a new client for the registrant. Form LD-2 is the standard periodic report that forms the basis for most of our analysis. Form LD-203 discloses direct financial contributions from registered lobbyists. LD-2 filings contain a variety of data, including the details of the registrant (including unique identification numbers for the U.S.~House and U.S.~Senate), the client's name, general report metadata, and specific lobbying income or expenses. Registrants are subsequently asked to report details of their activity. Each section of an LD-2 report reports activity for one of 79 ``issue codes'' during the report period. Registrants must identify the issue code, the government entities contacted, the specific details of the lobbying activity, and the lobbyists assigned to this issue in the time period. The specific details are provided as a free text block. There is no particular format for this text block, which means analyzing these records requires careful extraction and structuring of the text. The content of these records often includes a general description of the policy being lobbied (e.g., ``Corporate Tax Reform'', ``Accessibility'', ``Open Internet''), and/or a list of specific bills (e.g., ``S. 698 - Marketplace Fairness Act'', ``H.R. 2315 Mobile Workforce State Income Tax Simplification Act''). Filings for 2008--present are natively filed digitally, so the information can be directly extracted as text. Filings from 1999--2007 were filed on paper, and in some cases handwritten, so extraction of data requires text recognition methods (OCR). To extract computerized text, we make use of regular expressions and other methods, which we describe further below. Lobbyists specified as part of an LD-2 section supply two pieces of information: their name, and if changed since last reporting, any prior employment connection they had to ``covered entities'' (executive or legislative branch entities).

\begin{figure}[h]%
    \captionsetup[subfigure]{labelformat=empty} 
    \centering
    \subfloat[Shell H1 2004]{{\includegraphics[width=0.47\textwidth]{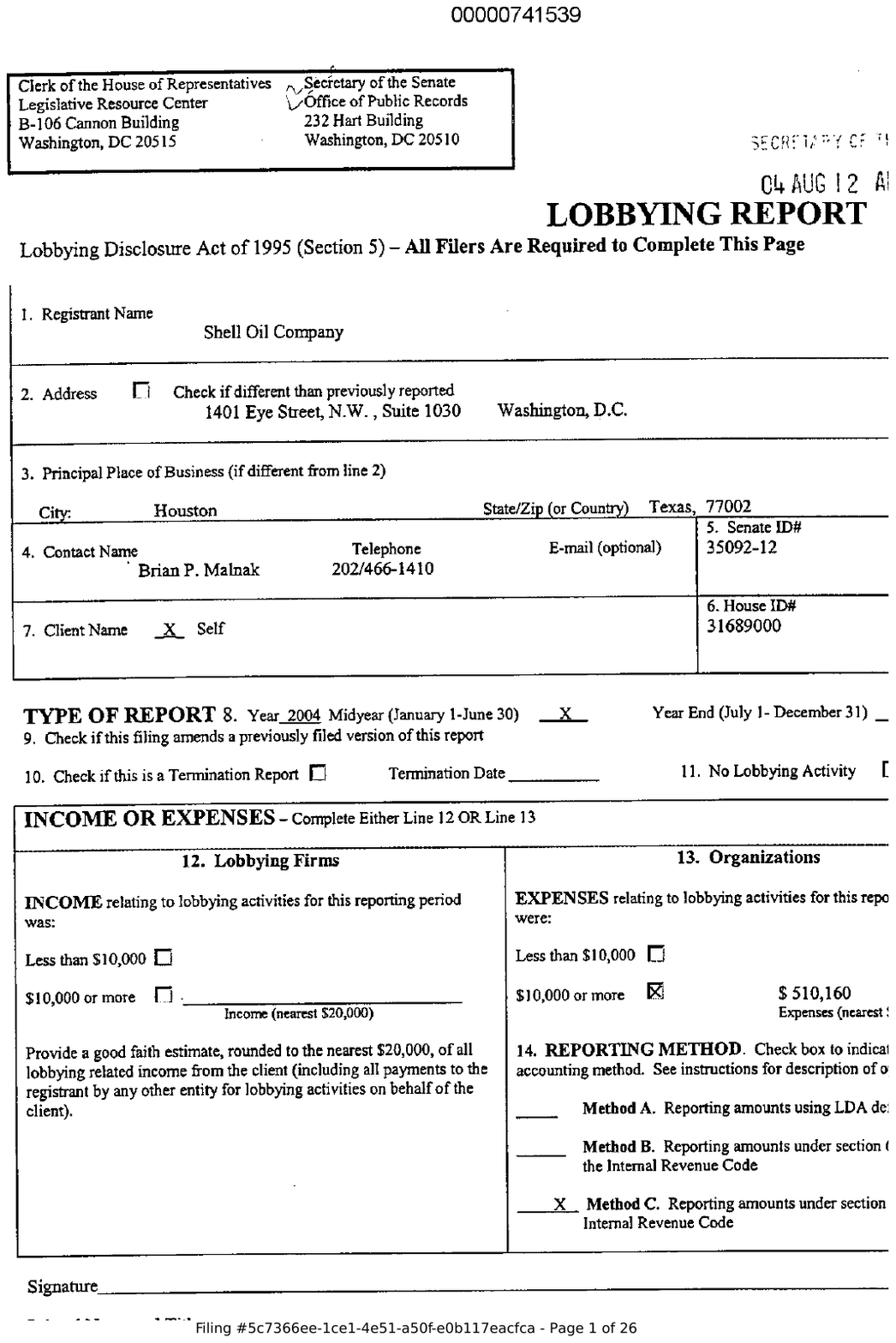} }}%
    \qquad
    \subfloat[Boeing Q3 2022]{{\includegraphics[width=0.47\textwidth]{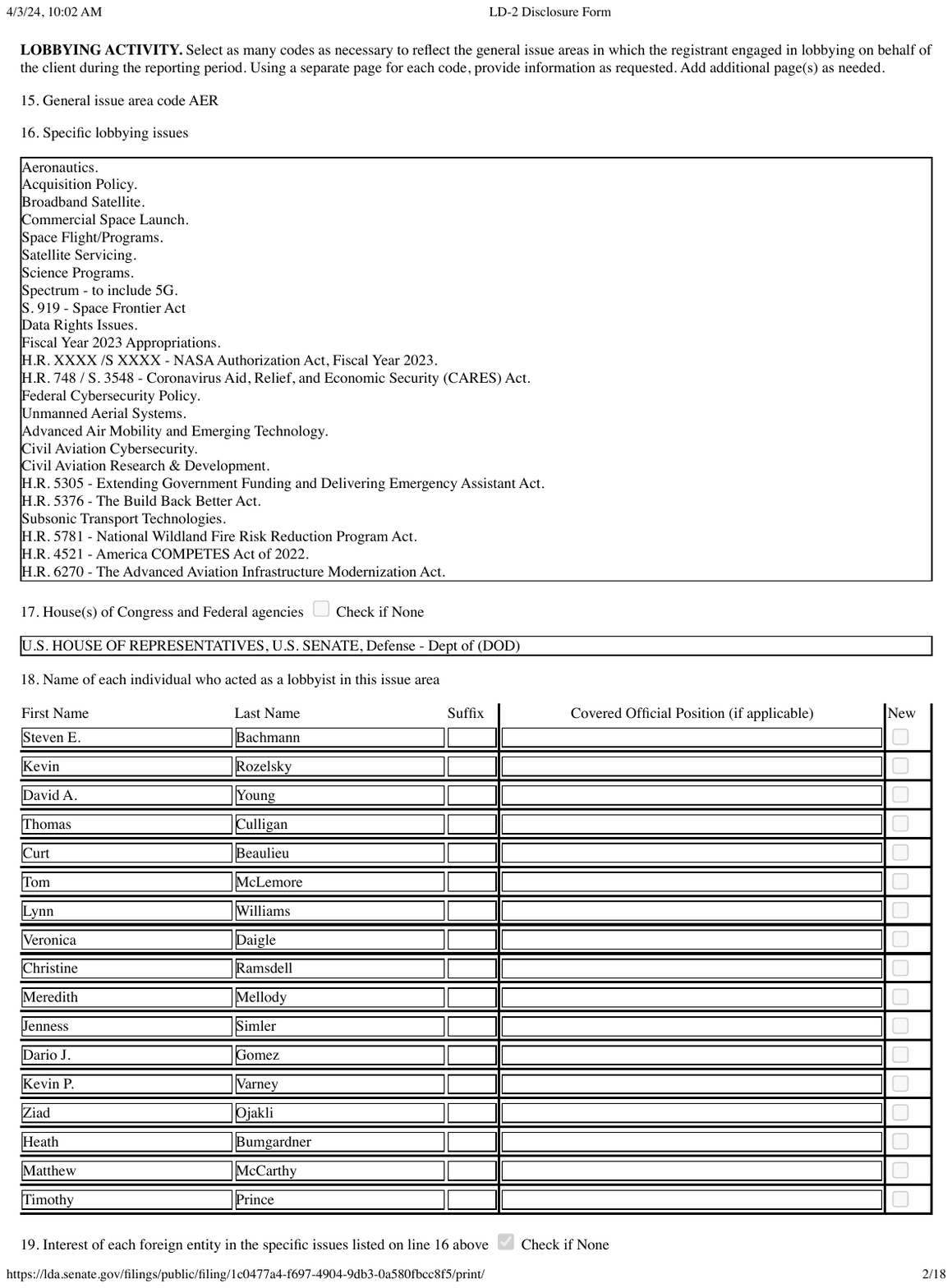} }}%
    \caption{Sample pages from LDA filings.
    \textbf{(Left)}~Filing from Shell Oil for the H1 2004 period. It is a digitized paper copy and depicts the registrant and client information as well as the total monetary expenditure for the period. Shell is filing as a self-filer, reporting lobbying activity done by lobbyists directly employed by Shell on its behalf. 
    \textbf{(Right)}~The right page is a filing from Boeing for the Q3 2022 period. This page depicts disclosure for a particular issue: Boeing is lobbying Aerospace issues.  This page also reports the names of Lobbyists employed by Boeing to lobby for Aerospace issues.}%
    \label{fig:sm:sample_lobbying_report}%
\end{figure}

Below, in Fig.~\ref{fig:sm:sample_lobbying_report}, we provide examples of individual pages taken from two real-world LDA filings. The leftmost page is a filing from Shell Oil for the first half of 2004 (supplied as a digitized paper copy). This depicts the basic registrant, client, and filing data. Here, Shell is acting as an in-house registrant or self-filer (as opposed to a registrant representing an external client). The rightmost page is a filing from Boeing for the third quarter of 2022 (as we describe below, the filing requirements changed from being biannual to quarterly as the LDA was ultimately amended). This page depicts an issue disclosure: Boeing has lobbied a number of bills and general issues under the AER (Aerospace) issue code. The variation in textual representation and the presence of incomplete metadata (e.g., describing the annual NASA Authorization Act as `H.R. XXXX/S. XXXX') depicts typical difficulties associated with processing data -- at the time the lobbying was occurring, Boeing did not have detailed information about what bill numbers would be assigned to recurring appropriation legislation in the given session. Boeing reports the names of many lobbyists who worked on this issue, none of whom have new covered position data to report.

\subsection{\textsf{LobbyView} database}

The \textsf{LobbyView} database is a large, frequently updated, relational database that consists of three major components: (1) all available lobbying reports, registrants, lobbyists (including lobbyist donation activity), and clients. For each of these groups, we retain raw data but also produce cleaned, de-duplicated, standardized versions of the data which add unique identifiers to lobbyist and client entities, as well as providing commonly used firm identifiers necessary for researchers to attach \textsf{LobbyView} data to external business datasets including S\&P CompuStat~\cite{compustat} and Moody's Orbis~\cite{orbis}; (2) all political donations made by members of the public, which can be linked to firms via firm-level identifiers and donation `employer' fields (we omit a more detailed treatment of donation data because it is not discussed in this paper); and (3) a complete database of all U.S.~Congress legislative activity, legislators, and committee assignments which can be linked to lobbying activity through bill identifiers and bill-committee assignments. 

In this section, we describe the general approach to gathering, ingesting, and storing this data. We provide more details about the ingestion and cleaning of specific portions of the data as they are discussed below in the context of the data used in this paper.

\textsf{LobbyView} begins by gathering data from public and private sources. Lobbying data is gathered from the U.S.~Senate Lobbying Disclosure website \href{https://lda.senate.gov/}{(lda.senate.gov)}. The U.S.~Senate makes available a data API, which we query exhaustively before ingesting. Reports include unique identifiers for registrants, which we use internally. Because clients and lobbyists are identified only by a textual representation, we need to disambiguate (resolve) them to canonical entities. In the relevant sections below, we describe our efforts to disambiguate clients and lobbyists. We extract issue text descriptions from these filings using regular expressions in order to identify specific legislative bills. We match registrants to associated private sector firms (using CompuStat and Orbis data). We then download relevant congressional data from the United States digital service~\cite{usdigitalservice_congress}, including a complete record of legislators who have served in the U.S.~Congress, their committee assignments, bill sponsorship and cosponsorship data, and bill histories. Thus, we are able to link a corporate firm to a lobby registrant to an LDA report to a bill debated by Congress to the committee that debated it to the legislators on that committee at the time, allowing for some of the rich data flows illustrated in this paper. We use Python to automate downloading the data, which is ingested into a PostgreSQL relational database and indexed appropriately. 

The data in \textsf{LobbyView} are made available to researchers in a variety of forms -- the access options are described in Section~\ref{sec:data_availability}.

\subsection{Lobbying data}
\label{sec:sm:lobbying_data}

We will now describe more precisely the subset of the \textsf{LobbyView} data that we analyze in this study. 
Here, we use only the LD-2 reports (see Sec.~\ref{sec:sm:report_structure}), and we exclude the reports that indicate 'no lobbying activity' in line 11, as well as the reports that are superseded by updated filings, based on line 9 of LD-2 filings.\footnote{When we refer to specific lines of the LD-2 form, the reader is welcome to consult Fig.~\ref{fig:sm:sample_lobbying_report}.}
Each lobbying report that we analyze is assigned a unique filing ID $f$, as well as nine other features of interest:
\begin{enumerate}
    \item Filing year $y(f) \in \{1999,2000,\dots,2023 \}$,
    \item Self-filing flag $s(f) \in \{ \text{True}, \text{False} \}$, 
    \item Monetary amount $m(f) \in \mathbb{R}^+$,
    \item Client ID $c(f) \in \{c_1,c_2,\dots \}$, 
    \item Registrant ID $r(f) \in \{r_1,r_2,\dots \} $, 
    \item Set of lobbyist IDs $L(f) \subseteq \{l_1,l_2,\dots \} $,
    \item Set of approached government entities (House(s) of Congress and Federal agencies) $G(f) \subseteq \{g_1,g_2,\dots \} $ 
    \item Set of general issue areas $A(f) \subseteq \{a_1,a_2,\dots, a_{79} \} $ (c.f.~Fig.~4 of the main text).
    \item Set of bills $B(f)$ 
\end{enumerate}

Combined into a tuple,
\begin{equation}
    \mathcal{LD}(f) = ( f,~y(f),~s(f),~m(f),~c(f),~r(f),~L(f),~G(f),~A(f)),
\label{eq:ld_def}
\end{equation}

they form a \textit{lobbying datum}, which is the smallest atom of our \textit{lobbying data}. We will now describe the basic characteristics and technical challenges associated with each one of the eight features. The data provided to support replicating this study anonymizes filing IDs, client IDs, registrant IDs, and lobbyist IDs.

\subsubsection*{Filing year}
\begin{figure}[h]
    \centering
    \includegraphics[width = .75\textwidth]{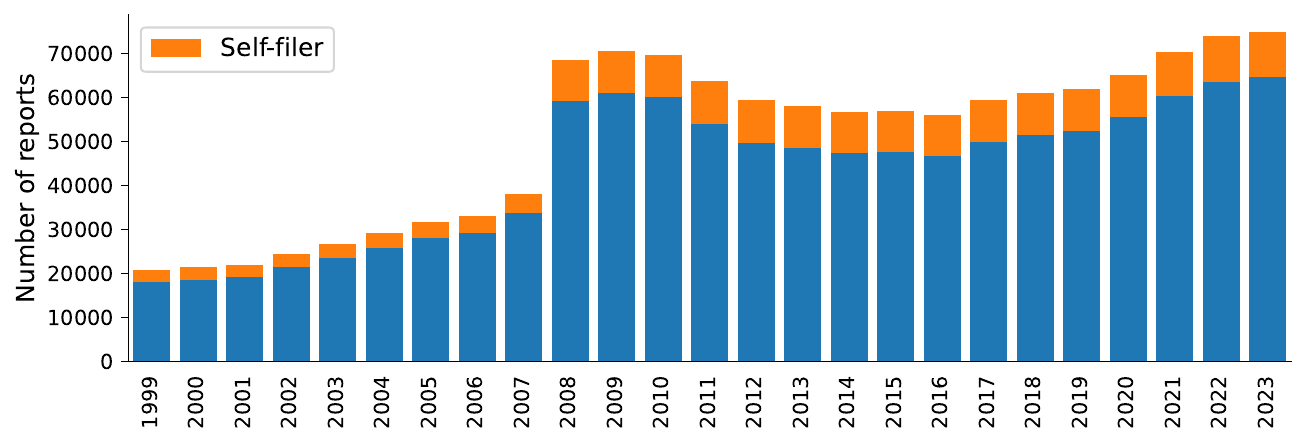}
    \caption{Number of lobbying reports used in our analysis. Most of the reports are filed by lobbying firms lobbying on behalf of a client (as opposed to self-filers). Note that the quantity of filings before 2008 is not directly comparable to those since 2008 due to a legislative change in reporting frequency. In total, we include in our analysis 1,277,411 filings.}
    \label{fig:sm:filings_per_year}
\end{figure}

Reports analyzed in this study cover the period 1999--2023. The LDA was signed into law on December 19, 1995, but first came into force for the calendar year 1999.
Figure~\ref{fig:sm:filings_per_year} presents the number of reports considered in our analysis for each year $y \in \{1999,2000,\dots,2023 \}$.
We immediately notice an abrupt increase in the number of reports in 2008. In 2007, Congress passed the Honest Leadership and Open Government Act~\cite{HLOGAct}, which, among other provisions, altered the mandatory reporting frequency provisions of the LDA 1995. Prior to 2008, reports were submitted biannually (twice a year). Since 2008, reports have been required each quarter. This is reflected in our data, where the number of reports doubled beginning in 2008. Filing and clerical errors (as well as late filings) during the transition period contribute to a small number of reports that have invalid filing period codes (e.g., reporting for Q2 in 2006 or for H2 in 2009). In order to unify our data across this policy transition, we aggregate the periodic data into annualized data, thus discarding the exact filing code and considering only the reporting year $y$ (line 8 of LD-2 form).

\subsubsection*{Self-filing}
Line 7 of LD-2 provides information on whether the filing pertains to an organization lobbying on its own behalf (self-filer flag $s(f) = \text{True}$) or a lobbying firm lobbying on behalf of a different client (self-filer flag $s(f) = \text{False}$). Figure~\ref{fig:sm:filings_per_year} shows that most of the reports are filed on behalf of someone else. 

\subsubsection*{Monetary amount}
\label{sec:sm:amount}
If the lobbying costs exceed a certain threshold (Table~\ref{tab:sm:min_reporting_amounts}), the lobbying report must report a dollar value (lines 12 and 13 of LD-2 filings). 
This field differs depending on the identity of the filer. Registrants who are self-filers (for whom $s(f) = \text{True}$) report the \textit{cost} of their lobbying activities (line 13 of LD-2 filings), while registrants lobbying on behalf of a client ($s(f) = \text{False}$) report the \textit{income} they received for lobbying (line 12 of LD-2 filings). We expect that firms book income that meets or exceeds the actual costs of their activities. Additionally, self-filers are subject to a higher numeric threshold for reporting lobbying costs than non-self-filers are for reporting income (Tab.~\ref{tab:sm:min_reporting_amounts}).

\begin{table}[t]
    \centering
    \caption{Lobbying activity minimum reporting amount 
    in terms of the year and registrant type.
    Extracted from~\cite{LDA1995,LDAamend1998,LDAguideAmend2008,HLOGAct}.
    }
    \begin{tabular}{lcc}
    \toprule
    Reporting year $y$ & Not a self-filer & Self-filer \\
    \midrule
    $<2008$ & \$5,000.00 & \$10,000.00 \\
    $ \geq 2008$   & \$3,000.00 & \$5,000.00 \\
    \bottomrule
    \end{tabular}
    \label{tab:sm:min_reporting_amounts}
\end{table}

\begin{figure}[h]
    \centering
    \includegraphics[width = .75\textwidth]{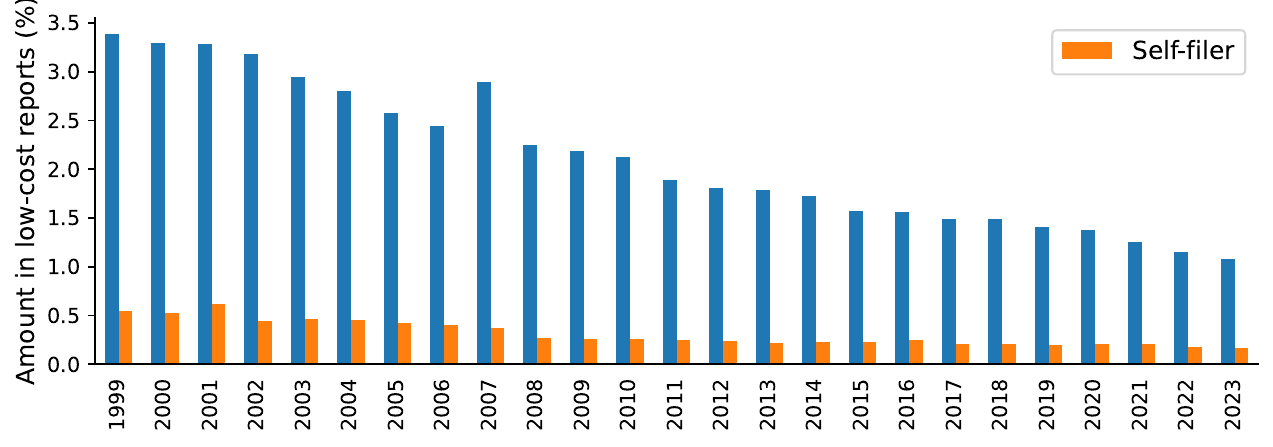}
    \caption{The upper bound of the total monetary amount associated with the `low-cost reports', i.e., the reports where the lobbying income/expenditure falls below the threshold value. Note a steady decrease in the proportion of the lobbying market taken by the low-cost activities.}
    \label{fig:sm:amount_low_cost}
\end{figure}

To estimate the money flow in lobbying, we neglect some of these subtleties, and we attach a single monetary amount $m(f)$ to each filing. 
For most filings, $m(f)$ is simply the amount reported in either line 12 or line 13. 
If the amount is below the reporting threshold  (listed in Table~\ref{tab:sm:min_reporting_amounts}), we set $m(f)$ to this threshold value (for low-cost lobbying, this is an upper bound of the actual unknown value).
Figure~\ref{fig:sm:amount_low_cost} compares the amount of this low-cost lobbying,
\[
\sum_f \mathbbm{1}\left[ m(f) \text{~bounded heuristically} \land y(f) = y \right] m(f)
\]
to the total amount and shows that this lobbying accounts for a very small percentage of our data (in part because lobbying firms would prefer to only report when they are legally compelled to do so, and in part because lobbying is expensive). We do not believe that this heuristic adjustment has any material impact on our results.
Fundamentally, there are two main cases where activity would not meet this threshold: the first is that the activity is highly scope-limited or highly informal, or both.
The LDA does not capture this type of advocacy, which is one of the scope conditions of our analysis.
The second case is one where a registrant-client pair does have an ongoing, formal, and in-scope lobbying arrangement, but there was simply not enough business in a given time period. Once registered, registrants must report or terminate their registration, and we think the option of reporting (but checking ``no activity'' or ``activity below reporting threshold'' as appropriate) is more realistic than repeatedly terminating and re-registering according to ebbs in activity.

\begin{figure}[t]
    \centering
    \includegraphics[width = .75\textwidth]{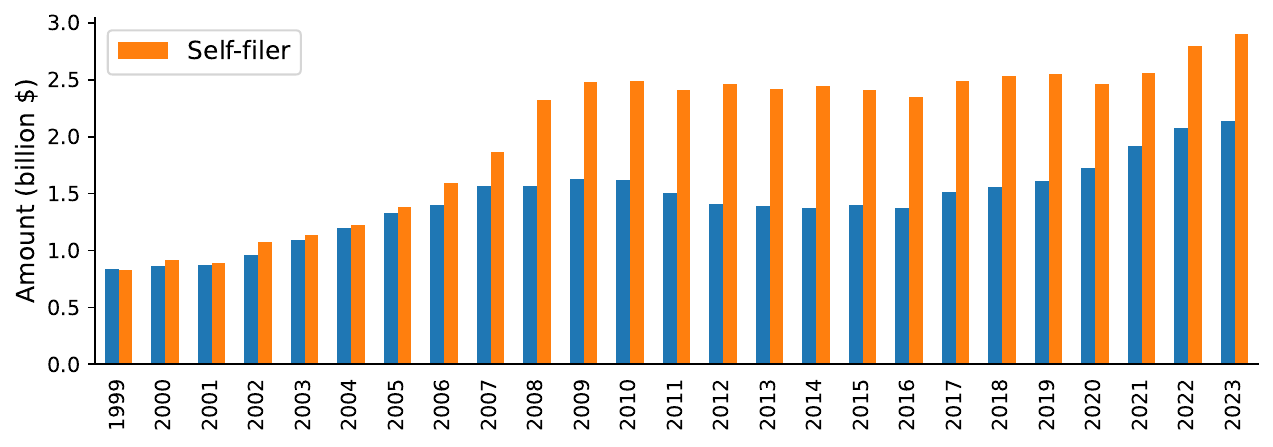}
    \caption{Total lobbying amount for all the reports in a given year. Due to a different definition of the lobbying amount, the figures for self-filers and not-self-filers are not exactly comparable.
    The total amount reported in these reports is \$87.5 billion.}
    \label{fig:sm:amount_per_year}
\end{figure}

In Fig.~\ref{fig:sm:amount_per_year}, we show the total annual amount associated with self-lobbying 
\[
\sum_f \mathbbm{1}\left[ s(f) = \text{True} \land y(f) = y \right] m(f)
\]
and the total annual amount associated with third-party lobbying
\[
\sum_f \mathbbm{1}\left[ s(f) = \text{False} \land y(f) = y \right] m(f).
\]
We note that the latter amount is often larger,
even though third-party lobbying is responsible for a larger number of reports (Fig.~\ref{fig:sm:filings_per_year}).
This difference might be partly due to the differences in filing requirements, but also due to structural differences between externally lobbying firms and in-house lobbying departments.

\subsubsection*{Client}
Each LD-2 filing discloses the name of the associated client in line 7. Clients include private firms, NGOs and interest groups, municipalities, and more. Because this is a free-form text label, it is non-trivial to isolate the exact real-world entity associated with a client name. The same firm might be a lobbying client on many filings under a variety of different representations. These issues range from minor ('APPLE, INC.' and 'APPLE INC.' are the same firm) to more complex ('APPLE COMPUTER' and 'APPLE INC.' are the same firm). When firms acquire, merge, or change names, variations can be much more extreme (e.g., `META PLATFORMS INC.' and `FACEBOOK' are the same firm). Clients are not assigned unique identifiers under the LDA because the LDA is intended to regulate lobbying firms, not interested clients. As a result, naively using the entity representations without resolving the underlying entities will lead to a failure to accurately assess each client's expenditure or level of activity. We use a process which we call \textit{search-assisted disambiguation} to solve the problem. In summary, using an online search engine, we perform a search for the representation of a client name. From the results, we identify the website that the search engine associates with the client name. When two different textual representations are attached to the same website, we consider them the same entity. This approach empirically performs very well for current entities, and perfectly for S\&P 500 companies. Effectively, search engines are designed to perform disambiguation of ambiguous search queries, so this should not be a surprise. Although this approach has some limitations,\footnote{Key limitations include the fact that long-bankrupt firms typically have very poor internet presence and that certain classes of non-public-facing firms are more likely to resolve to third-party directories and resources than to have their own internet presence. We take measures to mitigate these limitations: we attempt to use historical data taken from Wikipedia to identify cases where firms are defunct in order to identify the risk of a false positive, and we use a blocklist to remove a wide swath of online firm directories.} It improves on rules-based approaches (e.g., string distance, cosine similarity, etc.) typically used in these settings.

The total number of distinct clients that we identify in any given year 
will be presented in Fig.~\ref{fig:sm:order_per_year}A, alongside other entities. 

\subsubsection*{Registrant}
The lobbying report is submitted by the registrant $r$, which could be an organization, lobbying firm, or a self-employed individual. Registrants are identified by unique U.S.~Senate and U.S.~House IDs, which must be disclosed on lines 5 and 6 of the LD-2 filing, and so no further disambiguation is required.\footnote{As with clients, we are also able to connect registrants to external corporate identifiers by the same search engine-driven disambiguation method described above; although this is not directly relevant to this paper, we make linking identifiers available in our broader dataset.} Some registrants file only as self-filers. We call these registrants \textit{In-House} registrants -- as one example, consider industry-based interest groups such as the Chamber of Commerce of the United States of America, which has reported \$1.73 billion in lobbying expenditures over the life of our dataset, all self-filed; or larger corporate entities such as General Electric, which has reported \$364 million in in-house lobbying expenditures over the life of our dataset. Other registrants file also on behalf of other (as non-self-files), in which case we call them \textit{K-Street} registrants. We classify each registrant as \textit{In-House} or \textit{K-Street}, reviewing all the lobbying reports in the database, such that 
\begin{align}
\begin{split}
    r \in \text{In-House} &  \Leftrightarrow \forall_{f:~r(f) = r}~~s(f) = \text{True},\\
    r \in \text{K-Street} &  \Leftrightarrow \exists_{f:~r(f) = r}~~s(f) = \text{False}.
\end{split}
\label{eq:inhouse_kstreet_def}
\end{align}

The evolution of the total number of registrants is presented in Fig.~\ref{fig:sm:order_per_year}B, and an in-depth analysis of the dynamics of K-Street registrants is shown in Fig.~2B of the main article. 

\begin{figure}[h]
    \centering
    \includegraphics[width = .75\textwidth]{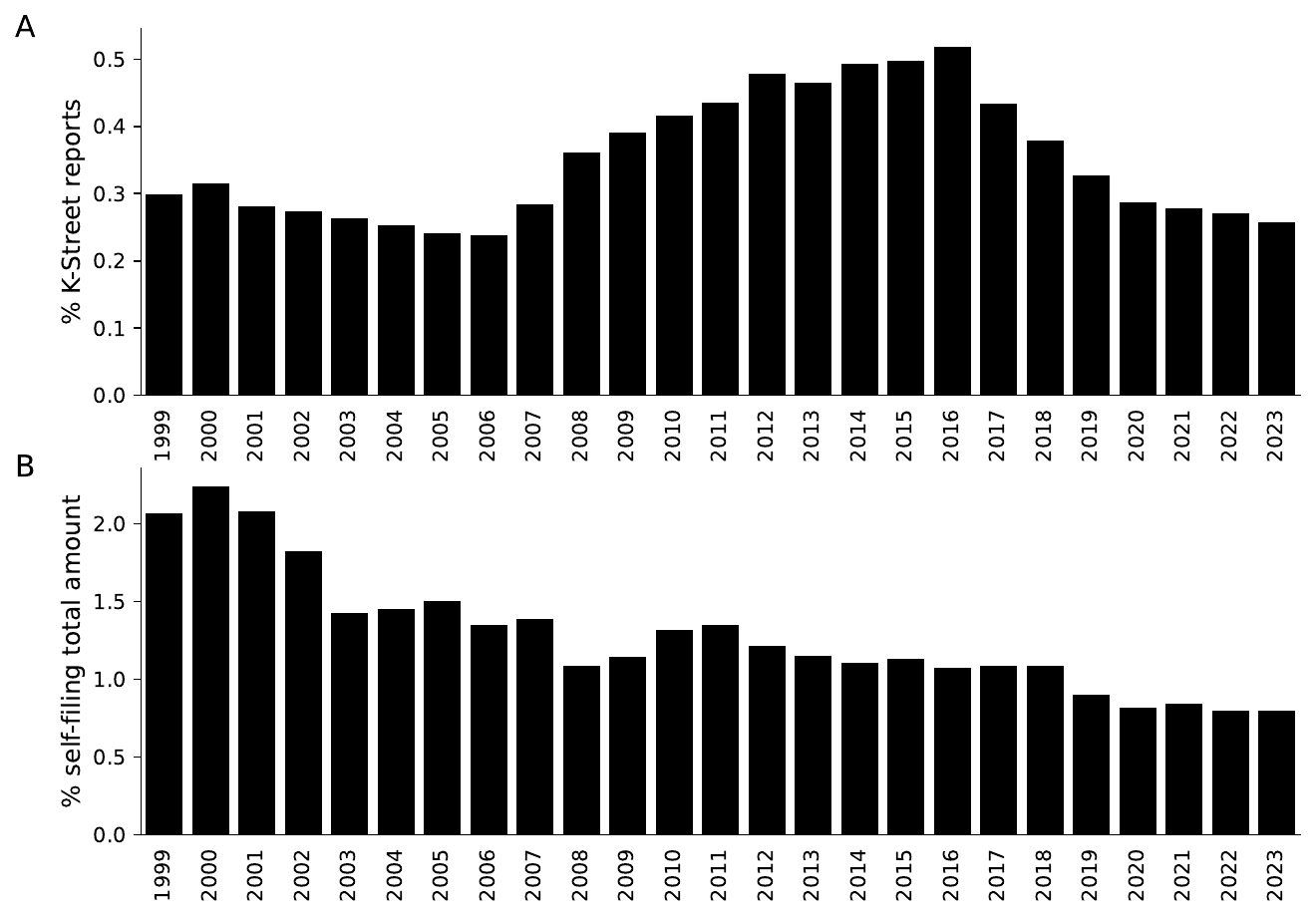}
    \caption{  (A) Percentage of lobbying reports submitted by K-Street registrants (according to definition~\eqref{eq:inhouse_kstreet_def}) on their own behalf (as self-filers).  (B) Percentage of the total self-filing amount generated by self-reports of K-street registrants decreases over time. }
    \label{fig:sm:selffiling}
\end{figure}

Some K-street registrants also lobby on their own behalf and thus occasionally submit reports as self-filers, but Fig.~\ref{fig:sm:selffiling} shows that such cases constitute only a modicum of all the lobbying data.

\subsubsection*{Lobbyists}

In addition to client $c$ and registrant $r$, the lobbying report also provides information about lobbyists $L(f)$ (line 18 of LD-2 filings). Note that each report contains exactly one client and exactly one registrant, but there is no limit to the number of lobbyists, and some reports mention no lobbyists at all.

Because lobbyists are identified uniquely by names, and not any other identifier, they also require disambiguation. If a lobbyist named Maria Gutierrez appears on multiple filings across multiple registrants, it may be the case that she is a single lobbyist who worked for multiple registrants. Or it may be the case that there are multiple lobbyists with the same name. Other cases are more complex: Are Jon Smith, John Smith, and Jonathan Smith the same lobbyist? Disambiguating individual names using our firm-name approach is currently infeasible, so we use a score-based entity resolution approach. We block on characteristics of lobbyists (name, past covered position history, registrants, years of operation) and agglomerate high-probability matches to yield distinct lobbyists from ambiguous name entries. Figure~\ref{fig:sm:order_per_year}C shows how the total number of active lobbyists has been changing over time.

\subsubsection*{Government entities}
A lobbying report provides information about the government entities (House(s) of Congress and Federal agencies) $G(f)$ (line 17) approached during the course of lobbying. Similarly to the set of lobbyists $L(f)$, the set of approached government entities can be empty or arbitrarily large. To be precise, a report can mention one government entity multiple times, in relation to different issues, but we disregard this multiplicity. In modern reports, entities are selected from a closed set; in earlier reports, entities are extracted and resolved from an open text submission. In general, because U.S.~agencies are a closed set, the matching task is simpler than the previously described fields: we do so fairly directly using string distance methods. The set of entities includes a number of offices that overlap (e.g., the Office of President vs.~the President; the White House Office vs. the Executive Office of the President), so in some applications one may wish to consolidate the list of entities based on the specific needs. We do that for example in Fig. 4 of the main text, where we consolidate White House Office and the Executive Office of the President (c.f.~\ref{sec:sm:probability}). The full set of government entities considered will be discussed in Sec.~\ref{sec:sm:target_portfolio}.

\subsubsection*{General issue areas}
The set of general issue areas $A(f)$ is derived from their abbreviations/codes on line 15 of the Lobbying Disclosure Form. 
The issue areas come from a standardized list of 79 items.
All of them can be found on the y-axis of Fig.~4A of the main article.

\subsubsection*{Bills}
This set includes all the bills that the lobbyists reportedly lobbied on. 
The bills are referenced with their ID number, which allows us to relate the lobbying activity to the Congressional data. 
In this study, we use the bill data in the example of the probabilistic analysis~\ref{sec:sm:probability}).

\subsection{Lobbyist association data}
\label{sec:sm:connections_data}

From mandatory disclosures made in LD-2 filings (line 18), we also gather \textit{association data}, which describes the past professional relationships of registered lobbyists.

When registering, a lobbyist discloses if they have a past history as a \textit{covered official}, which is a legal term under the LDA that includes a variety of positions and offices. In brief, covered officials include all elected officials in the federal government, their assistants and Chiefs of Staff, civil servants who are political appointees, executives of federal agencies or departments (this requirement is broad enough to cover U.S.~Parole Commission commissioners, National Parks Service directors, and federally-operated regional electrical agencies), high-ranking military officers, and more. The LDA requires that officials disclose such past positions as part of the public interest in understanding ``revolving-door'' lobbying. Once disclosed in a filing, the positions need not be disclosed again. Thus, an individual lobbyist will typically disclose their past positions when they begin lobbying and never again, though some lobbyists do routinely re-report the same relationships with each filing. Should the lobbyist return to public service and serve in other covered positions, they are obligated to disclose if they subsequently lobby. The lobbyists do not need to provide detailed timeline information for covered positions, and so it is impossible to identify from the filings whether a lobbyist's employment was recent or in the distant past. We apply text processing methods, including regular expressions, to identify the exact position from the free-form text and match the official to agencies, departments, or individual legislators. 

We compile our findings in a \textit{government association function} $g(l,y)$ and the \textit{political association function} $p(l,y)$, which take as an input lobbyist ID $l$ and year $y$. 

The government association function outputs the set of government entities that the lobbyist worked for that have been mentioned in the LDA reports in year $y$ or earlier. 
Controlling for the year, i.e., using $g(l,y)$, instead of $g(l)$, allows us to avoid using associations reported in the future to make conclusions about the past. 
By construction,
\[
y_1<y_2 \quad \Rightarrow \quad g(l,y_1) \subseteq  g(l,y_2).
\]
The political association function $p(l,y)$ is an analogous set of all the present legislators (members of the Congress in year $y$) that lobbyist $l$ reported working for (in year $y$ or earlier). 
Note that this time
\[
y_1<y_2 \quad \not\Rightarrow \quad p(l,y_1) \subseteq  p(l,y_2),
\]
as the legislator might not be in office anymore.

\subsection{Political data}
\label{sec:sm:political_data}

Our data on legislators is sourced directly from the United States digital service congressional legislator database~\cite{usdigitalservice_congress}. We retrieve and ingest this \textit{political data} for each legislator (politician) $p$, and we construct the \textit{affiliation functions} $\text{Party}(p, y)$ and $\text{Chamber}(p,y)$
providing information about the partisan affiliation and Congress chamber of the legislator $p$ in year $y$.

Party affiliations of the President, House majority, and Senate majority (shown in Fig.~2--4 of the main text) are taken from
\url{https://history.house.gov/Institution/Presidents-Coinciding/Party-Government/}.

\section{Lobbying network construction}
\label{sec:sm:network_construction}
After the initial processing described in Sec.~\ref{sec:sm:sources}, we use the lobbying data (Sec.~\ref{sec:sm:lobbying_data}), affiliation data (Sec.~\ref{sec:sm:connections_data}), and the political data (Sec.~\ref{sec:sm:political_data}) to construct year-indexed multi-layer directed graphs $\mathcal{G}_{1999}, \mathcal{G}_{2000}, \dots, \mathcal{G}_{2023}$, that we call \textit{lobbying networks}.
The nodes and edges of $\mathcal{G}_y$ are denoted by $V(\mathcal{G}_y)$ and $E(\mathcal{G}_y)$ respectively.
The lobbying networks are constructed for each year separately. 
In this Section, we describe the details of their construction. 

\subsection{Nodes}
\label{sec:sm:network_construction:nodes}
The lobbying network $\mathcal{G}_y$ has 5 different sets of nodes, organized in layers:
\begin{enumerate}  
    \item The first (most upstream) layer of $\mathcal{G}_y$ corresponds to \textit{client nodes}
    \begin{equation}
        \text{Clients}(y) =  \bigcup_{f:~y(f)=y} c(f). 
        \label{eq:clients_def}
    \end{equation}
   
    \item The second layer of $\mathcal{G}_y$ comprises \textit{registrant nodes}
    \begin{equation}
    \text{Registrants}(y) = \bigcup_{f:~y(f)=y} r(f)
        \label{eq:registrants_def}
    \end{equation}
    which are further divided based on the registrant type (eq.~\eqref{eq:inhouse_kstreet_def}) into two disjoint subsets
    \begin{equation}
        \text{In-House Registrants}(y) =  \{ r \in \text{Registrants}(y) : r \in \text{In-House} \},
        \label{eq:registrants_inhouse_def}
    \end{equation}
    and
    \begin{equation}
        \text{K-Street Registrants}(y) =  \{ r \in \text{Registrants}(y) : r \in \text{K-Street} \}.
        \label{eq:registrants_kstreet_def}
    \end{equation}

    \item The third layer of $\mathcal{G}_y$ consists of \textit{lobbyist nodes}
    \begin{equation}
    \text{Lobbyists}(y) = \bigcup_{f:~y(f)=y} L(f). 
        \label{eq:lobbyists_def}
    \end{equation}
    These first three layers use only the lobbying data $\mathcal{LD}$ defined in eq.~\eqref{eq:ld_def}. 
    
    \item To construct the fourth layer of $\mathcal{G}_y$ representing \textit{government entity nodes}, we also use the government association function  $g(l,y)$ defined in Sec.~\ref{sec:sm:connections_data}. Thus,
    \begin{equation}
    \text{Government entities}(y)  = \bigcup_{f:~y(f)=f} \quad \bigcup_{l \in  L(f)}  g(l,y),  
        \label{eq:gov_entities_def}
    \end{equation}
    is the set of all the government entities with a historical employment link to at least one of the lobbyists, with the restriction that the connection must have been mentioned in year $y$ or earlier. 
    
    \item The fifth layer of $\mathcal{G}_y$ corresponding to \textit{legislator nodes}, is constructed similarly by using the political association function $p(l,y)$ defined in Sec.~\ref{sec:sm:connections_data}. 
    Thus,
    \begin{equation}
    \text{Legislators}(y)  = \bigcup_{f:~y(f)=f} \quad \bigcup_{l \in  L(f)}   p(l,y) 
        \label{eq:legislators_def}
    \end{equation}
    is the set of all active legislators (present in Congress in year $y$) with a historical employment link to at least one of the lobbyists, with the restriction that the connection must have been mentioned in year $y$ or earlier.
\end{enumerate}

\subsection{Edges}
\label{sec:sm:network_construction:edges}
The lobbying network $\mathcal{G}_y$ has 5 different sets of edges:
\begin{enumerate}  
    \item $\text{Clientships}(y)$: connections between clients and registrants (first to second layer),
    such that a directed edge $(c,r)$ linking client $c$ and registrant $r$ exists if and only if the registrant lobbied for the client, i.e.,
    \begin{equation}
        (c,r) \in \text{Clientships}(y) \qquad \Leftrightarrow \quad \exists f:~y=y(f) \land c=c(f) \land r= r(f). 
        \label{eq:cr_edge_def}
    \end{equation}
    which are further divided based on the registrant type (eq.~\eqref{eq:inhouse_kstreet_def}) into two disjoint subsets
    \begin{equation}
        \text{In-House Clientships}(y) =  \{ (c,r) \in \text{Clientships}(y) : r \in \text{In-House} \},
        \label{eq:cr_edge_inhouse_def}
    \end{equation}
    and
    \begin{equation}
        \text{K-Street Clientships}(y) =  \{ (c,r) \in \text{Clientships}(y) : r \in \text{K-Street} \}.
        \label{eq:cr_edge_kstreet_def}
    \end{equation}

    \item $\text{Lobbyist contracts}(y)$: connections between registrants and lobbyists (second to third layer), 
    such that a directed edge $(r,l)$ linking registrant $r$ and lobbyist $l$ exists if and only if the lobbyist was mentioned in a report filed by the registrant, i.e.,
    \begin{equation}
    (r,l) \in \text{Lobbyist contracts}(y) \quad \Leftrightarrow \quad \exists f:~y=y(f) \land r= r(f) \land l \in L(f). 
        \label{eq:rl_edge_def}
    \end{equation}
    which are further divided based on the registrant type (eq.~\eqref{eq:inhouse_kstreet_def}) into two disjoint subsets
    \begin{equation}
        \text{In-House Lobbyist contracts}(y) =  \{ (r,l) \in \text{Lobbyist contracts}(y) : r \in \text{In-House} \},
        \label{eq:rl_edge_inhouse_def}
    \end{equation}
    and
    \begin{equation}
        \text{K-Street Lobbyist contracts}(y) =  \{ (r,l) \in \text{Lobbyist contracts}(y) : r \in \text{K-Street} \}.
        \label{eq:rl_edge_kstreet_def}
    \end{equation}
    
    \item $\text{Government associations}(y)$: connections between lobbyists and government entities (third to fourth layer), 
    such that a directed edge $(l,g)$ linking lobbyist $l$ and government entity $g$ exists if and only if an employment connection between the two has been documented in year $y$ or earlier, i.e., 
    \begin{equation}
    (l,g) \in \text{Government associations}(y) \quad \Leftrightarrow \quad
    l \in \text{Lobbyists}(y) \land g \in g(l,y),   
        \label{eq:lg_edge_def}
    \end{equation}
    where $g(l, y)$ is the government association function defined in Sec.~\ref{sec:sm:connections_data}. 
    
    \item $\text{Legislator associations}(y)$: connections between lobbyists and legislators (third to fifth layer), 
    such that a directed edge $(l,p)$ linking lobbyist $l$ and legislator $p$ exists if and only if an employment connection between the two has been documented in year $y$ or earlier, and the legislator is a Congress member in year $y$, i.e.,
    \begin{equation}
    (l,p) \in \text{Legislator associations}(y) \quad \Leftrightarrow \quad 
    l \in \text{Lobbyists}(y) \land p \in p(l,y), 
        \label{eq:lp_edge_def}
    \end{equation}
    where $p(l,y)$ is the political association function defined in Sec.~\ref{sec:sm:connections_data}.
\end{enumerate}
All the edges are directed and unweighted (c.f.~Sec.~\ref{sec:sm:network_construction:weights}), and we disregard any multi-edges.

\subsection{Weights}
\label{sec:sm:network_construction:weights}
Although by default our lobbying network is unweighted, we could attach weights $w$ to the client-registrant connections by adding all the monetary values in corresponding filings: 
\begin{equation}
    w[(c,r) \in E(\mathcal{G}_y)] = \sum_f \mathbbm{1}\left[ y(f) = y \land c(f) = c \land r(f) = r \right] m(f).
    \label{eq:weight_cr_def}
\end{equation}


\section{Order and size of the lobbying network}
\begin{figure}[h]
    \centering
    \includegraphics[width=.75\textwidth]{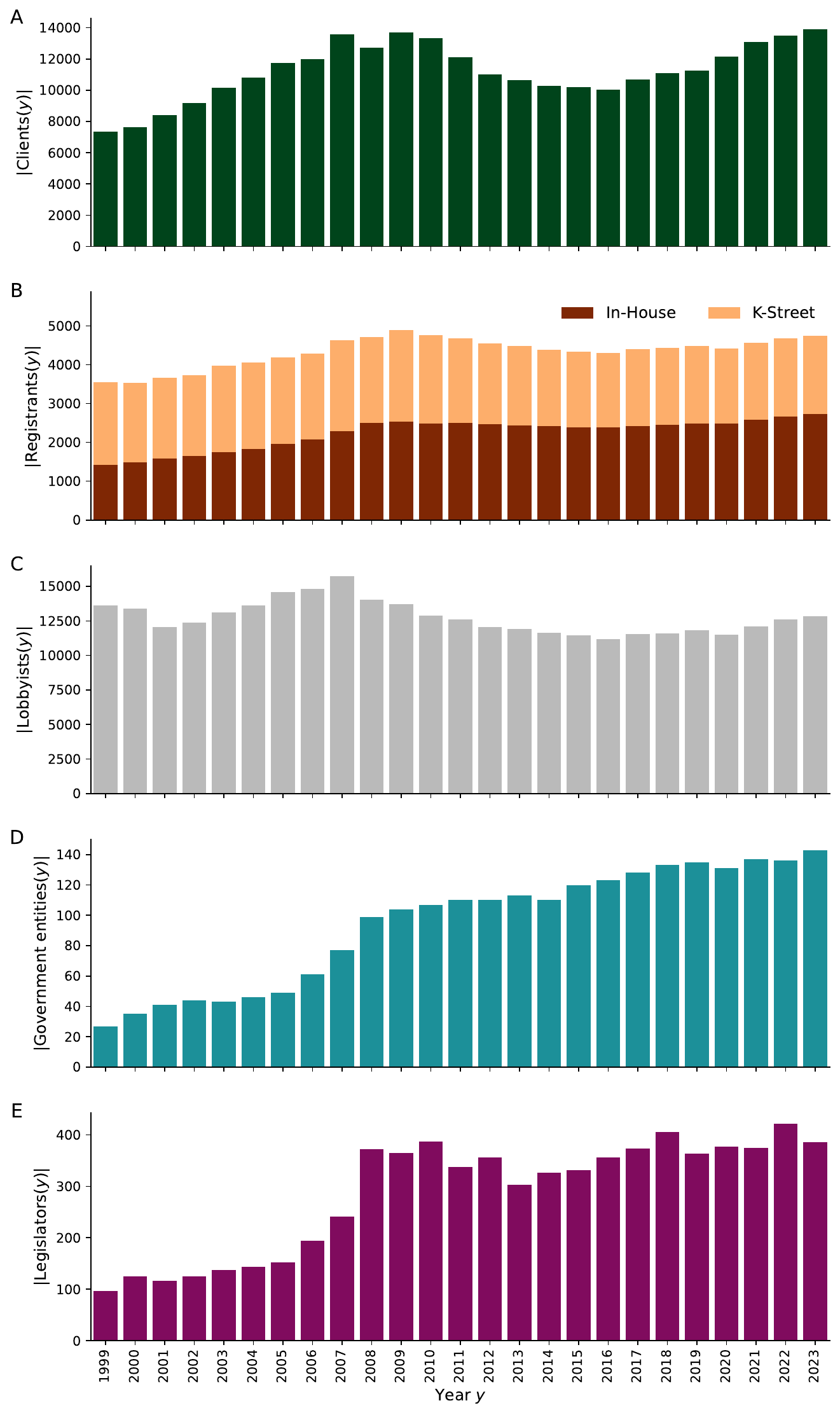}
    \caption{Order of the lobbying network each year. 
    (A) The number of distinct clients in a given year. 
    (B) The number of distinct K-Street and In-House registrants in a given year. 
    (C) The number of distinct lobbyists in a given year. 
    (D) The number of distinct government entities in a given year. 
    (E) The number of distinct legislators in a given year. 
    Note that the set of registrants is significantly smaller than the sets of clients and lobbyists, which are of comparable size.}
    \label{fig:sm:order_per_year}
\end{figure}

\begin{figure}[h]
    \centering
    \includegraphics[width=.75\textwidth]{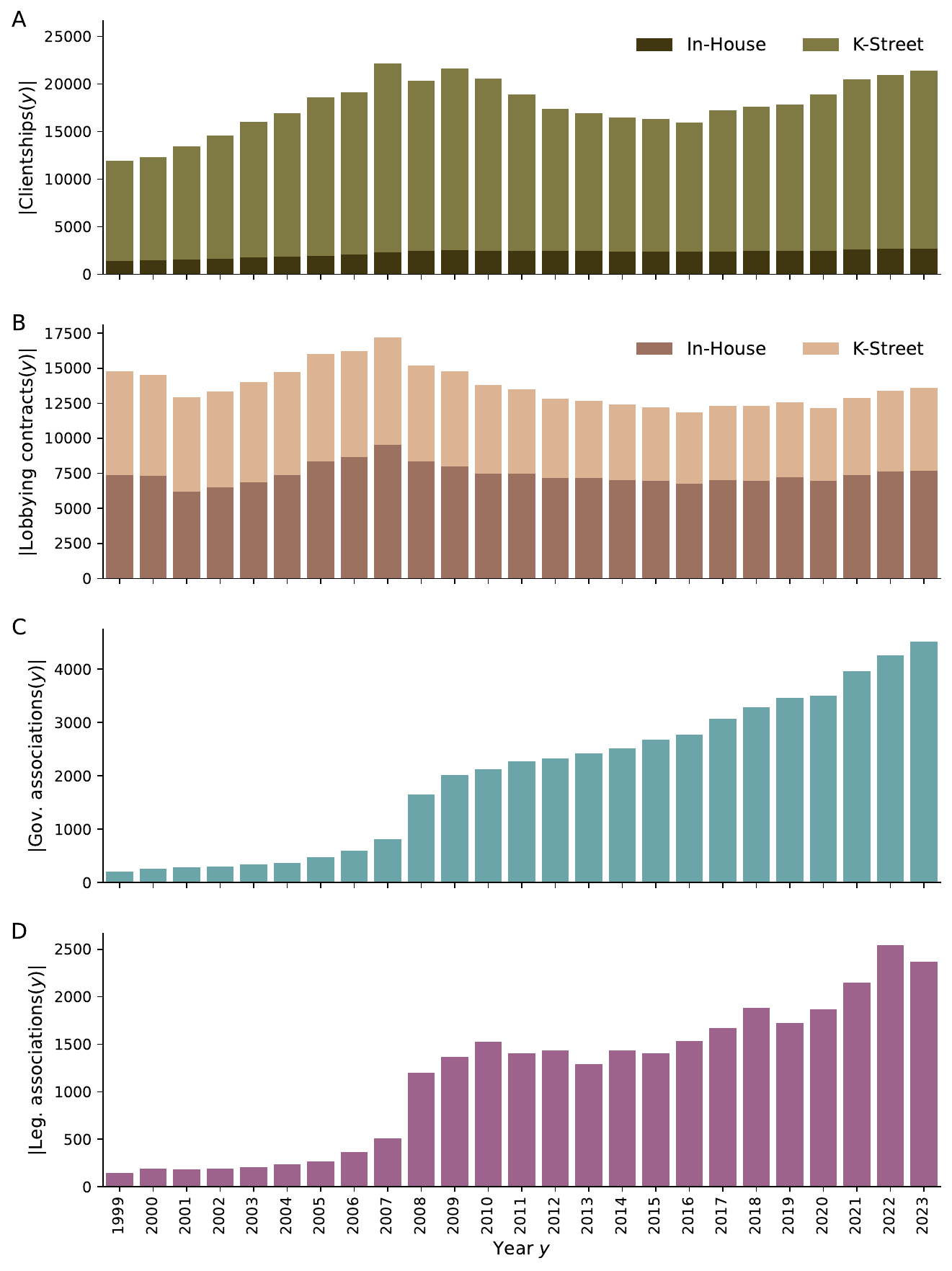}
    \caption{Size of the lobbying network each year. 
    (A) The number of distinct clientships (client-registrant edges) in a given year. 
    (B) The number of distinct lobbying contracts (registrant-lobbyist edges) in a given year. 
    (C) The number of distinct government associations (lobbyist-government entity edges) in a given year. 
    (D) The number of distinct legislator associations (lobbyist-legislator edges) in a given year.
    }
    \label{fig:sm:size_per_year}
\end{figure}

In this Section, we discuss the number of nodes (order) and the number of edges (size) of the lobbying network. 
The year-to-year evolutions of order and size are presented in Fig.~\ref{fig:sm:order_per_year} and Fig.~\ref{fig:sm:size_per_year}, respectively.

The number of clients (c.f.~eq.~\eqref{eq:clients_def}) steadily increases until the financial crisis (2007--2008), after which it declines for about the next 7 years before the steady increase is recovered.
A similar pattern can be observed for the total number of registrants and lobbyists  (c.f.~eq.~\eqref{eq:registrants_def} and~\eqref{eq:lobbyists_def}), as well as the number of connections between the three upstream layers (clientships and lobbyist contracts, c.f.~eq.~\eqref{eq:cr_edge_def} and~\eqref{eq:rl_edge_def}). 

Note that the  K-Street registrant count (c.f.~eq.~\eqref{eq:registrants_kstreet_def}) dominates in the early years while the In-House registrant count (c.f.~eq.~\eqref{eq:registrants_inhouse_def}) dominates in the more recent years.
Moreover, the clientships involving K-Street registrants outnumber the clientships involving In-House registrants. 
Nevertheless, lobbyist contracts with In-House registrants slightly outnumber those with K-Street registrants, especially in the more recent years.

\begin{figure}[h]
    \centering
    \includegraphics[width = .9\textwidth]{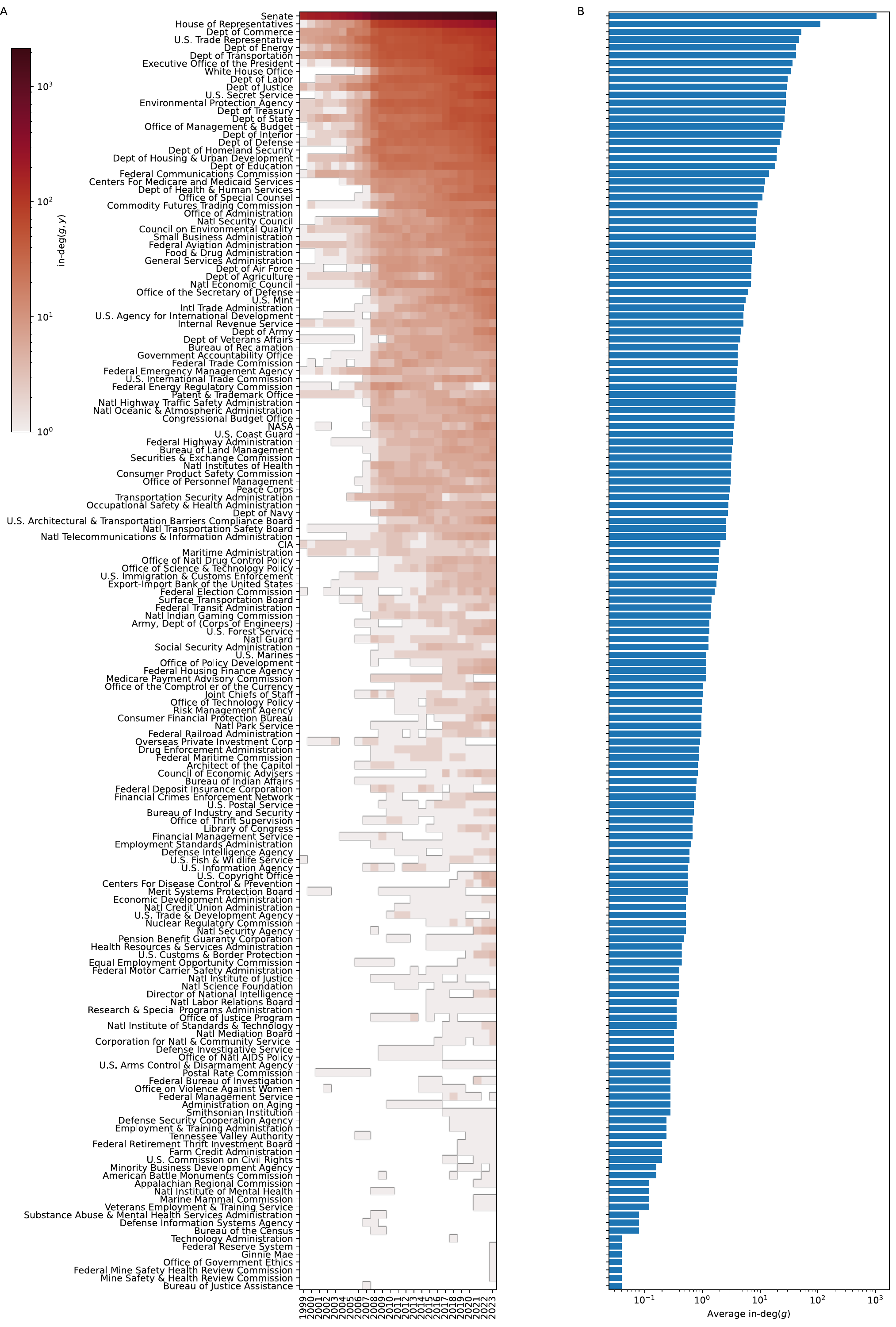}
    \caption{In-degree (number of identified lobbyist associations) of specific government entities.
    (A) By year. (B) Time-averaged. 
    The entities are sorted by the average government entity in-degree.} 
    \label{fig:sm:gov_indegree}
\end{figure}

The number of government entities (c.f.~eq.~\eqref{eq:gov_entities_def}) included in the lobbying network, and the corresponding number of lobbyist-government associations (c.f.~eq.~\eqref{eq:lg_edge_def}) grow steadily over time.
In Fig.~\ref{fig:sm:gov_indegree}, we show which government entities have the best known ties to the lobbyists (highest in-degree) and how this quantity varies in time. 

\begin{figure}[h]
    \centering
    \includegraphics[width = .55\textwidth]{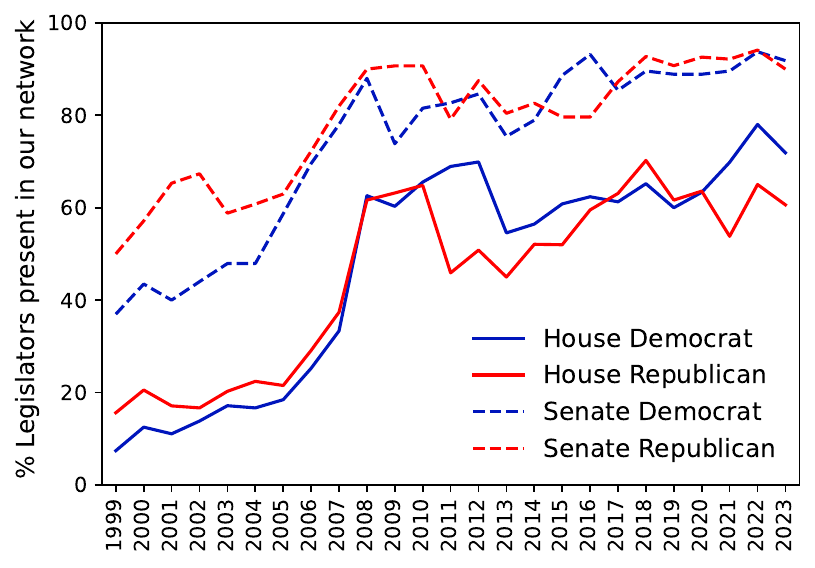}
    \caption{Percentage of legislators included in the lobbying network, in terms of the legislator's chamber and party affiliations.}
    \label{fig:sm:legislator_connections}
\end{figure}

The number of legislators (c.f.~eq.~\eqref{eq:legislators_def}) and the lobbyist-legislator associations (c.f.~eq.~\eqref{eq:lp_edge_def}) also steadily grow in time. 
In Fig.~\ref{fig:sm:legislator_connections}, we show that since 2008, around 80\% of Senators and 50\% of Representatives, from both the Democratic and Republican parties, are included in our lobbying network every year.
By construction, the legislator is included if there is at least one active lobbyist who, according to our database, worked for this legislator in the past. 

\afterpage{\clearpage}

\section{Degree distributions}
\label{sec:sm:degree_distributions}
In a given year $y$, let the out-degree (the number of outgoing edges) of a node type $i$ and the in-degree (the number of incoming edges) of a node type $j$ be
\begin{equation}
    \text{out-deg}(i,y) = |\left\{ (i,j) \in E(\mathcal{G}_y) \right\} |,
    \quad \text{and} \quad
    \text{in-deg}(j,y) = |\left\{ (i,j) \in E(\mathcal{G}_y) \right\} |.    
    \label{eq:c_in_out_deg_def}
\end{equation}

For client nodes, we can further distinguish their out-degree to K-Street and In-House registrants as 
\begin{equation}
    \text{out-deg}_{K}(c,y) = |\left\{ (c,r) \in E(\mathcal{G}_y) : r \in \text{K-Street} \right\} |,
    \quad \text{and} \quad
    \text{out-deg}_{I}(c,y) = |\left\{ (c,r) \in E(\mathcal{G}_y) : r \in \text{In-House} \right\} |.
    \label{eq:c_out_deg_by_reg_type_def}
\end{equation}

\subsection{Average degree}
\begin{figure}[h]
    \centering
    \hspace{34mm} 
    \includegraphics[width=0.8\textwidth]{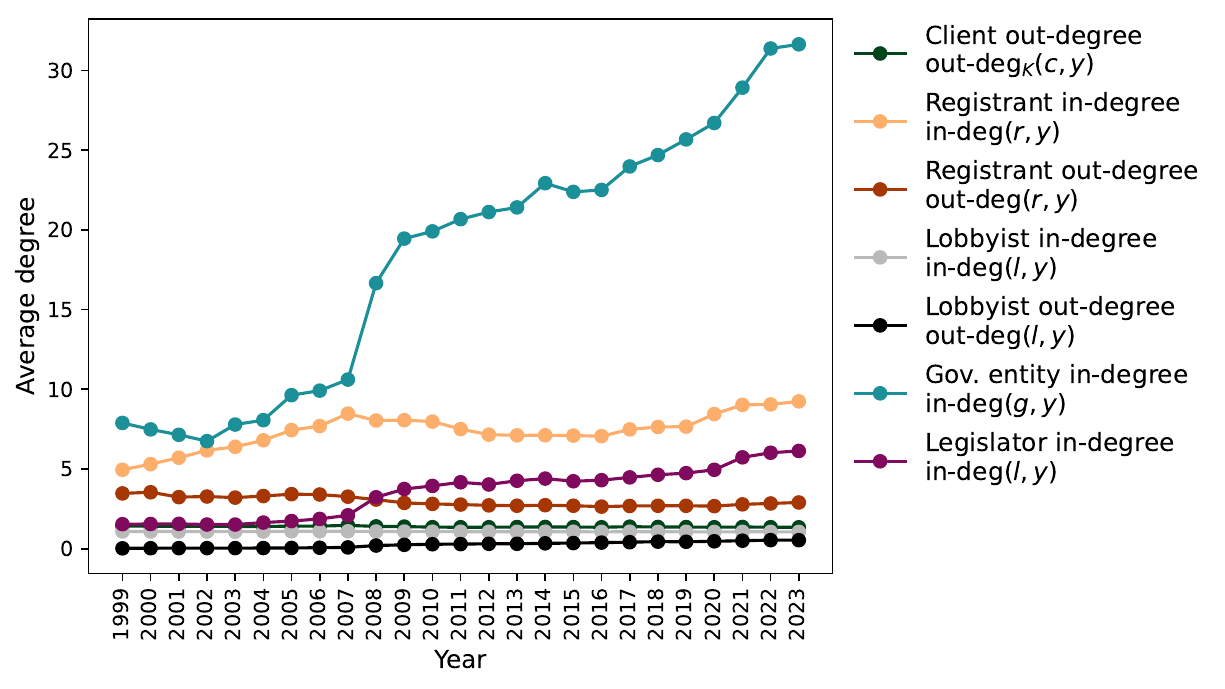}
    \caption{Average in/out-degree of each lobbying network layer in years 1999--2023.
    The client out-degree, registrant in-degree, and registrant out-degree are all computed for K-Street registrants.
    }
    \label{fig:sm:average_degrees_per_year}
\end{figure}

We begin by analyzing the average in- and out-degree of each network layer.
These data are plotted in Fig.~\ref{fig:sm:average_degrees_per_year}.
The most dynamic series presented here is the average in-degree of government entities and registrants, indicating that, at least as far as our database is concerned, government entities become connected to increasingly more lobbyists and registrants to increasingly more clients.
The average degrees for the remaining layer-to-layer connections are relatively constant over time.

\subsection{Distribution inference}
Following the methodology of Ref.~\cite{Barabasi_book}, in this study, we characterize the degree distribution by computing the complementary cumulative distribution function (CCDF).
For example, in Fig.~2F of the main article, we plot the CCDFs of $\text{in-deg}(r,y)$ for $r \in \text{K-Street}$.
To be precise, this CCDF $\rho$ for this particular distribution is defined as
\begin{equation}
    \rho(d,y) = \frac{\sum_{r \in \text{K-Street Registrants}(y)} \, \mathbbm{1} [\text{in-deg}(r) \geq d ] } {|\text{K-Street Registrants}(y)| }.
\end{equation}
In a similar way, in Fig.~\ref{fig:sm:remaining_degree_distributions} we plot the CCDFs of other degree distributions.

\begin{figure}[h]
    \centering
    \includegraphics[width=0.6\textwidth]{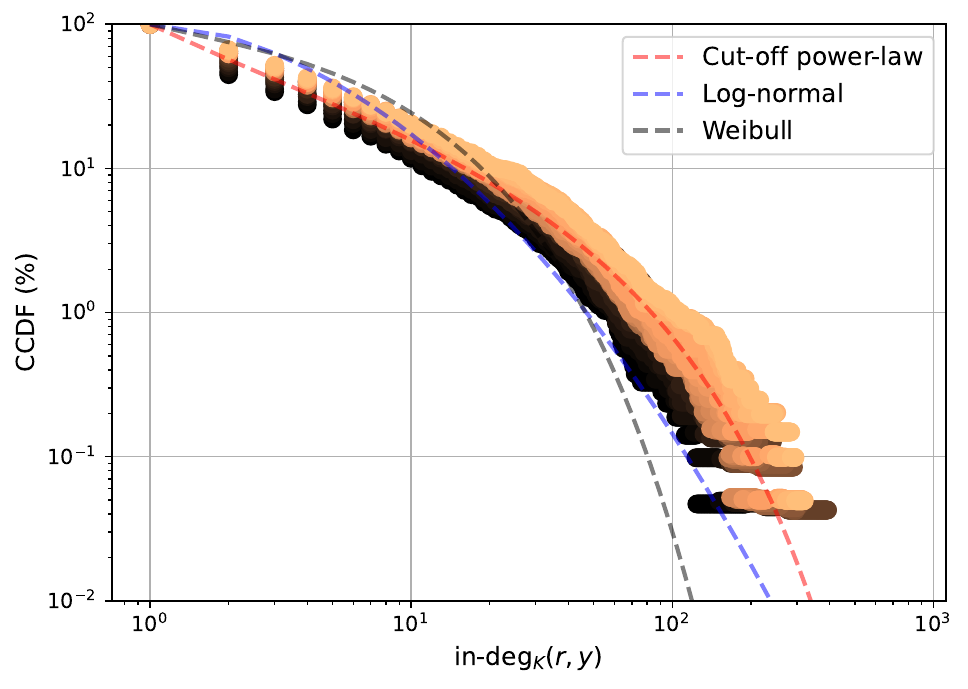}
    \caption{K-Street registrant in-degree distribution CCDF, overlaid with MLE fits of standard distributions: (red) augmented power law distribution with saturation value $c_{sat}=0$, cut-off value $c_{cut} = 86$, and exponent $d = 1.54$, (blue) log-normal distribution with shape parameter $s= 0.18$, and (gray) Weibull distribution with shape parameter $s=0.73$.
    The fits were performed using aggregated data from all years between 1999 and 2023. 
    }
    \label{fig:sm:distribution_fits}
\end{figure}

\begin{figure}[h]
    \centering
    \includegraphics[width=0.95\textwidth,trim={3.5cm 0 3.5cm 0},clip]{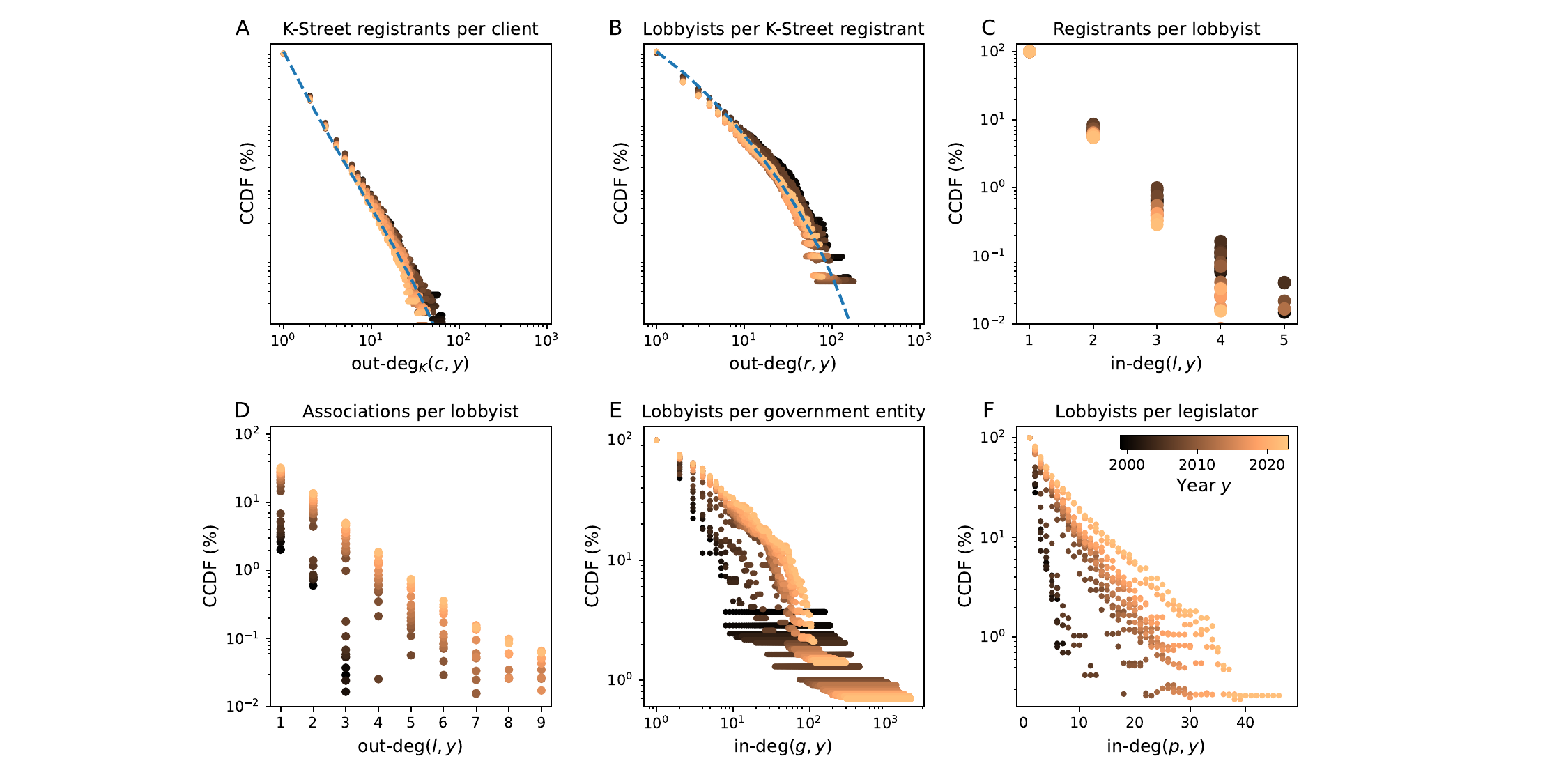}
    \caption{Complementary cumulative distributions (CCDFs) of node in/out-degrees.
    (A) CCDF of the client out-degree $\text{out-deg}_{K}(c,y)$: \% of clients with $\text{out-deg}_{K}(c,y)$ or more K-Street registrants
    The blue dashed line is the MLE fit of an augmented power-law distribution with saturation value $r_{sat}=0$, cut-off value $r_{cut} = 80$, and exponent $d=2.88$.
    (B) CCDF of K-Street registrant out-degree $\text{out-deg}(r,y)$: \% of K-street registrants with $\text{out-deg}(r,y)$ or more lobbyists.
    The blue dashed line is the MLE fit of an augmented power-law distribution with saturation value $l_{sat}=1$, cut-off value $l_{cut} = 84$, and exponent $d=2.36$.
    (C) CCDF of lobbyist in-degree $\text{in-deg}(l,y)$: \% of lobbyists employed by $\text{in-deg}(l,y)$ or more registrants.
    (D) CCDF of lobbyist out-degree $\text{out-deg}(l,y)$: \% of lobbyists with $\text{out-deg}(l,y)$ or more associations (government entities and legislators). 
    (E) CCDF of the government entity in-degree $\text{in-deg}(g,y)$: \% of government entities with $\text{in-deg}(g,y)$ or more lobbyists.
    (F) CCDF of the legislator in-degree $\text{in-deg}(p,y)$: \% of legislators with $\text{in-deg}(p,y)$ or more lobbyists.
    All CCDFs are computed separately for each year in the range 1999--2023.
    }
    \label{fig:sm:remaining_degree_distributions}
\end{figure}

Some of the distributions, e.g., the out-degree of clients (Fig.~\ref{fig:sm:remaining_degree_distributions}A) or registrants (Fig.~\ref{fig:sm:remaining_degree_distributions}B) can be classified as heavy-tail distributions with probability mass functions decaying slower than exponentially.
To gain further insight into the problem, we can fit one of the standard distributions.
In Fig.~\ref{fig:sm:distribution_fits}, we show the results of the fitting procedure for the distribution of clients per K-Street registrant (c.f.~Fig.~2F of the manuscript). 
The three families of distributions we tested are:
\begin{enumerate}
    \item Augmented power-law distribution with exponent $d$, low-end \textit{saturation value} $c_{sat}$ and a high-end \textit{cut-off value} $c_{cut}$:
    $$
    p(c) \propto (c+c_{sat})^d \exp \left( \frac{c}{c_{cut}}\right).
    $$
    This customized distribution is recommended as particularly suitable for degree distributions in networks in Chapter 4 of Ref.~\citep{Barabasi_book}. 
    \item Log-normal distribution with \textit{shape parameter} $s$
    $$
    p(c) = \frac{1}{\sqrt{2 \pi}sc} \exp \left(-\frac{\log^2(c)}{2s^2} \right).
    $$
    \item Weibull distribution with \textit{shape parameter}  $s$ 

    $$
    p(c) = s c^{s-1} \exp(-c^s).
    $$
\end{enumerate}

All the parameter fits are based on maximum likelihood estimation (MLE); the less standard distribution (1) was fitted by customized parameter sweeping, and the distributions (2) and (3) were fitted using standard MLE routines. 

In Fig.~2F of the paper, we show the best fit of the distribution (1), which, besides being the closest fit to the CCDF, also gives us some interpretable insights. Even though we allowed a non-zero saturation, the optimal fit has $c_{sat} = 0$, indicating no low-count saturation effect. 
The cut-off value $c_{cut} = 86$ unravels the characteristic size of a client base, beyond which the growth of a lobbying firm becomes limited.
Finally, the fitted exponent $d = 1.54$ allows us to make comparisons with other distributions within the lobbying network, as well as distributions in other complex systems~\cite{Barabasi_book}.

For example, the best fit of distribution (1) for client out-degree distribution (Fig.~\ref{fig:sm:remaining_degree_distributions}A) has the exponent $d = 2.88$.
This value is on par with $\gamma = 3$ found for the Barabasi-Albert model~\cite{barabasi2000scale}, and in the range of real-world networks~\cite{Barabasi_book}. 
We need to remember, however, that all the degree distributions we compute appear in the context of a multipartite graph, so the Barabasi-Albert model is not exactly applicable. 
The generative process of bipartite network connectivity might be more akin to the Pitman-Yor process~\cite{courtney2018dense,bassetti2009statistical}, which can yield exponents $\gamma \in (1,2]$. 
Nevertheless, fitting a network evolution model to our lobbying network is beyond the scope of this work.

The heavy-tail degree distributions of clients and registrants are in stark contrast
with the exponential distribution of contracts and associations among
lobbyists. Most lobbyists work for only one registrant, and the number
of lobbyists with multiple contracts decays exponentially with the
number of contracts
(Fig.~\ref{fig:sm:remaining_degree_distributions}C).
In Sec.~VI of~\cite{mm}, we show that the distribution of reports (jobs) among individual lobbyists in large lobbying firms is also exponential (except for a small set of `super-lobbyists' who work with more than 200 clients). 
Furthermore, most lobbyists (ca.~90\%) do not have any documented
historical associations to government entities and legislators, and
the number of lobbyists with $a>0$ associations decays exponentially
with~$a$
(Fig.~\ref{fig:sm:remaining_degree_distributions}D). 
Our results indicate, therefore, that the distribution of connections among individual lobbyists is more egalitarian than the distribution of influence among the lobbying firms. 

The structural differences between the registrant layer and the lobbyist layer are likely a consequence of distinct evolutionary principles, which govern the development of the lobbying network.  
Hierarchical degree distributions can spontaneously arise as a consequence of accumulated advantage (Matthew Principle)~\cite{Barabasi1999,Bianconi2001,bassetti2009statistical}, and the exponential distribution is reminiscent of random Erd\"os-Renyi graphs, where each new link is added independently~\cite{Erdos1960}.  

For completeness, in (Fig.~\ref{fig:sm:remaining_degree_distributions}E-F) we show the in-degree distribution of government entities and legislators. 
The graph of the former has a characteristic kink indicating the possibility of two qualitatively different sets of government entities (large and small). 
The latter distribution can be well approximated by an exponential function in the entire range.

\subsection{Weighted degree}

As explained in Sec.~\ref{sec:sm:network_construction:weights}, we can attach weights (monetary values) to the client-registrant edges $(c,r)$.
We thus define the weighted out-degree of a client $c$ to K-Street and In-House registrants as 
\begin{equation}
    \text{w-out-deg}_{K}(c,y) = \sum_{\substack{(c,r) \in E(\mathcal{G}_y),\\r \in \text{K-Street}}} w[(c,r)]. \qquad \text{and} \qquad 
    \text{w-out-deg}_{I}(c,y) = \sum_{\substack{(c,r) \in E(\mathcal{G}_y),\\r \in \text{In-House}}} w[(c,r)].
    \label{eq:c_weighted_out_deg_def}
\end{equation}
which represent the total \textit{lobbying expenses} of the client $c$ via  K-Street and In-House registrants in year $y$.
The overall weighted out-degree of a client $c$ is then
\begin{equation}
    \text{w-out-deg}(c,y) = \text{w-out-deg}_{K}(c,y) + \text{w-out-deg}_{I}(c,y).
    \label{eq:c_weighted_out_deg_both_def}
\end{equation}

Similarly, we let the weighted in-degree of a K-Street or In-House registrant $r$ be
\begin{equation}
    \text{w-in-deg}(r,y) = \sum_{(c,r) \in E(\mathcal{G}_y)} w[(c,r)],
    \label{eq:c_weighted_in_deg_def}
\end{equation}
which represents the total \textit{monetary flow} of the K-Street or In-House registrant $r$ in year $y$.

\begin{figure}[h]
    \centering
    \includegraphics[width=.9\textwidth]{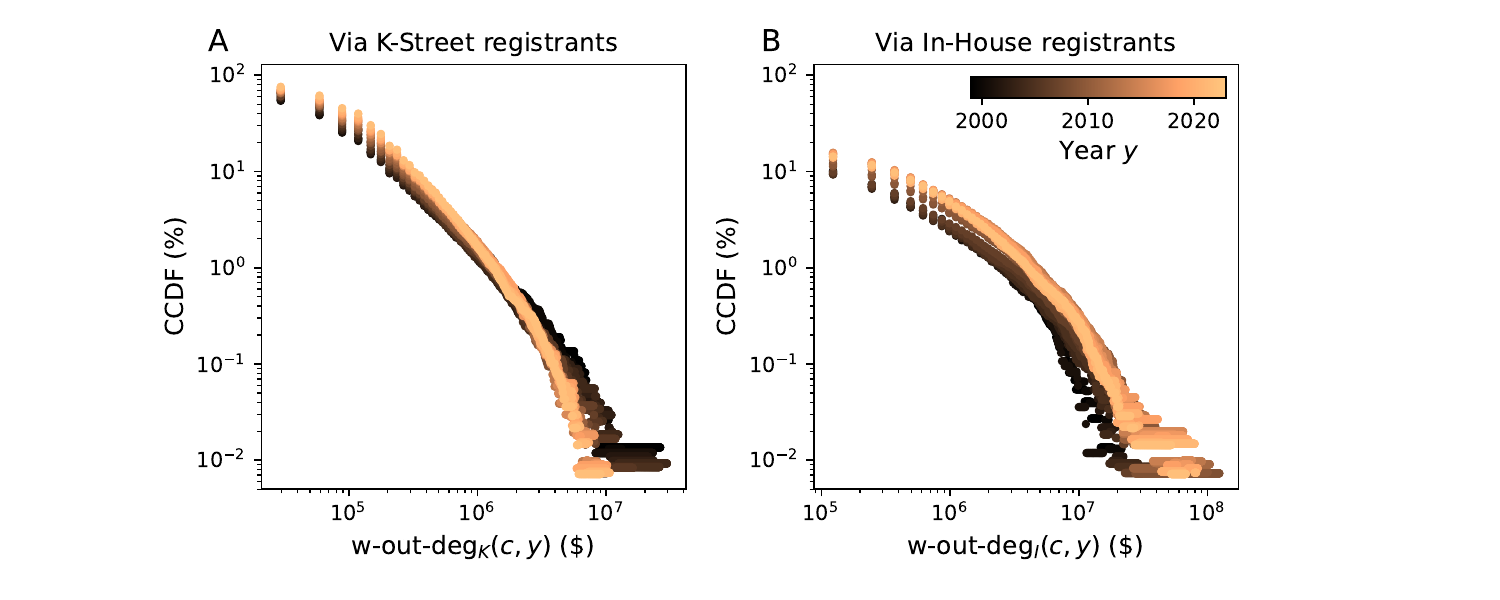}
    \\
    \includegraphics[width=.9\textwidth]{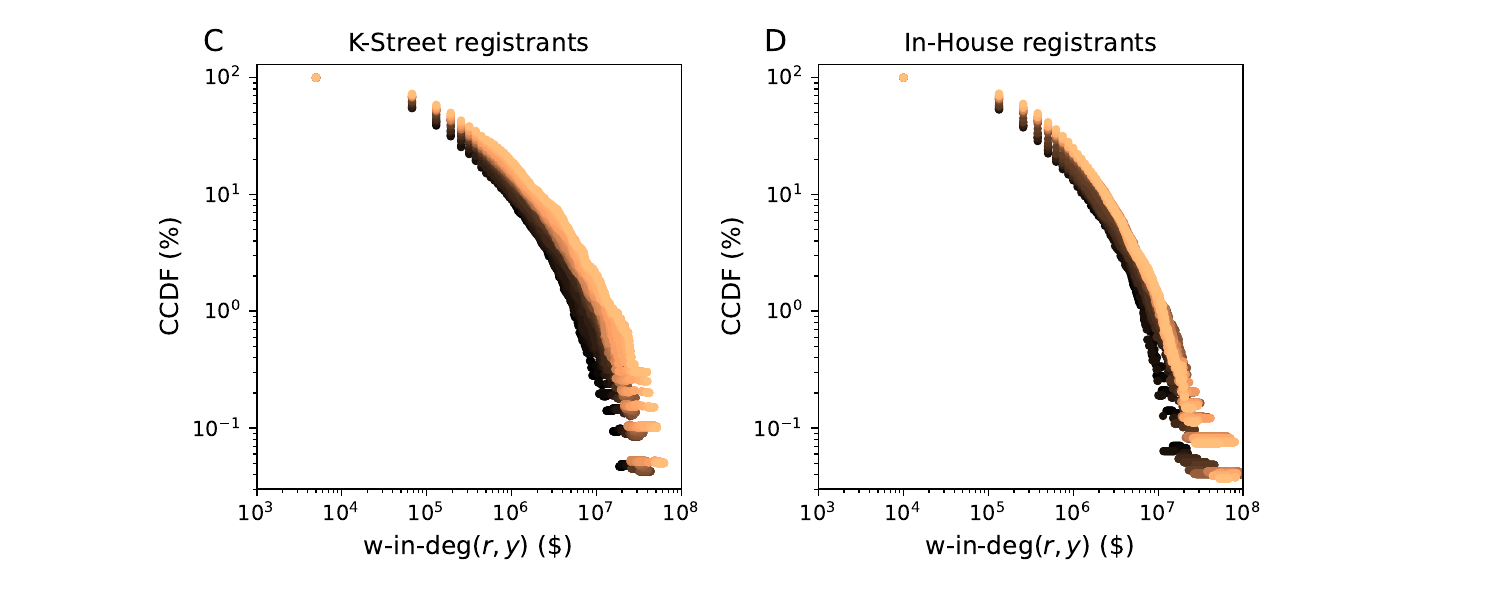}
    \caption{Complementary cumulative distributions (CCDFs) of weighted node in/out-degrees.
    (A) CCDF of the weighted client out-degree $\text{w-out-deg}_{K}(c,y)$: \% of clients with lobbying expenses via K-Street registrants of $\text{w-out-deg}_{K}(c,y)$ or more.
    (B) CCDF of the weighted client out-degree $\text{w-out-deg}_{I}(c,y)$: \% of clients with lobbying expenses via In-House registrants of $\text{w-out-deg}_{I}(c,y)$ or more.
    (C) CCDF of the weighted K-Street registrant in-degree $\text{w-in-deg}(r,y)$: \% of K-Street registrants with monetary flow of $\text{w-in-deg}(r,y)$ or more.
    (D) CCDF of the weighted In-House registrant in-degree $\text{w-in-deg}(r,y)$: \% of In-House registrants with monetary flow of $\text{w-in-deg}(r,y)$ or more.
    All CCDFs are computed separately for each year in the range 1999--2023.
    }
    \label{fig:sm:weighted_degree_distributions}
\end{figure}

The CCDFs of the weighted degrees $\text{w-out-deg}_{K}(c,y)$, $\text{w-out-deg}_{I}(c,y)$, and $\text{w-in-deg}(r,y)$ are shown in Fig.~\ref{fig:sm:weighted_degree_distributions}.
The weighted in/out-degree distributions also follow heavy-tailed distributions, thereby corroborating the existence of 'hyper-influential' clients and registrants (not only in terms of the number of connections but also in terms of money flow).
Comparing the client lobbying expenses via K-Street vs.~In-House registrants (Fig.~\ref{fig:sm:weighted_degree_distributions}A--B), it is evident that lobbying expenses via In-House registrants have higher monetary amounts, while there are almost 10 times more clients who lobby via K-Street registrants.
We also observe interesting temporal trends, e.g., clients with the highest lobbying expenses via K-Street registrants have reduced their spending.

\subsection{Degree correlation}
\begin{figure}[h]
    \centering
    \includegraphics[height=.35\textwidth]{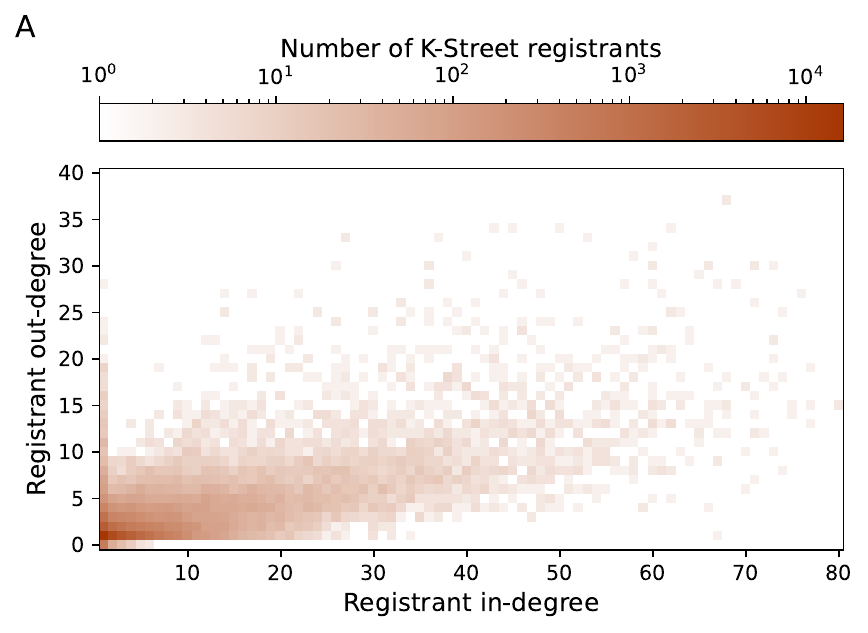}
    \includegraphics[height=.35\textwidth]{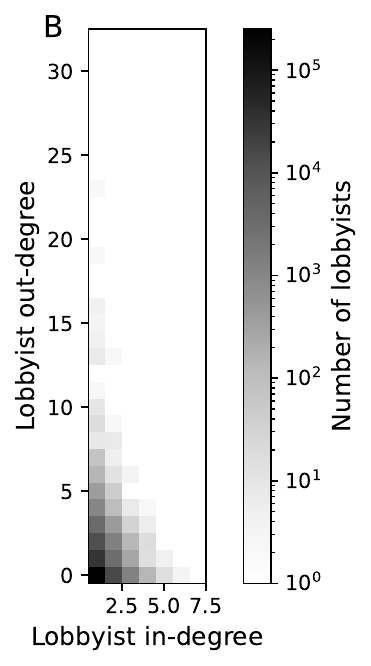}
    \caption{Correlation between in- and out-degrees of same-layer nodes for the lobbying network's inner layers.
    (A) Number of K-Street registrants in terms of their in- and out-degrees.
    (B) Number of lobbyists in terms of their in- and out-degrees.
    The 2D histograms are computed such that one datapoint represents one K-Street registrant/lobbyist per year, spanning the years 1999--2023.
    }
    \label{fig:sm:in_out_degree_correlations}
\end{figure}

For the lobbying network's inner layers (K-Street registrants and lobbyists), we investigate the correlation between the in- and out-degrees of nodes of the same layer.
We compute 2D histograms of K-Street registrant/lobbyist counts such that one datapoint corresponds to one registrant/lobbyist per year.
Figure~\ref{fig:sm:in_out_degree_correlations}A shows that K-Street registrants with more clients tend to be connected to more lobbyists, supporting the notion of well-connected influential K-Street registrants.
By contrast, Fig.~\ref{fig:sm:in_out_degree_correlations}B shows that the lobbyists with a very large number of government and legislator associations (ca.~more than 10) work for only one registrant.

\section{Concentration analysis}

One corollary of the hierarchical organization of a heavy-tail degree distribution is the Pareto $X$-$Y$ (e.g., 80-20) rule.
We can illustrate this property with a so-called \textit{Pareto chart}, or quantify it with a \textit{Gini index}.

\subsection{Pareto charts}
\begin{figure}[h]
    \centering
    \includegraphics[width=.65\textwidth]{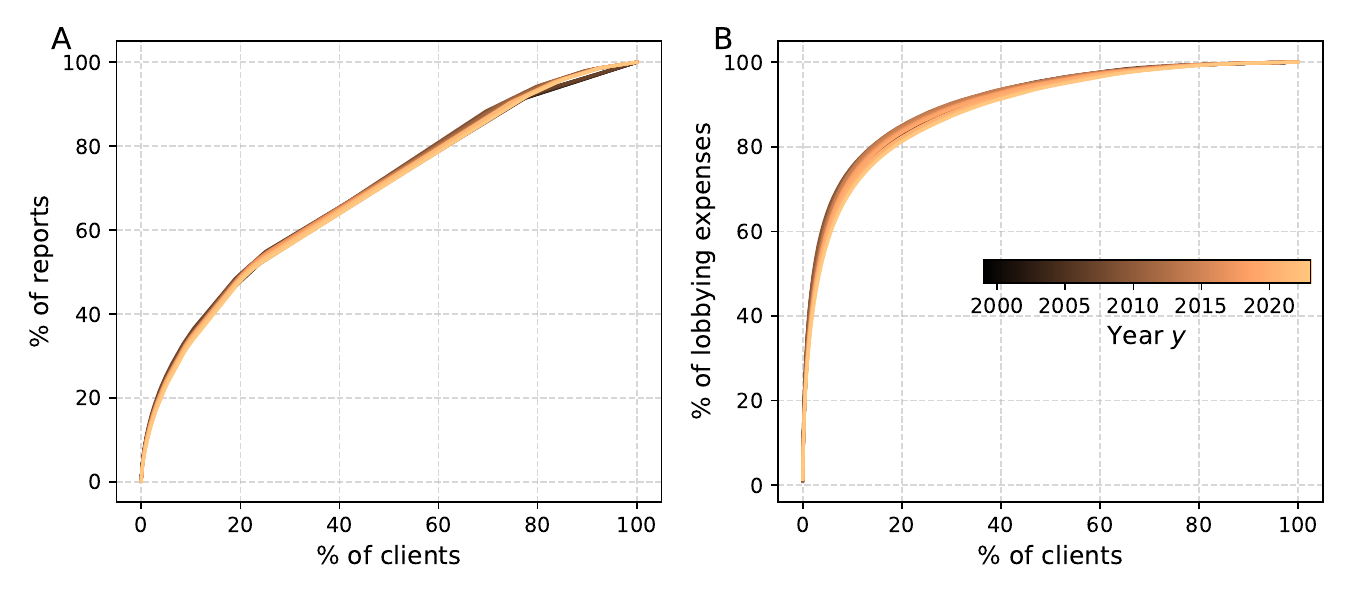}
    \caption{Pareto charts for clients in 1999--2023. 
    (A) The number of reports filed by clients. The top quartile of clients files around half of the LDA reports. 
    (B) The distribution of the lobbying expenses among clients follows the 80-20 rule. 
    }
    \label{fig:sm:client_pareto}
\end{figure}

We will now introduce the Pareto chart (Lorentz curve) constructed by using the distribution of reports among clients in a given year as an example.
First, we sort all the clients $c$ by the number of filed reports $n(c)$ (from largest to smallest): $n_1, \dots, n_N$. 
The Pareto chart is then obtained by plotting the normalized ranking score $R(i) = i/N$ on the x-axis, against the cumulative number of reports
\begin{equation}
    C(i) = \frac{\sum_{j = 1}^{i}  n_j }{\sum_{j = 1}^{N}  n_j}
\end{equation}
on the y-axis. 
This specific plot is shown in Fig.~\ref{fig:sm:client_pareto}A. 

\begin{figure}[h]
    \centering
    \includegraphics[width=.65\textwidth]{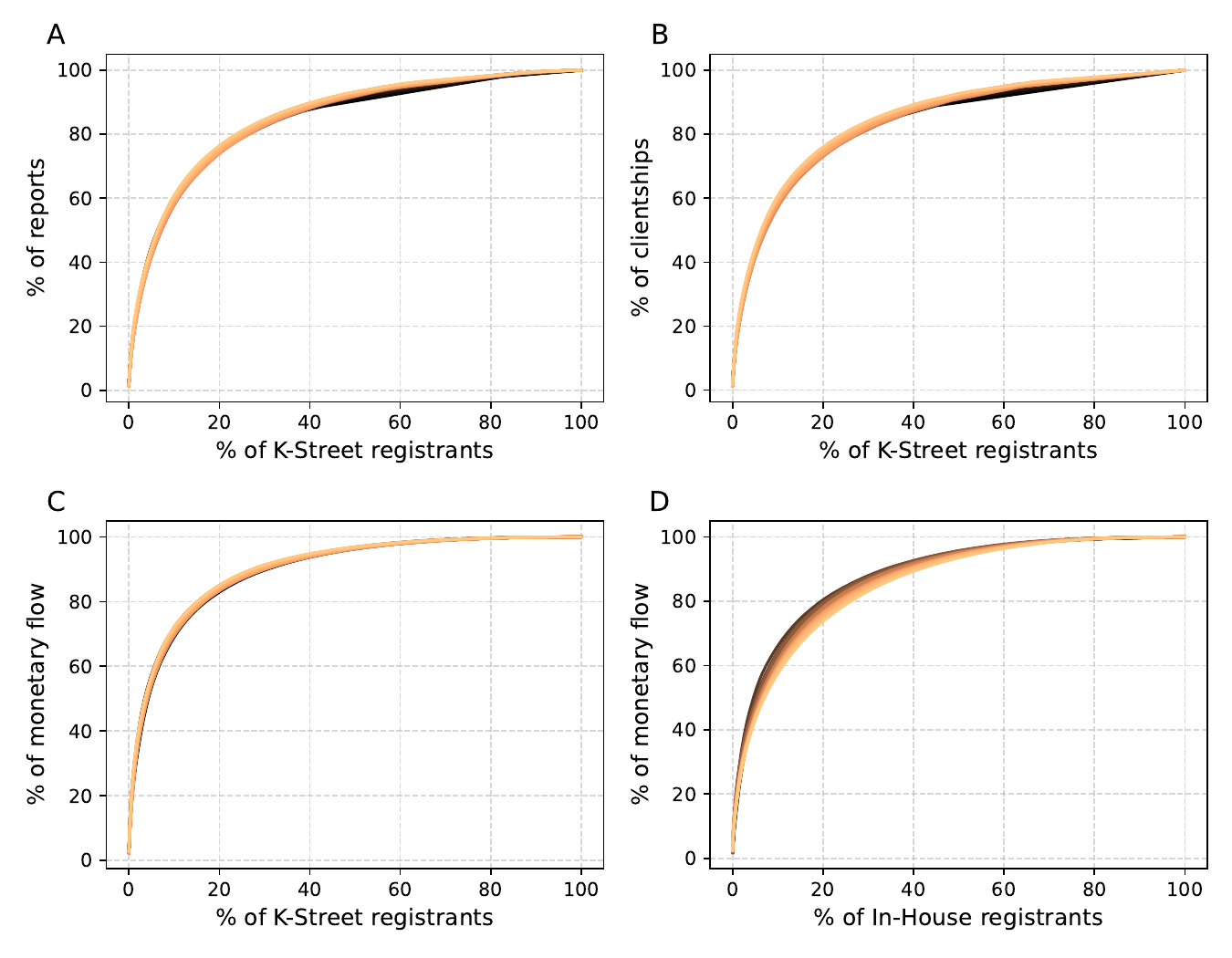}
    \caption{Pareto charts for registrants in 1999--2023. 
    (A) The number of reports per K-Street registrant. 
    (B) The number of clientships per K-Street registrant. 
    (C) Monetary flow per K-Street registrant. 
    (D) Monetary flow per In-House registrant. 
    }
    \label{fig:sm:registrant_pareto}
\end{figure}

More relevant to the problem of disparity is Figure~\ref{fig:sm:client_pareto}B, which shows the Pareto chart based on the total lobbying budget (monetary value of the associated reports). 
It confirms the hierarchical organization of the first layer of the lobbying network, and displays almost exactly the archetypal 80-20 rule, where $80\%$ of lobbying expenses are due to $20\%$ of clients. 
In Fig.~\ref{fig:sm:registrant_pareto}, we perform an analogous analysis for registrants, based on the number of reports they file, their monetary flow, and their number of clients. 
In all cases, we find a variant of the Pareto rule. 

\begin{figure}[h]
    \centering
    \includegraphics[width=.65\textwidth]{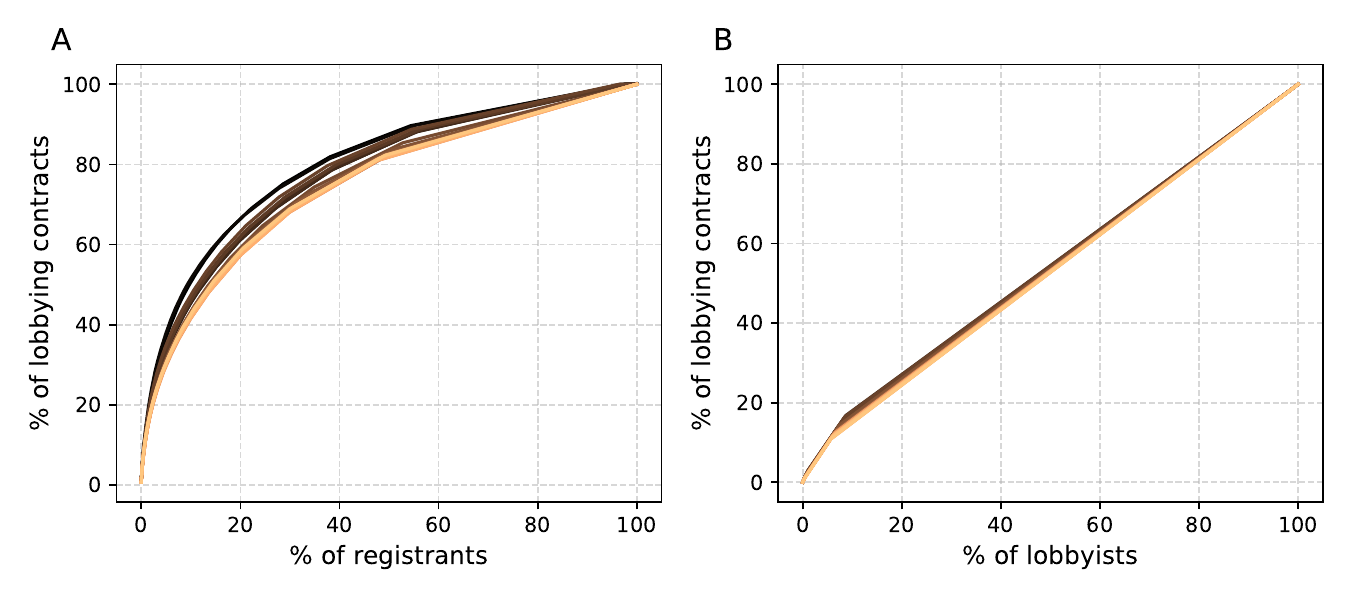}
    \caption{Pareto charts for registrant-lobbyist connections (lobbying contracts) in 1999--2023.  
    (A) The distribution of lobbying contracts among registrants follows the Pareto rule (hierarchy).  
    (B) The number of lobbying contracts among the lobbyists does not follow the Pareto rule (egalitarianism). 
    }
    \label{fig:sm:lobbyist_pareto}
\end{figure}

In Fig.~\ref{fig:sm:lobbyist_pareto}, we contrast the hierarchical organization of registrants with a random organization of lobbyists. 
While $20\%$ of K-Street registrants are responsible for more than $70\%$ of the lobbyist contracts (registrant-lobbyist connection in the network), the connections are distributed among the lobbyists in an egalitarian fashion.

\subsection{Gini index}

The shape of the Pareto chart is often characterized by a \textit{Gini index}, a measure of disparity which can be computed by looking at the area under the Pareto curve. 
For the example of Fig.~\ref{fig:sm:lobbyist_pareto}A introduced in the previous section, 
\begin{equation}
    \text{Gini} = \frac{2}{N} \sum_{i=1}^N  C(i) - R(i).
\end{equation}
For a uniform distribution of reports per clients, $C(i) = R(i)$, so the Gini index would be zero. 
As the disparity between large and small clients increases, the Gini index approaches 1.

\begin{figure}[h]
    \centering
    \includegraphics[width=.45\textwidth]{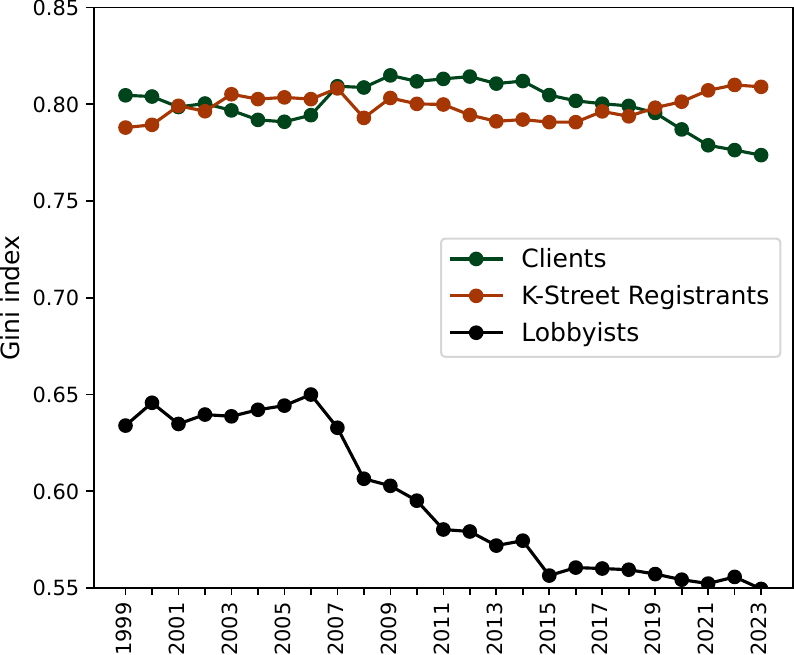}
    \caption{Gini index evolution in time. 
    }
    \label{fig:sm:gini_in_time}
\end{figure}

In Fig.~\ref{fig:sm:gini_in_time}, we show the monetary Gini indices quantifying the distribution of the lobbying budget. 
We find that, over the last two decades, the cumulative Gini indices (based on lobbying activities across all issues and agencies) for both clients and K-street lobbying firms have remained approximately constant at a remarkably high value $\sim 0.8$, indicating that lobbying activities have been persistently dominated by affluent clients and a few large external lobbying firms.
The Gini index for lobbyists' income is relatively lower ($\sim 0.6$) but still markedly higher than the Gini index for U.S. income distribution ($\sim 0.4$ according to the World Bank estimate). 
In other words, the lobbying industry appears to have a higher degree of wealth concentration than the U.S. society as a whole. 

\begin{figure}[h]
    \centering
    \includegraphics[width=.95\textwidth]{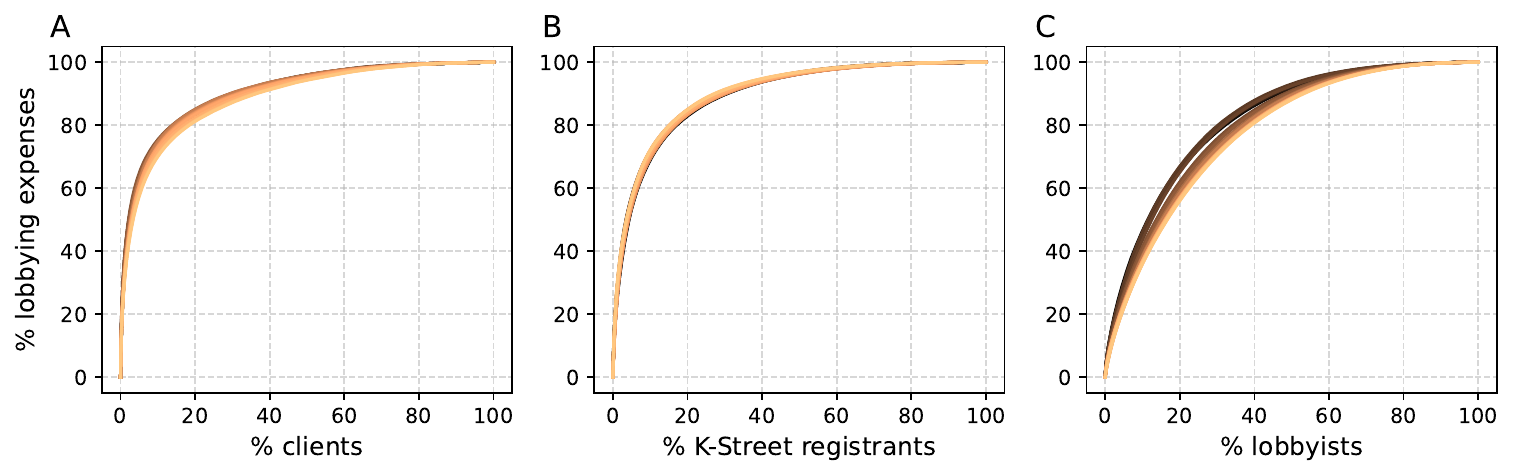}
    \caption{Pareto charts for the budget distribution in 1999--2023.  
    (A) The Pareto chart of the distribution of the total lobbying budget (monetary value of the filings) among the clients.  
    (B) The Pareto chart of the distribution of the total lobbying budget among K-Street registrants.
    (C) The Pareto chart of the distribution of the total lobbying budget among the lobbyists. 
    }
    \label{fig:sm:pareto_for_gini}
\end{figure}

In Fig.~\ref{fig:sm:pareto_for_gini}, we show all the Pareto curves that were used to calculate the Gini indices. 
One important technical remark is that the lobbyist income is computed by dividing the monetary value of a given filing equally among all the lobbyists mentioned therein.

\begin{figure}[h]
    \centering
    \includegraphics[width=.95\textwidth]{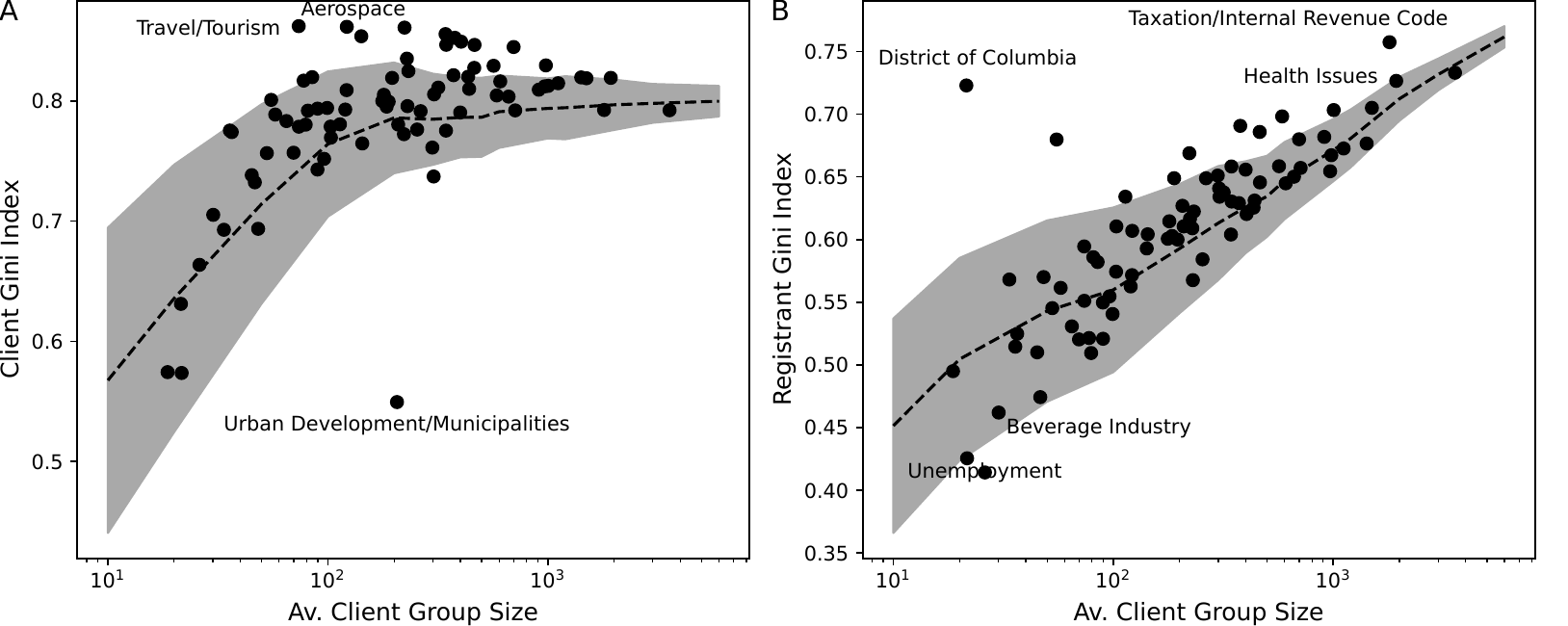}
    \caption{Concentration in area-specific subnetworks.  
    (A) Client Gini Index.  
    (B) Registrant Gini Index. 
    The dashed line and the shaded region correspond to the randomized network with a given number of clients. 
    }
    \label{fig:sm:gini_by_issue}
\end{figure}

\begin{figure}[h]
    \centering
    \includegraphics[width=.45\textwidth]{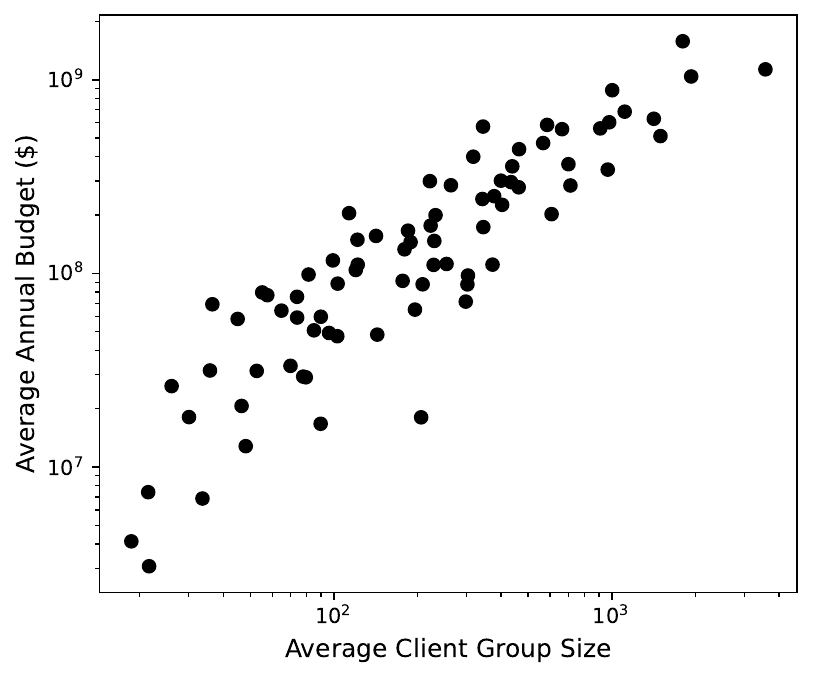}
    \caption{Average value of the client group size interested in one of the standard issue areas in years 1999--2023 plotted against the average annual budget in this area.
    }
    \label{fig:sm:interest_group_vs_budget}
\end{figure}

\begin{figure}[h]
    \centering
    \includegraphics[width=.95\textwidth]{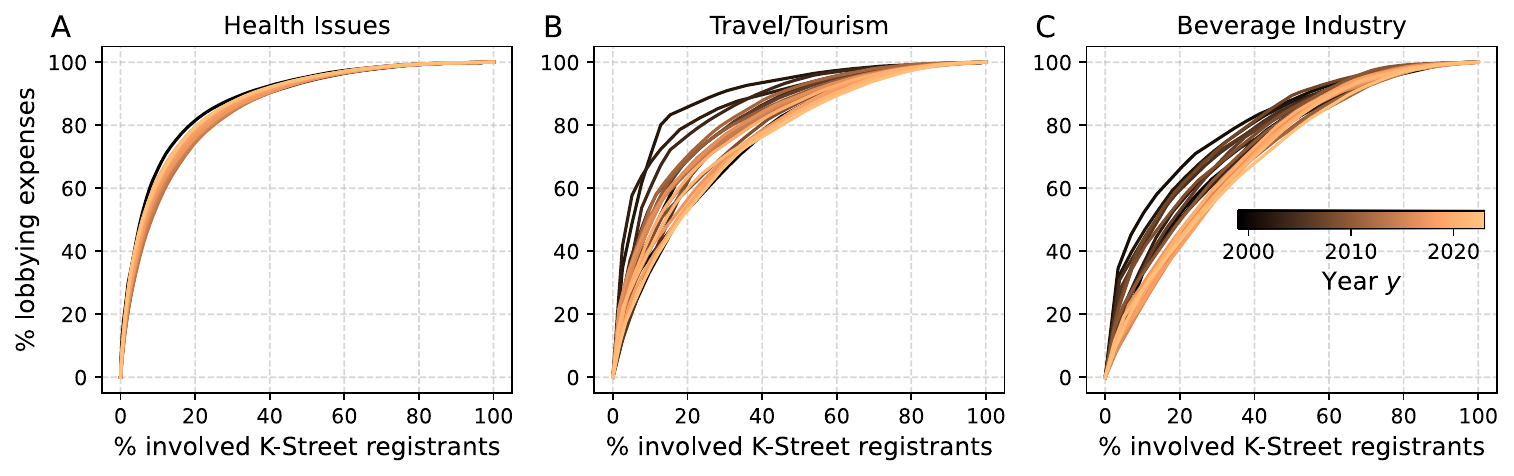}
    \caption{Example Pareto charts for the area-specific budget distribution among K-Street registrants in 1999--2023.  
    (A) Health is an example of a popular (high-budget) issue area. The fact that the Pareto curve is steep indicates a high degree of concentration. 
    (B) Travel/Tourism is an example of an average-sized issue area. Note that in response to the 9/11 crisis, the Pareto curve got steeper. 
    (C) Beverage Industry is an example of a small issue area with a handful of interested clients. In such categories, the budget distributions tend to be most egalitarian. 
    }
    \label{fig:sm:pareto_examples}
\end{figure}

It is also interesting to compute issue-specific Gini indices (for clients and K-Street registrants) are computed by using only the filings $f$ mentioning the given general issue area $a$ ($a \in A(f)$).
Fig.~\ref{fig:sm:pareto_examples} shows K-Street registrant Pareto charts for three selected issue areas with different client group size. 
Generally, we find that the larger the issue, the steeper the Pareto chart, and so the larger the registrant Gini index (Fig.~\ref{fig:sm:gini_by_issue}).
The size of the issue can be measured either by the client group size or the total budget, which are correlated (Fig.~\ref{fig:sm:interest_group_vs_budget}.

The Gini Index and the lobbying network size are also correlated for randomly generated subnetworks (dashed line and shading in Fig.~\ref{fig:sm:gini_by_issue}).
For the Registrant Gini Index (Fig.~\ref{fig:sm:gini_by_issue}B), we choose a random year and a random set of $n$ clients connected to K-Street registrants. 
We then look at all their filings through K-Street registrants, and we use them to construct a partial budget distribution among K-Street registrants. 
This procedure is repeated 100 times for each $n \in \{ 10, 20, 50, 100, 200, 300, 400, 500, 600, 1000, 1200, 2000, 3000, 6000 \}$, and each time we use the partial distribution to compute the K-Street Registrant Gini Index.  
The dashed line and the shade in Fig.~\ref{fig:sm:gini_by_issue}B report the mean and standard deviation of their values, respectively. 
In Fig.~\ref{fig:sm:gini_by_issue}A, we use an analogous procedure to compute the Client Gini Index in a randomized sample. 
This time, we do not restrict our sample to K-Street Clients, but we use the same values of $n$. 
The relationship between the size of the subnetwork and the Gini Index underscores the importance of working with a complete dataset. 
Indeed, subsampling could introduce systematic biases.

\clearpage
\section{Higher order interactions}
So far, we have analyzed the ties between clients and registrants, as well as the ties between registrants and lobbyists. 
We can also more directly ask about the connections between clients and lobbyists by introducing the notion of \textit{client-registrant-lobbyist triads} 
\begin{equation}
    \text{Lobbying triads}(y) = \{ (c,r,l): \exists f: y = y(f) \land c = c(f) \land r= r(f) \land l \in L(f)\}.
\end{equation}

\begin{figure}[h]
    \centering
    \includegraphics[width=.85\textwidth]{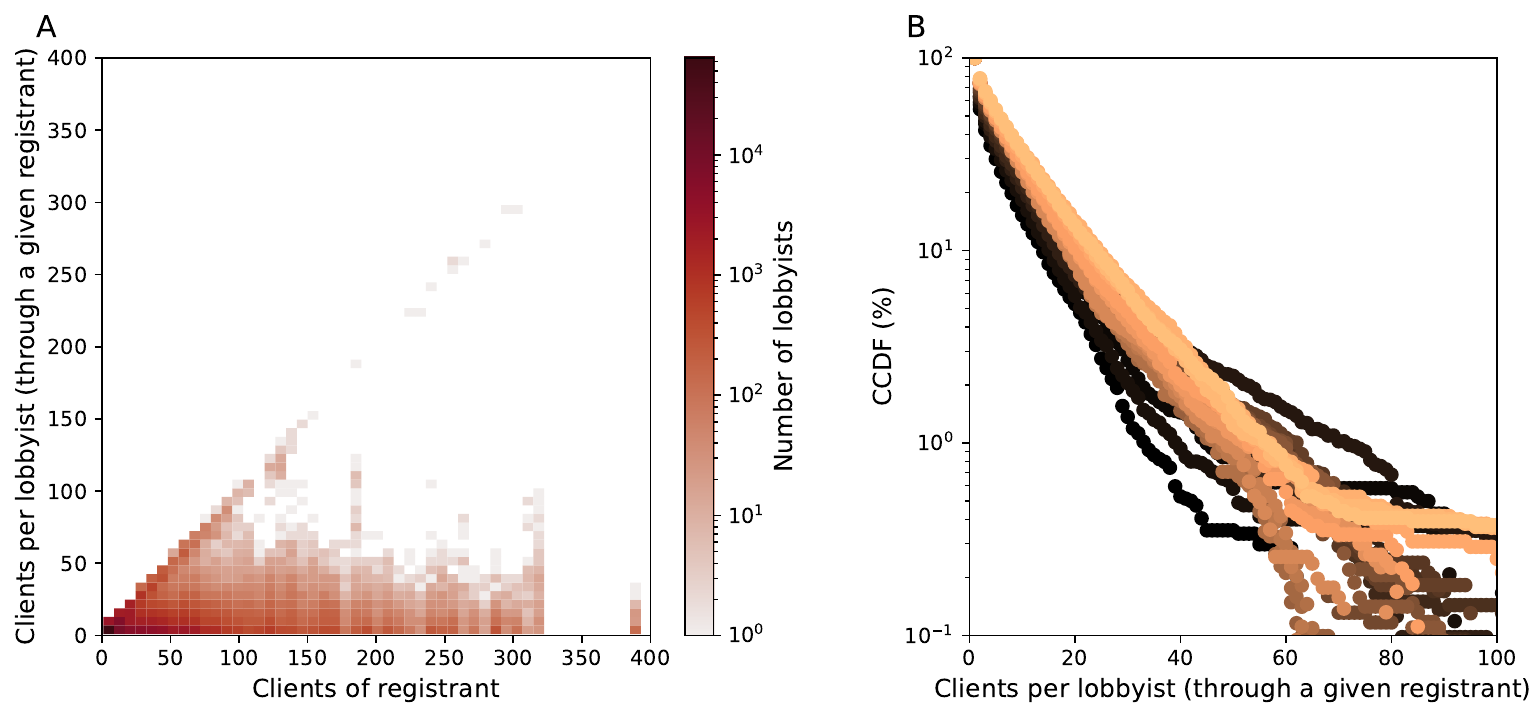}
    \caption{Assignment of clients of a given registrant to their lobbyists.   
    (A) The distribution of the number of clients of registrant $r$ that lobbyist $l$ worked for, as a function of the total number of clients of a given registrant.
    (B) The marginal distribution of the number of clients of registrant $r$ that lobbyist $l$ worked for, broken by year. 
    Data from years 1999--2023.
    }
    \label{fig:sm:lobbyist_jobs}
\end{figure}

We will now use the notion of lobbying triads to answer questions about the distribution of jobs in a given lobbying firm. 
Specifically, we can analyze the number $n(r,l)$ of clients of registrant $r$ that lobbyist $l$ worked for, i.e., all the lobbying triads of the form $(\ast,l,r)$.
By construction, $n(r,l)$ cannot be larger than the number of clients of $r$ (i.e., $\text{in-deg}(r)$).
In Fig.~\ref{fig:sm:lobbyist_jobs}A, we show that the number of clients a given lobbyist works for almost never exceeds 100, even if the number of clients of a given registrant is much larger. 
Nevertheless, we also find examples of lobbyists whose names appear on virtually all the reports of a given registrant, up to 300 clients per lobbyist. 
For most registrant-lobbyist pairs, the number of clients per lobbyist $n(r,l)$ decays exponentially (Fig.~\ref{fig:sm:lobbyist_jobs}B). 

On the one hand, this finding corroborates the general picture that the distribution of lobbying cases per individual lobbyist is more egalitarian than the distribution of lobbying cases per lobbying firms that have the capacity to accumulate advantage. 
On the other hand, we also find a small number of \textit{super-lobbyists}, who appear to be an exception to this rule. 
Further analysis of the profiles of individual lobbyists (generalists vs.~specialists, etc.) can be the subject of an interesting future study.

\section{Lobbying network visualization}

To visualize the entire network graph for one year (as shown in Fig.1A for 2017), we use a customized multilayer layout, with a particular embedding strategy for each set of nodes. 
We will now describe the placement of nodes in each layer separately, and then the way they are assembled together. 
The final visualization is facilitated with the \emph{PyGraphistry} library and its cloud-based visualization tool~\cite{pygraphistry}.

\subsection{Node placement}

The $x$-coordinate of a client node $c$ represents its lobbying expenses in year $y$ and is proportional to $-\log \left[ \text{w-out-deg}(c,y) \right]$,
where $\text{w-out-deg}(c,y)$ has been defined in eq.~\eqref{eq:c_weighted_out_deg_both_def}. 
The $y$-coordinate of node $c$ is proportional to the client's inclination $i_C(c,y)$ to lobby through In-House registrants such that
\[
i_C(c,y) =\frac{ \text{w-out-deg}_{I}(c,y) - \text{w-out-deg}_{K}(c,y) }{ \text{w-out-deg}(c,y) }.
\]

The $x$-coordinate of a registrant node $r$ represents its monetary flow in year $y$ and is proportional to $
-\log \left[ \text{w-in-deg}(r,y) \right]$.
The $y$-coordinate of node $r$ represents its out-degree and is proportional to $
\chi(r) \log \left[ \text{out-deg}(r,y) \right]
$
where
\[
\chi(r) = \begin{cases}
  1 & \text{if} \quad r \in \text{In-House} \\
  -1 & \text{if} \quad r \in \text{K-Street}. \\
\end{cases}
\]

The $x$-coordinate of a lobbyist node $l$ represents its in-degree in year $y$ and is proportional to $
- \text{in-deg}(l,y)$.
The $y$-coordinate of node $l$ is proportional to the lobbyist's inclination $i_L(l,y)$ to lobby through In-House registrants such that
\[
i_L(l,y) =\frac{\sum_{(r,l) \in E(\mathcal{G}_y)} \chi(r) }{\text{in-deg}(l,y)}.
\]

\label{sec:sm:node_placement:gov}
The $x$-coordinate of a government entity node $g$ represents its monetary flow $\mu(g,y)$ in year $y$. 
Specifically, it is a piece-wise linear, order-preserving  transformation of $-\log \left[ \mu(g,y) \right]$,
where
\[
\mu(g,y) = \sum_{f} \mathbbm{1}\left[ y(f)=y \land g \in G(f) \right] m(f).
\] 
The $y$-coordinate of node $g$ reflects its similarity (measured by co-mention count in filings) to other government entity nodes.
In particular, we construct virtual weighted edges between government entity nodes such that the edge weight is set to the normalized government entity co-mention count in filings, and the edges with a small co-mention count are removed.
We then run the Fruchterman--Reingold force-directed algorithm~\cite{fruchterman1991graph} on the subgraph consisting of government entity nodes and their internal edges and use the obtained close-to-equilibrium node $x$-coordinates to set the $y$-coordinates of the government entity nodes.
The two largest government entity nodes (Senate and House of Representatives) are positioned manually to align them with the corresponding division in the last (legislators) layer.

The legislator nodes are arranged on semicircles with the radii highlighting the difference in size between the Senate and House of Representatives.
On each semicircle, the legislator nodes are spaced/distributed evenly, grouped by party affiliation (such that Democrats have higher $y$-coordinates than Republicans), and ordered randomly within each group.
In 2017, our data also contained one Independent Senator, but we neglected the corresponding node in the visualization.

\begin{figure}[h]
    \centering
    \includegraphics[width=.8\textwidth]{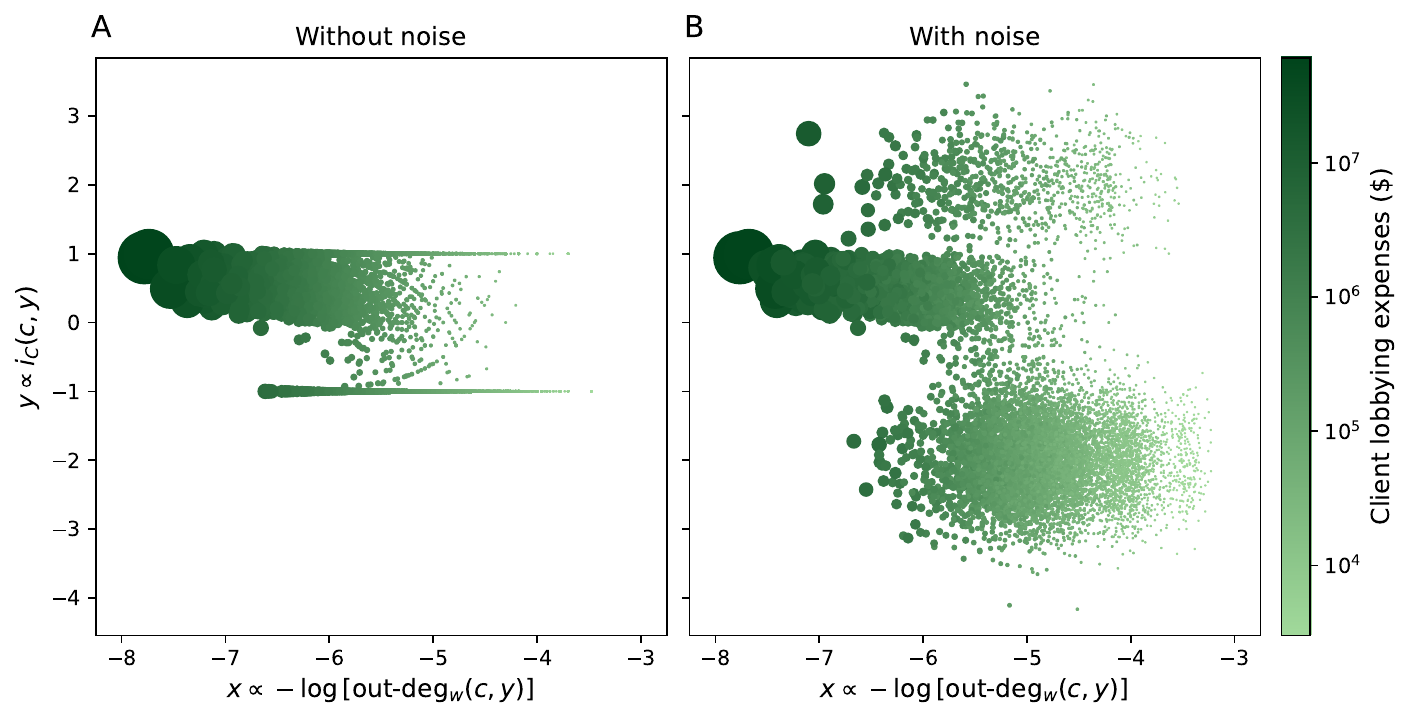}
    \\
    \includegraphics[width=.8\textwidth]{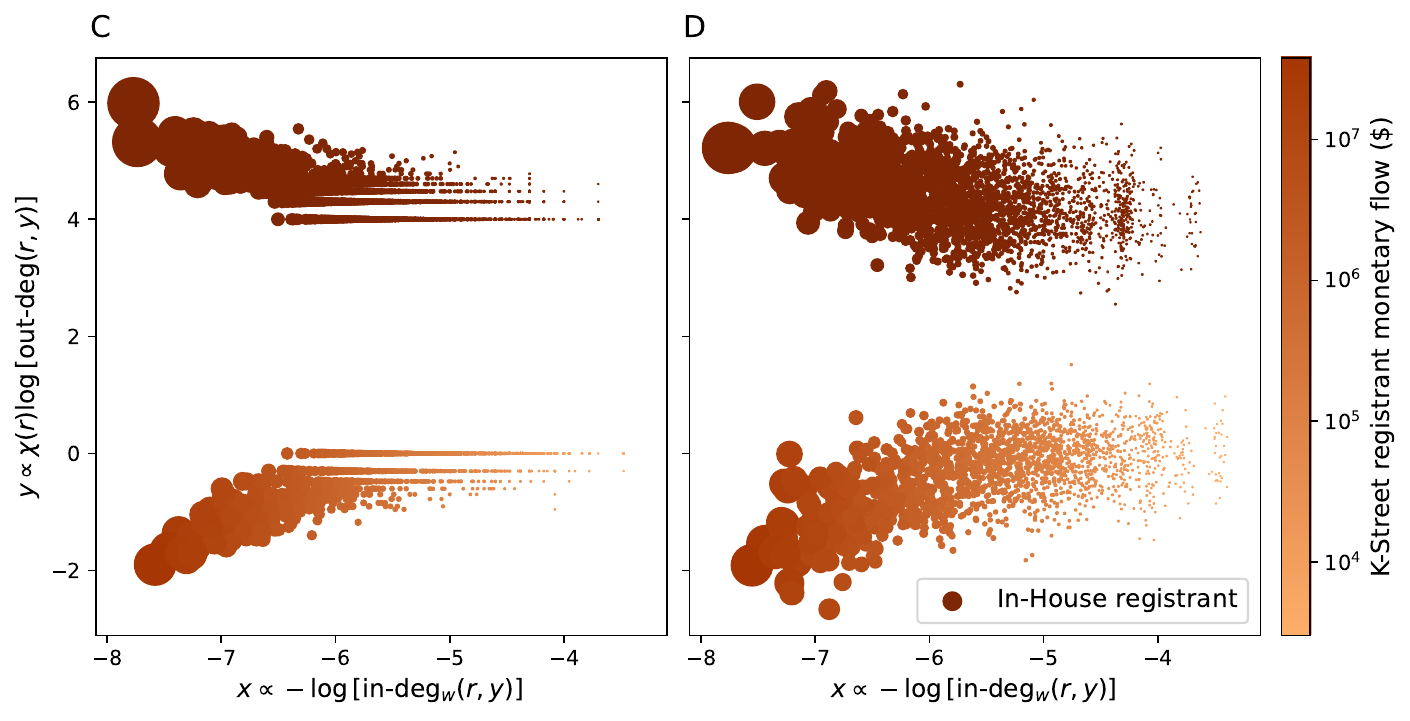}
    \caption{Node placement based on network graph measures without (left) and with (right) the white noise perturbation. 
    Used to construct the whole lobbying network graph for the year 2017 in Fig.~1A.
    (A, B) Clients.
    (C, D) Registrants.
    }
    \label{fig:sm:node_placement_noise_perturbation:c_r}
\end{figure}

\begin{figure}[h]
    \centering
    \includegraphics[width=.8\textwidth]{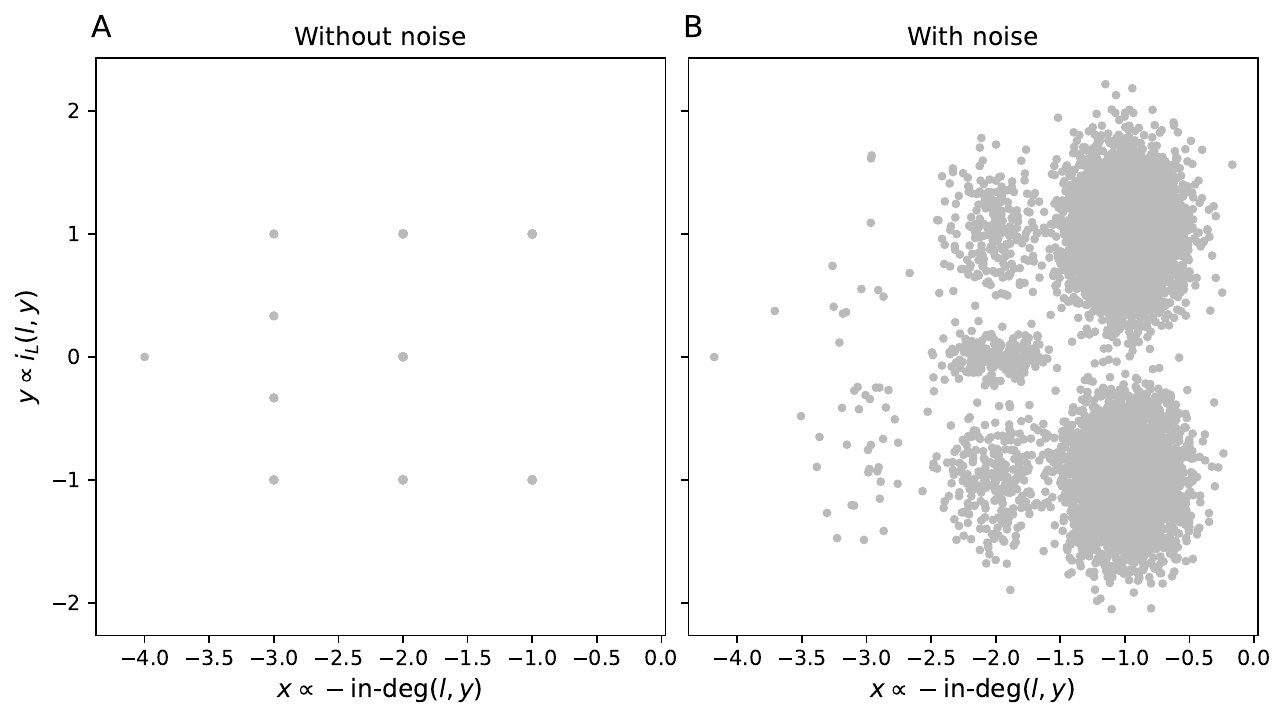}
    \\
    \includegraphics[width=.819\textwidth]{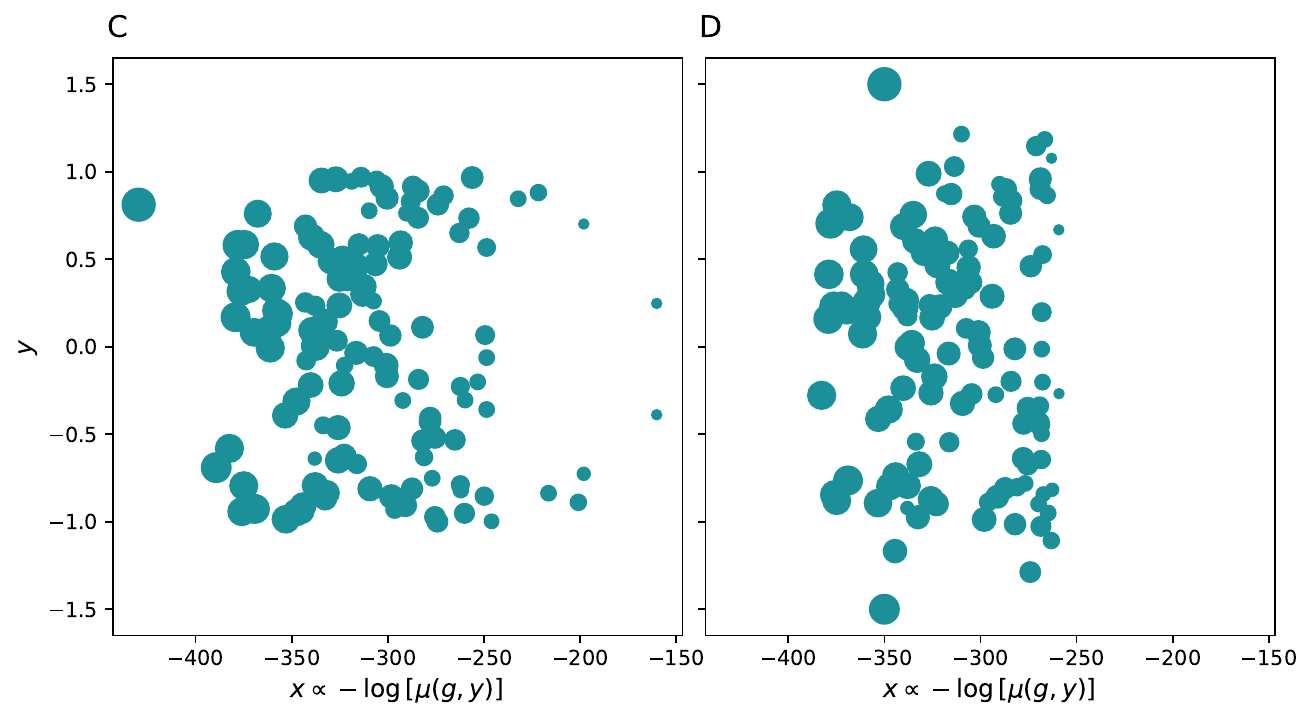}
    \caption{Node placement based on network graph measures without (left) and with (right) the white noise perturbation.
    Used to construct the whole lobbying network graph for the year 2017 in Fig.~1A.
    (A, B) Lobbyists.
    (C, D) Government entities: Panel (C) does not include the piece-wise linear, order-preserving transformation of $x$-coordinates nor the manual re-positioning of the two largest nodes (Senate and House of Representatives). For $y$-coordinate computation see Sec.~\ref{sec:sm:node_placement:gov}.
    }
    \label{fig:sm:node_placement_noise_perturbation:l_g}
\end{figure}

Except for the legislator nodes, we perturb all node positions with white noise for better node separation in graph visualization.
The differences between node placements without and with the white noise perturbation for all affected types of nodes are shown in Fig.~\ref{fig:sm:node_placement_noise_perturbation:c_r}--\ref{fig:sm:node_placement_noise_perturbation:l_g}.
As we can see, the white noise perturbation reduces the overlapping of nodes, while minimally disturbing/violating the above-described node placement based on network graph measures.

\subsection{Node size and color}
The client node size is proportional to the client's lobbying expenses $\text{w-out-deg}(c,y)$.
The registrant node size is proportional to the registrant's monetary flow $\text{w-in-deg}(r,y)$.
The size of the government entity node $g$ represents its total mention count $n(g,y)$ in filings in year $y$ and is proportional to $\log \left[ n(g,y) \right]$,
where 
\[
n(g,y) = \sum_{f} \mathbbm{1}\left[ y(f) = y \land g \in G(f) \right].
\]
The lobbyist and legislator node sizes are fixed.
Note that the above-described node sizes are passed to the \textit{PyGraphistry} library, where a further order-preserving transformation is applied~\cite{pygraphistry}.

The color-coding of nodes is summarized in Tab.~\ref{tab:sm:node_colors}.
The color of client and K-Street registrant nodes ranges between the specified minimum and maximum values in proportion to the corresponding scaling law.
Specifically, the client node color represents the client's lobbying expenses, and the K-Street registrant node color reflects the registrant's monetary flow.
The colors of other node types are all fixed.

\begin{table}[h]
    \centering
    \caption{Lobbying network node colors. Used in Fig.~1A and Fig.~2B of the main text.}
    \begin{tabular}{lccc}
    \toprule
    & \multicolumn{2}{c}{HEX color} & \\
    Node type & Min & Max & Scaling law \\
    \midrule
    Client & \hexcolorbox{A1D99B} & \hexcolorbox{00441B} & $\log \left[ \text{w-out-deg}(c,y) \right]$ \\
    K-Street registrant & \hexcolorbox{FDAE6B} & \hexcolorbox{A63603} & $\log \left[ \text{w-in-deg}(r,y) \right]$ \\
    In-House registrant & \multicolumn{2}{c}{\hexcolorbox{7F2704}} & -- \\
    Lobbyist & \multicolumn{2}{c}{\hexcolorbox{BABABA}} & -- \\
    Government entity & \multicolumn{2}{c}{\hexcolorbox{1C9099}} & -- \\
    Democratic legislator & \multicolumn{2}{c}{\hexcolorbox{0015BC}} & -- \\
    Republican legislator & \multicolumn{2}{c}{\hexcolorbox{FF0000}} & -- \\
    \bottomrule
    \end{tabular}
    \label{tab:sm:node_colors}
\end{table}

\subsection{Final assembly}
We embed the above-described sets of nodes in a common 2D space using translation and scaling operations as needed for a reasonable layered network arrangement.
The resulting $(x,y)$-coordinates, along with the color and size of all nodes, constitute the final \textit{node data}.
We then construct unweighted edges as described in Sec.~\ref{sec:sm:network_construction:edges} such that the obtained list of $(\text{source node}, \text{destination node})$ pairs forms \textit{edge data}.
We also specify the following \textit{global properties}:
node opacity = 100\% (default),
edge opacity = 5\%,
edge color interpolates the colors of the connecting nodes along the edge (default),
edge size/width = 50 (default),
and edge curvature = 20\% (default).

The \textit{node data}, \textit{edge data}, and \textit{global properties} are fed to the \emph{PyGraphistry} library to render the lobbying network visualization in the browser using the cloud-based visualization tool~\cite{pygraphistry}.

\section{Network evolution}
In this Section, we describe the technical aspects of our network evolution analysis.

\subsection{Rejuvenation}
\label{sec:sm:rejuvenate}
By comparing the nodes between the lobbying networks $\mathcal{G}_{y_1}$ and $\mathcal{G}_{y_2}$ in two different years $y_1$ and $y_2$, we can quantify the degree to which the lobbying network refreshes and rejuvenates. 

\begin{figure}[h]
    \centering
    \includegraphics[width = \textwidth]{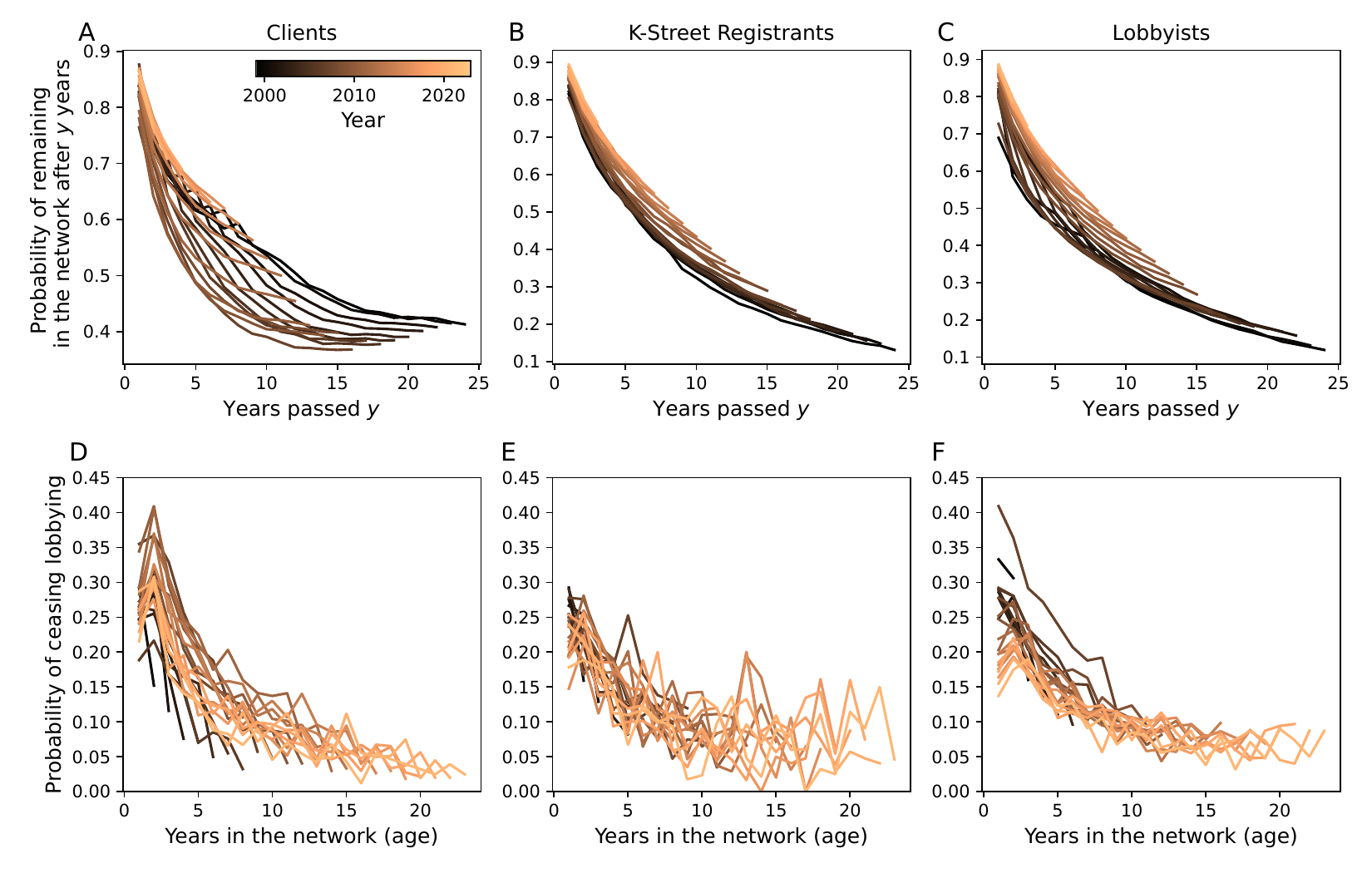}
    \caption{Node removal probability.
    (A) Client retention probability. (B) K-Street registrant retention probability. (C) Lobbyist retention probability. 
    (D) Probability of client ceasing lobbying conditioned on their `age'. 
    (E) Probability of K-Street registrant ceasing lobbying conditioned on their `age'.
    (F) Probability of lobbyist ceasing lobbying conditioned on their `age'.}
    \label{fig:sm:survival}
\end{figure}

In Fig.~\ref{fig:sm:survival}A-C, we estimate the empirical (frequentist) probability of a node remaining in the network after $y$ years.  
For clients (Fig.~\ref{fig:sm:survival}A), this quantity is defined as
\begin{equation}
    \mathbb{P}\left[ \text{Client lobbying in year } y_0 \text{ remains in the network after } y \text{ years}  \right]  = \frac{|\text{Clients}(y_0) \cap \text{Clients}(y_0 + y) |  }{|\text{Clients}(y_0) |},
\end{equation}
and the definition is analogous for K-Street registrants (Fig.~\ref{fig:sm:survival}B, and Fig.~2 of the main text) and lobbyists (Fig.~\ref{fig:sm:survival}C). 
The different curves in this figure correspond to different starting years $y_0 \in \{1999, 2000, \dots, 2022\}$. 
For clients, these curves do not collapse, most likely due to economical interruptions. 
For registrants and lobbyists, the decay of operation probability to first approximation follows a universal law, but we also observe a steady increase in longevity in recent years. 

The probability of leaving the lobbying network is not the same for all the clients/registrants/lobbyists, but it depends on individual characteristics. One particularly interesting predictor of a node ceasing to operate is age,
defined as the maximum number of consecutive years for which node $x$ was present in the network prior to year $y$,
\begin{equation}
\text{Age}(x,y) = \sum_{y'<y} \mathbbm{1}\left[ \forall_{y'\leq y'' \leq y} :  
x \in V(\mathcal{G}_{y''})    \right].
\end{equation}
Figure~\ref{fig:sm:survival}D shows that the conditional probability that a client $c$ of a given age ceases lobbying in year $y_0$ (i.e., $y_0$ is the last year the client is observed in the lobbying network) conditioned on its age $y$ 
\begin{equation}
\mathbb{P}\left[c \text{ ceases lobbying in year } y_0 |\text{Age}(c,y_0) = y \right] = 
\frac{
    \sum_{c \in V(\mathcal{G}_{y_0})} \mathbbm{1}\left[ \text{Age}(c,y_0) = y \land c \notin V(\mathcal{G}_{y_0+1}) \right]
}{
    \sum_{c \in V(\mathcal{G}_{y_0})} \mathbbm{1}\left[ \text{Age}(c,y_0) = y \right]
}    
\end{equation}
is a decaying function of age. 
The same is true for an analogous quantity for K-Street registrants (Fig.~\ref{fig:sm:survival}E), and lobbyists (Fig.~\ref{fig:sm:survival}F).
In other words, long-operating lobbying actors are less likely to leave the lobbying network. 
This trend over time may lead to `aging' and accumulation of expertise. 
Interestingly, the \textit{aging curve} has a similar shape for clients, K-street registrants, and lobbyists, with a characteristic timescale of approximately 10 years. 

\subsection{Preferential attachment/detachment inference}
\label{sec:sm:preferential}
Following the works of Barabasi et al.~\cite{Barabasi1999}, we analyze the dynamics of registrant connections through the lens of preferential attachment.
In the main text, we present the results of this analysis for clientships (client-registrant connections) in Fig.~2E-F. 
The specific quantity that we look at is defined as follows. 
First, we identify K-Street registrants with a degree $d$ in year $y-1$:
\begin{equation}
    \text{K-Street Registrants}_{\text{in} = d}  (y-1) =  \{r \in  \text{K-Street Registrants}(y-1) : \text{in-deg}(r,y-1) = d \}.
\end{equation}
Then, we find all the new clientships that involve an existing registrant of in-degree $d$
\begin{equation}
 \text{Attached clientships}(y; d) = \left \{ (c,r) \in E(\mathcal{G}_{y}) \setminus E(\mathcal{G}_{y-1}) : r \in  \text{K-Street} \land r \in V(\mathcal{G}_{y-1}) \cap V(\mathcal{G}_{y}) \land \text{in-deg}(r, y-1) = d  \right \}
\end{equation}
and removed clientships
\begin{equation}
 \text{Detached clientships}(y; d) = \left \{ (c,r) \in E(\mathcal{G}_{y-1}) \setminus E(\mathcal{G}_{y}) : r \in  \text{K-Street} \land r \in V(\mathcal{G}_{y-1}) \cap V(\mathcal{G}_{y}) \land \text{in-deg}(r, y-1) = d  \right \}
\end{equation}

The absolute probability that one of the attached clientships involves an existing registrant of in-degree $d$ is simply the ratio
\begin{equation}
    p^+(d,y) = \frac{ | \text{Attached clientships}(y; d)| }{\sum_{d'} |\text{Attached clientships}(y; d')|}.
\end{equation}
Analogously, the absolute probability that one of the detached clientships involves an existing registrant of in-degree $d$ is
\begin{equation}
    p^-(d,y)  = \frac{ | \text{Detached clientships}(y; d)| }{\sum_{d'} |\text{Detached clientships}(y;  d')|}.
\end{equation}

\begin{figure}[h]
    \centering
    \includegraphics[width = .85\textwidth]{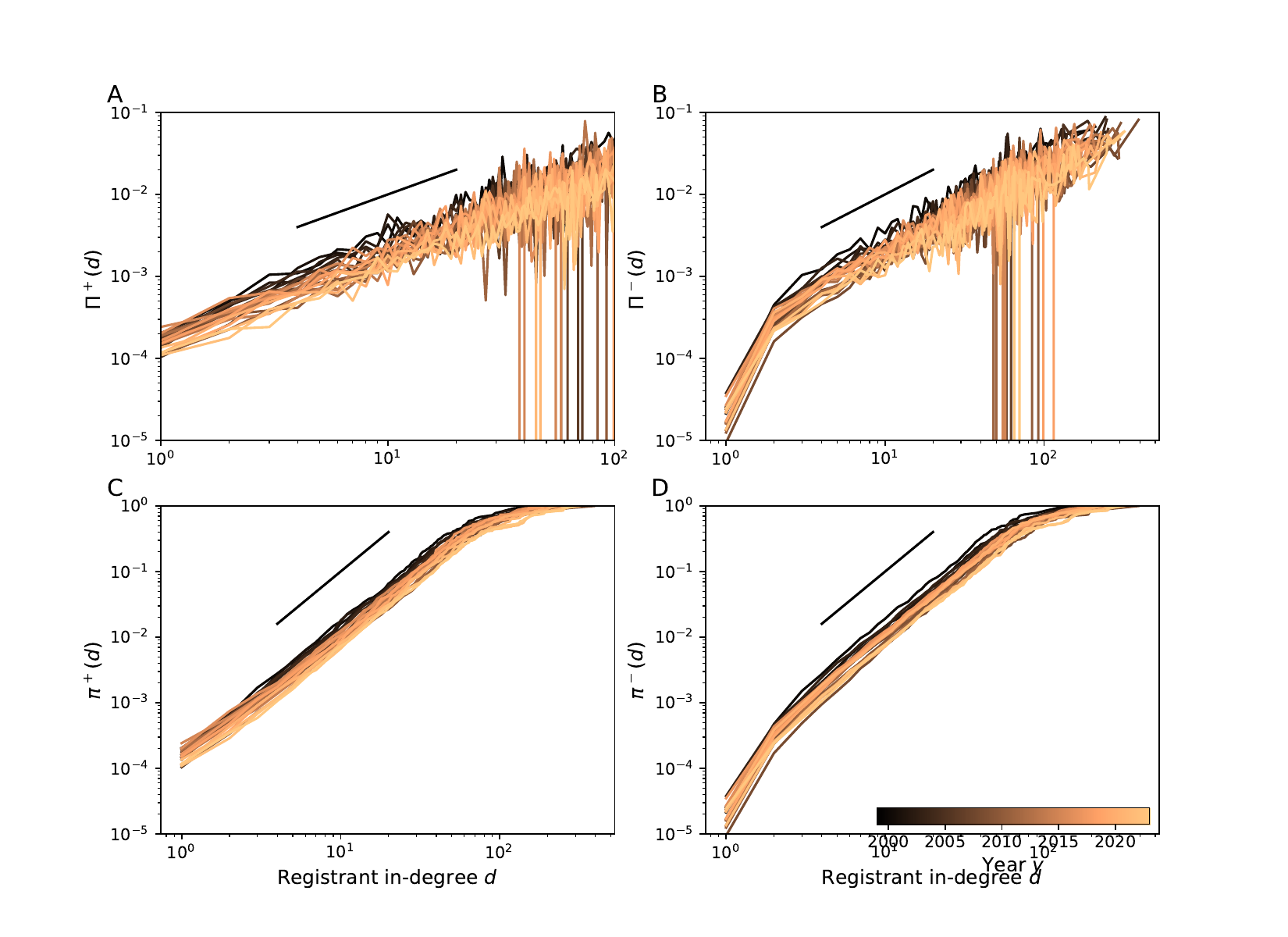}
    \caption{Clientship dynamics in terms of K-Street registrant in-degree $\text{in-deg}(r,y)$. (A) Preferential attachment function $\Pi^{+}(d,y)$. (B) Preferential detachment function $\Pi^{-}(d,y)$.
    (C) Cumulative preferential attachment function $\pi^{+}(d,y)$.
    (D) Cumulative preferential attachment function $\pi^{-}(d,y)$.
    For definitions of attachment functions see eq.~\eqref{eq:sm_pi_+_d}--\eqref{eq:sm_pi_cum}.
    All attachment functions are computed separately for each year in the range 1999--2023.
    }
    \label{fig:sm:preferential}
\end{figure}

Of more interest is the \textit{attachment function}
\begin{equation}
\Pi^{+}(d,y) = \left(\frac{| \text{Attached clientships}(y;  d)|  }{|\text{K-Street Registrants}_{\text{in} = d}  (y-1)| } \right)  \left( \sum_{d'} \frac{| \text{Attached clientships}(y;  d')|  }{|\text{K-Street Registrants}_{\text{in} = d'}  (y-1)| } \right)^{-1},
\label{eq:sm_pi_+_d}
\end{equation}
and the \textit{detachment function }
\begin{equation}
\Pi^{-}(d,y)= \left(\frac{| \text{Detached clientships}(y; d)|  }{|\text{K-Street Registrants}_{\text{in} = d}  (y-1)| } \right)  \left( \sum_{d'} \frac{| \text{Detached clientships}(y;  d')|  }{|\text{K-Street Registrants}_{\text{in} = {d'}}  (y-1)| } \right)^{-1},
\label{eq:sm_pi_-_d}
\end{equation}
which capture the preferential tendencies in network development.
Figure~\ref{fig:sm:preferential}A-B shows a direct evaluation of them for our lobbying network.
As the attachment function may be noisy, in Fig.~\ref{fig:sm:preferential}C-D we show the equivalent, but more robust to infer, \textit{cumulative attachment/detachment functions}~\cite{Barabasi_book}
\begin{equation}
    \pi^\pm(d,y) = \sum_{k=1}^d \Pi^\pm(k,y). 
    \label{eq:sm_pi_cum}
\end{equation}
Critically, functions $\Pi^{+}(d,y) $ are independent of $y$ and linear in $d$ (equivalently, $\pi^+(d,y) \sim d^2$), i.e., the clientship dynamics satisfy linear preferential attachment~\cite{Barabasi1999}. 
Interestingly, the detachment function  $\Pi^{-}(d,y) $ is also linear in $d$, so we also find linear preferential detachment.

\subsection{Registrant out-degree dynamics}
\begin{figure}[h]
    \centering
    \includegraphics[width = .85\textwidth]{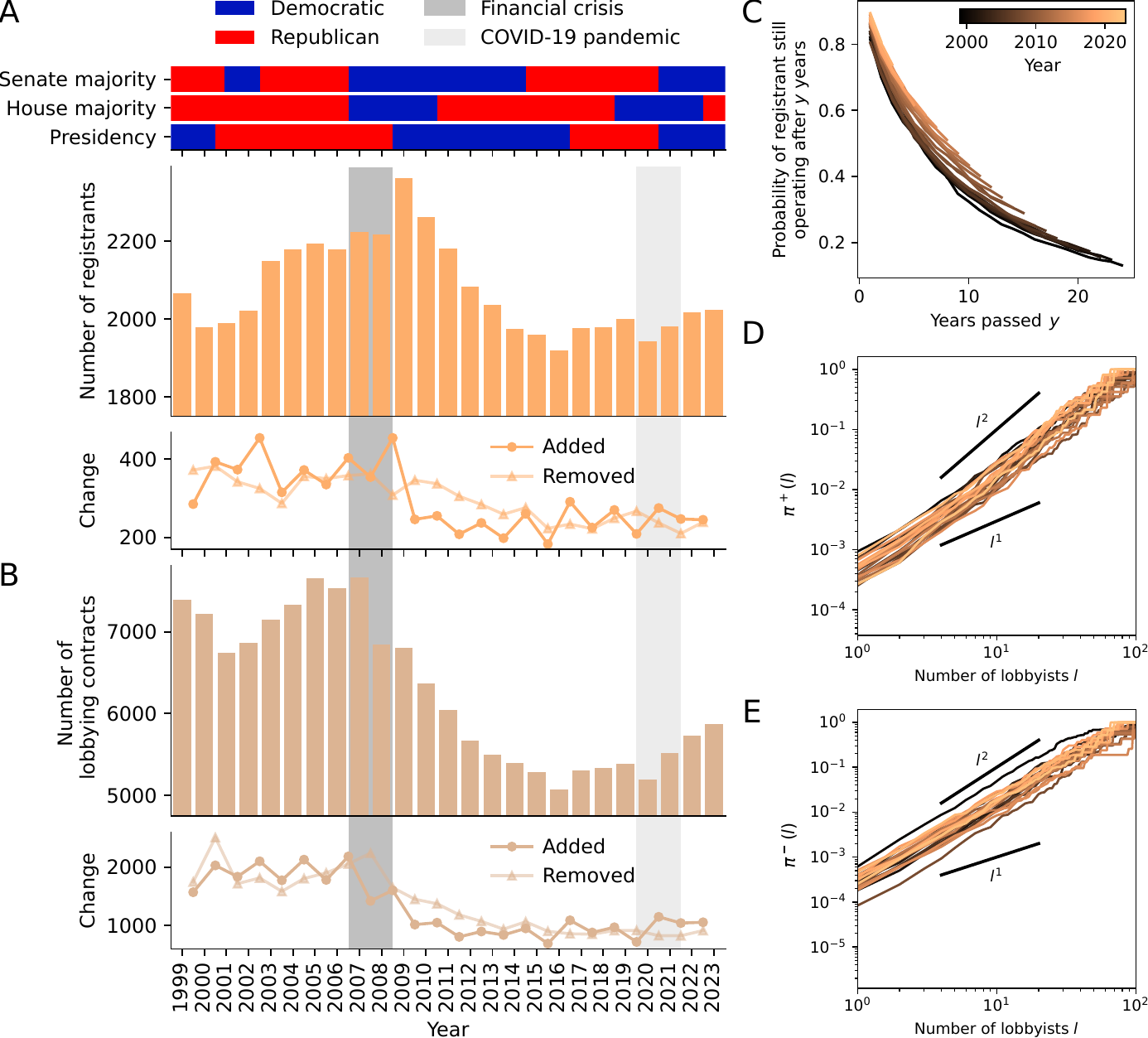}
    \caption{Lobbying contract dynamics for K-Street registrants. (A)
    Number of active K-Street registrants per year (bars) and their yearly increments (circles) and decrements (triangles).
    The top horizontal bars show the party (Democratic in blue and Republican in red) of Senate majority, House majority, and President by year.
    The financial crisis (2007--2008) and COVID-19 pandemic (2019--2021) are shown by gray overlays.
    (B)~Number of active lobbying contracts (registrant-lobbyist connections) per year (bars) and their yearly increments (circles) and decrements (triangles).
    (C)~K-Street registrant `survival probability'. 
    (D)~Cumulative preferential attachment function $\pi^+(l)$ in terms of K-Street registrant out-degree.
    (E)~Cumulative preferential detachment function $\pi^-(l)$.
    Quantities in (C--E) are computed separately for each year in the range 1999--2023.
    }
    \label{fig:sm:kstreet_lobbyist_evolution}
\end{figure}

\begin{figure}[h]
    \centering
    \includegraphics[width = .85\textwidth]{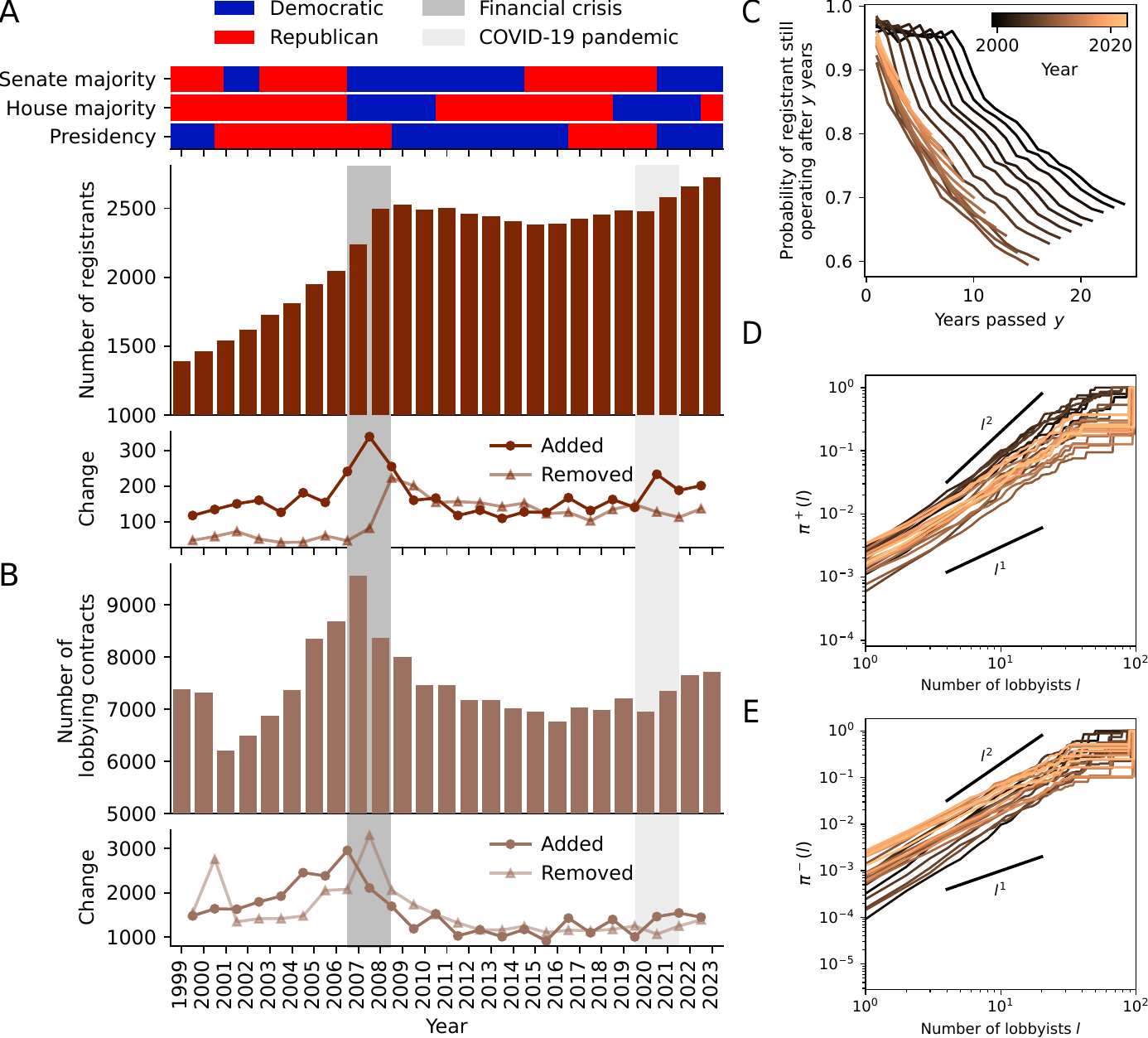}
    \caption{Lobbying contract dynamics for In-House registrants. (A)
    Number of active In-House registrants per year (bars) and their yearly increments (circles) and decrements (triangles).
    The top horizontal bars show the party (Democratic in blue and Republican in red) of Senate majority, House majority, and President by year.
    The financial crisis (2007--2008) and COVID-19 pandemic (2019--2021) are shown by gray overlays.
    (B)~Number of active lobbying contracts (registrant-lobbyist connections) per year (bars) and their yearly increments (circles) and decrements (triangles).
    (C)~In-House registrant `survival probability' is generally larger than the `survival probability' of K-Street registrants (Fig.~\ref{fig:sm:inhouse_lobbyist_evolution}C). We also see a qualitative shift around the financial crisis of 2007--08 to less stable In-House registrants (higher drop-out probability).  
    (D)~Cumulative preferential attachment function $\pi^+(l)$ in terms of In-House registrant out-degree.
    (E)~Cumulative preferential detachment function $\pi^-(l)$.
    Quantities in (C--E) are computed separately for each year in the range 1999--2023.
    }
    \label{fig:sm:inhouse_lobbyist_evolution}
\end{figure}

It is interesting to ask whether the preferential attachment is also present in the lobbying contract dynamics of registrant-lobbyist connections. 
In Figures~\ref{fig:sm:kstreet_lobbyist_evolution} and \ref{fig:sm:inhouse_lobbyist_evolution}, which are analogous to Fig.~2 of the main text, we analyze the evolutionary dynamics of the lobbying contracts involving K-Street and In-House registrants, respectively. 

Panels A and C of Fig.~\ref{fig:sm:kstreet_lobbyist_evolution} are identical to the respective panels of Fig.~2. 
Figure~\ref{fig:sm:kstreet_lobbyist_evolution}C, however, instead of focusing on registrants' incoming edges (clientships), reports the number of their out-going edges (lobbying contracts). 
Figure~\ref{fig:sm:kstreet_lobbyist_evolution}D,E presents the preferential attachment/detachment analysis, analogous to the one outlined in Sec.~\ref{sec:sm:preferential}.
Specifically, we define
\small
\begin{equation}
    \text{K-Street Registrants}_{\text{out} = d}  (y-1) =  \{r \in  \text{K-Street Registrants}(y-1) : \text{out-deg}(r,y-1) = d \},
\end{equation}
\begin{equation}
 \text{Attached lobbying contracts}(y; d) = \{ (r,l) \in E(\mathcal{G}_{y}) \setminus E(\mathcal{G}_{y-1}) : r \in \text{K-Street} \cap \in V(\mathcal{G}_{y-1}) \cap V(\mathcal{G}_{y}) \land \text{out-deg}(r, y-1) = d   \},
\end{equation}
\begin{equation}
\begin{split}
     \text{Detached lobbying contracts}(y; d) = \{ (r,l) \in E(\mathcal{G}_{y-1}) \setminus E(\mathcal{G}_{y}) : r \in  \text{K-Street} \cap r \in V(\mathcal{G}_{y-1}) \cap V(\mathcal{G}_{y}) \land \text{out-deg}(r, y-1) = d   \},
\end{split}
\end{equation}
\normalsize
and compute the cumulative preferential attachment/detachment function  analogous to eq.~\eqref{eq:sm_pi_cum}, except this time
\small
\begin{equation}
    \Pi^+(d,y) = \left(\frac{| \text{Attached lobbying contracts}(y; d)|  }{|\text{K-Street Registrants}_{\text{out} = d}  (y-1)| } \right)  \left( \sum_{d'} \frac{| \text{Attached lobbying contracts}(y; d')|  }{|\text{K-Street Registrants}_{\text{out} = d'}  (y-1)| } \right)^{-1},
\end{equation}
\normalsize
and
\small
\begin{equation}
    \Pi^-(d,y) = \left(\frac{| \text{Detached lobbying contracts}(y; = d)|  }{|\text{K-Street Registrants}_{\text{out} = d}  (y-1)| } \right)  \left( \sum_{d'} \frac{| \text{Detached lobbying contracts}(y; d')|  }{|\text{K-Street Registrants}_{\text{out} = d'}  (y-1)| } \right)^{-1}.
\end{equation}
\normalsize
As $\pi^-(l) \sim l^{2}$, we conclude that, similarly to the result of Sec.~\ref{sec:sm:preferential}, the registrant-lobbyist connections are removed according to linear preferential detachment. 
Nevertheless, this time the preferential attachment is sublinear, and the preference is more pronounced for larger (higher out-degree) registrants. 

Figure~\ref{fig:sm:inhouse_lobbyist_evolution} presents analogous analysis for the registrant-lobbyist connections, where the registrant is In-House. 
The preferential attachment/detachment (approximately linear) is still at play, but we find a different trend in the lobbying probability as a function of time (Fig.~\ref{fig:sm:inhouse_lobbyist_evolution}C). 
It appears that prior to the financial crisis of 2007--08, In-House registrants were much more stable, and nowadays they follow a similar drop-out dynamics to K-Street registrants.

\subsection{Small component example}
In Fig.~2A of the main text, we show the attachment and detachment processes in a sample client-registrant-lobbyist network component.
We have selected this component through the following principled search algorithm:

Given the graphs $\mathcal{G}_{y}$ for $y \in \{1999, 2000, \dots, 2023\}$ constructed according to Sec.~\ref{sec:sm:network_construction}, we first remove all In-House registrant, government entity, and legislator nodes along with their edges, obtaining the subgraphs 
\begin{alignat*}{2}
\mathcal{G}^*_{y} = \bigl(
    &V(\mathcal{G}_{y}) &&\setminus \text{In-House Registrants}(y) \setminus \text{Government entities}(y) \setminus \text{Legislators}(y),\\
    &E(\mathcal{G}_{y}) &&\setminus \text{In-House Clientships}(y) \setminus \text{In-House Lobbyist contracts}(y)\\
        & &&\setminus \text{Government associations}(y) \setminus \text{Legislator associations}(y)
\bigr).
\end{alignat*}
For technical reasons, we treat $\mathcal{G}^*_{y}$ as undirected.
For each year, we compute the set $\mathcal{W}_y$ of connected components of $\mathcal{G}^*_{y}$.
We denote the number of registrants in a component $C \in \mathcal{W}_y$ by $\nu(C,y) = \left|\{ n \in V(C) : n \in \text{Registrants}(y) \}\right|$ and select only the components that have $\nu(C,y) = 3$ registrant nodes, creating a set of \textit{candidate components} $\mathcal{C}_y = \{ C \in \mathcal{W}_y : \nu(C,y) = 3 \}$.
After sorting the components for a given year by the number of registrants $\nu(C,y)$ in descending order, we find that for all years the largest $\nu(C,y)$ is of the order of $10^3$ and the second-largest $\nu(C,y) \leq 4$ (c.f.~Sec.~\ref{sec:sm:component}).
We choose candidate components with $\nu(C,y) = 3$ as they illustrate the attachment and detachment processes better than those with $\nu(C,y) = 4$ and still have a reasonable (not too small) size.

Next, we track the evolution of each component $C \in \mathcal{C}_y$ two years forward and two years backward (for years $\max(y-2,1999) \leq y' \leq \min(y+2,2023) $). 
In particular, first, new client and lobbyist nodes from $\mathcal{G}^*_{y\pm1}$ are added to the component via existing registrant nodes, and new registrant nodes from $\mathcal{G}^*_{y\pm1}$ are added to the component via existing client and lobbyist nodes.
 Crucially, the obtained set of nodes is then connected with the edges present $\mathcal{G}^*_{y\pm1}$ to remove any nodes no longer present in year $y\pm1$.
 
This operation can be formalized by defining a \textit{one-year forward/backward evolution map} for vertices
\[
\epsilon^\pm(V(C),y) = \left[ V(\mathcal{G}^*_{y\pm1}) \cap V(C) \right] \cup  N(V(\mathcal{G}^*_{y\pm1}) \cap V(C))
\]
where $N(X, \mathcal{G})$ is the set of all the neighbors of nodes $n \in X$ in graph $\mathcal{G}$, 
and the evolution map for edges 
\[
\epsilon^\pm(E(C),y) = E(\mathcal{G}^*_{y\pm1}) \cap \left[ (n_1,n_2): n_1,n_2 \in \epsilon^\pm(V(C),y) \right].
\]
Thus, for each $C \in \mathcal{C}_y$, we then compute a sequence of 5 graphs
\[
\mathcal{E}(C,y) = \biggl(
\epsilon^-\bigl(\epsilon^-(C,y),y-1\bigr),~
\epsilon^-(C,y),~
C,~
\epsilon^+(C,y),~
\epsilon^+\bigl(\epsilon^+(C,y),y+1\bigr)
\biggr)
\]
and we select one of them to illustrate the graph evolution with a real example.

\section{Further analysis of bipartite subgraphs}
In this section, we present further analysis of the lobbying network topology.
Specifically, we look at various bipartite subgraphs of the network and compute their average clustering coefficient and the size of the largest connected component. 
The bipartite graphs that we consider are naturally constructed by restricting attention to two adjacent layers of the lobbying network $\mathcal{G}_y$ and the edges between them (c.f.~Sec.~\ref{sec:sm:network_construction}).
All of the bipartite subgraphs are listed in Table~\ref{tab:sm:bipartite_graphs}, 
where we make a distinction between the upstream set and the downstream set. 
As the lobbying network is directed, the induced subgroups are also directed. 
Nevertheless, for notational simplicity, in this section, we will treat the bipartite graphs as undirected. 
We also exclude In-House Registrants, as by definition their in-degree is fixed to 1. 

\begin{table}[h]
    \centering
    \caption{Bipartite subgraphs of our lobbying network, excluding In-House registrants.}
    \begin{tabular}{cc}
    \toprule
    Upstream Set & Downstream Set \\
    \midrule
    Clients & K-Street Registrants \\
    K-Street Registrants & Lobbyists \\
    Lobbyists & Government entities \\ 
    Lobbyists & Legislators \\
    \bottomrule
    \end{tabular}
    \label{tab:sm:bipartite_graphs}
\end{table}

\subsection{Clustering coefficient}
Let $\mathcal{B}$ be a bipartite graph with the upstream node-set $U$ and the downstream node-set $D$. 
Without loss of generality, we will now consider a node in the upstream set $u \in U$. 
We define the set of its first neighbors $N(u) \subseteq D $  and the set of its second neighbors $N(N(u)) \subseteq U$. 
The \textit{clustering coefficient} of node $u$ is defined as
\begin{equation}
    C(u) = \frac{1}{|N(N(u))|} \sum_{v \in N(N(u))} \frac{|N(u) \cap N(v) | }{|N(u) \cup N(v)|}, 
\end{equation}
and by averaging 
\begin{equation}
    C(U) = \frac{1}{|U|} \sum_{u \in U} C(u), 
\end{equation}
we can compute the \textit{average clustering coefficient} for the upstream set $U$ in graph $\mathcal{B}$. 
The clustering coefficient $C(D)$ for the downstream set $D$ is defined analogously. 

\begin{figure}[h]
    \centering
    \includegraphics[width = \textwidth]{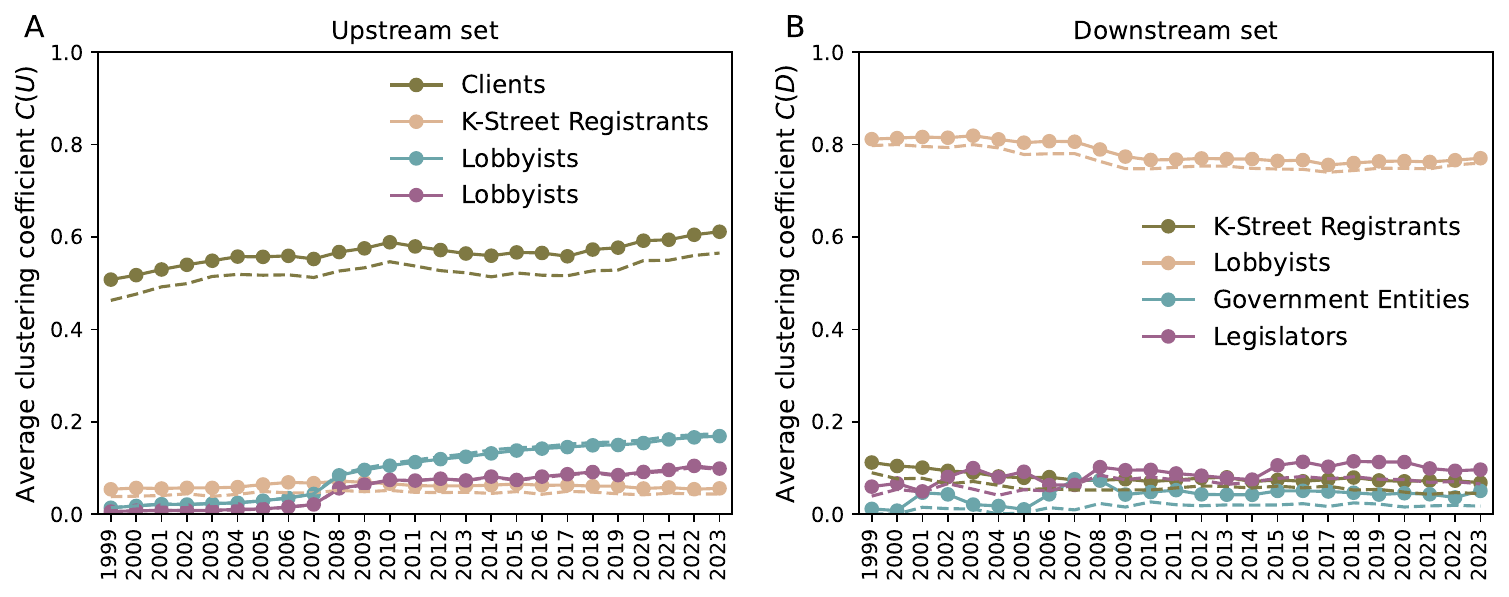}
    \caption{Clustering coefficient for bipartite subgraphs averaged over (A) the upstream set $U$ and (B) the downstream set $D$. The solid line with circles denotes empirical measurements, and the dashed line denotes the expected value of the configuration model. }
    \label{fig:sm:clustering}
\end{figure}

In Fig.~\ref{fig:sm:clustering} we report the average clustering coefficient for both the upstream set and the downstream set of $\mathcal{B}$, where $\mathcal{B}$ is one of the subgraphs (c.f.~Tab.~\ref{tab:sm:bipartite_graphs}) of the lobbying network $\mathcal{G}_y$ in year $y$. 
Two coefficients stand out due to their consistently high value: $C(U)$ for the Client--K-Street Registrant graph, and $C(D)$ for the K-Street Registrant--Lobbyist graph (noting that by construction the clustering coefficient is bounded between $0$ and $1$). 
Both of them reflect the focusing characteristic of the registrant layer.
Indeed, a typical registrant will have a cluster of clients and a cluster of lobbyists. 
We also note some slight temporal trends in the clustering coefficient (increasing for the upstream set), which might be related to the preferential attachment dynamics. 

The dashed lines in Fig.~\ref{fig:sm:clustering} correspond to the expected value of the corresponding coefficient in a bipartite configuration model, i.e., a randomized bipartite network with the same partition and degree sequence~\cite{Newman_book}.
We generally find a close match between this expected value and the empirically computed value.

\subsection{Largest component}
\label{sec:sm:component}
Another question we can ask about a bipartite subgraph is the size of its largest connected component. 
Recall that we treat the subgraph as undirected; alternatively, in a directed graph, we would consider a weakly connected component.

\begin{figure}[h]
    \centering
    \includegraphics[width = \textwidth]{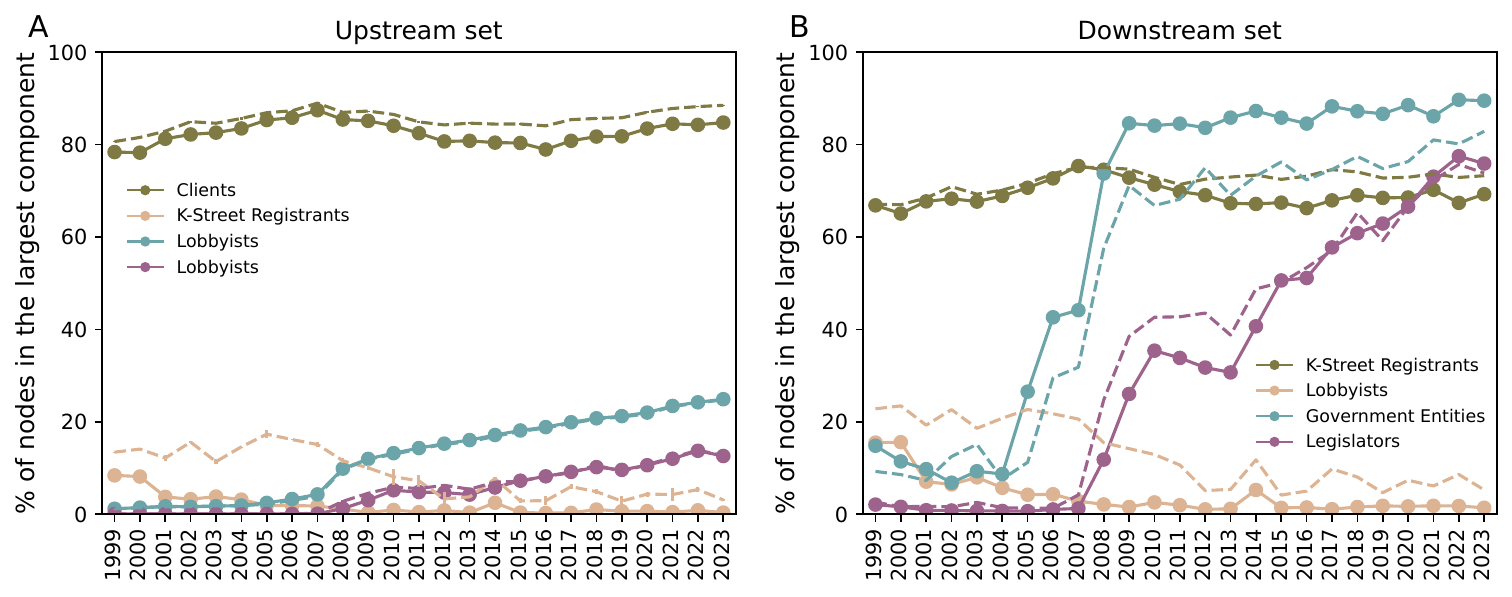}
    \caption{Largest component analysis of bipartite subgraphs. 
    (A) The fraction of the upstream set contained in the largest component.   
    (B) The fraction of the downstream set contained in the largest component. The dashed line denotes the expected value in the bipartite configuration model.  }
    \label{fig:sm:giant}
\end{figure}

In Fig.~\ref{fig:sm:giant}, we analyze what fraction of each set (upstream and downstream) the largest connected component captures. 
For all subgraphs, the results are consistent with the prediction of the bipartite configuration model, but we find that only the Client--K-Street Registrant subgraph possesses a large (giant) connected component at all times.
For the Lobbyist--Government Entity and Lobbyist--Legislator subgraphs, we observe a transition with a large component starting to appear as the number of associations (edges) increases (c.f.~Fig.~\ref{fig:sm:size_per_year}C,D).
The reader might notice these components contain almost all nodes in the downstream set (government entities or legislators), but only a minority of the nodes in the upstream set (lobbyists). 
This is because, even in recent years, only a minority of lobbyists have governmental or lobbyist associations disclosed via the covered position field of LD-2 filings (c.f.~Fig.~\ref{fig:sm:average_degrees_per_year}).

\section{Registrant centrality}
\label{sec:reach_centrality}
\begin{figure}[h]
    \centering
    \includegraphics[width = \textwidth]{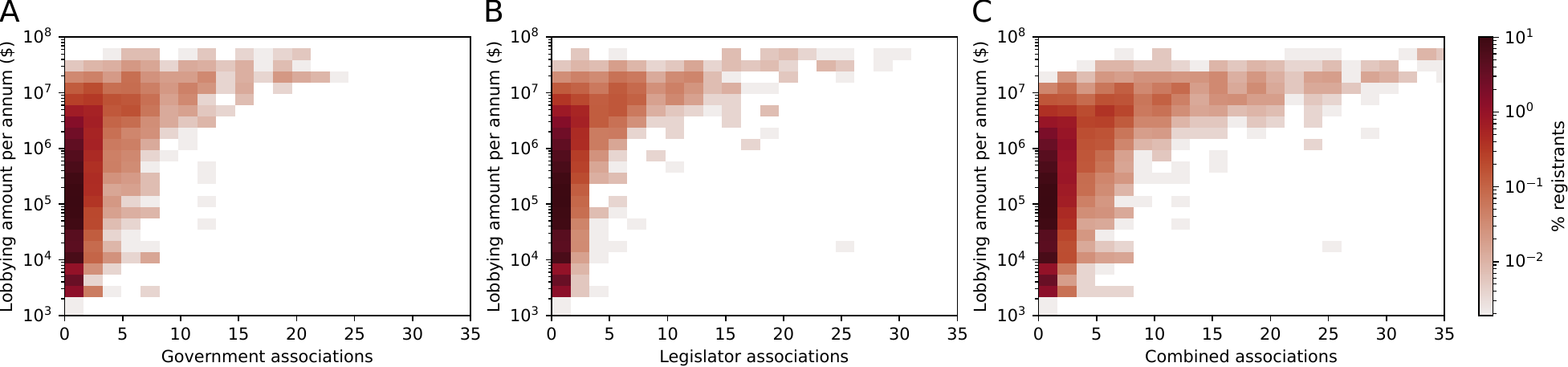}
    \caption{Total annual income of K-Street registrants in relation to their (A) number of government associations, (B) number of legislator associations, and (C) the total number of associations. The 2D histograms comprise all the data from 1999--2023, where one active registrant in one year contributes one data point. }
    \label{fig:sm:reach}
\end{figure}

In this section, we explain the methodology of generating Fig.~3D of the main text, which relates the total income of a K-Street registrant $r$ ($\text{w-in-deg} (r,y)$, as defined in eq.~\eqref{eq:c_weighted_in_deg_def}) to the number of \textit{reachable} associations (government entities and legislators).
This number of associations $\rho(r,y)$ can be thought of as a centrality score of a registrant node $r$ and is simply the total number of government entities and legislators that can be reached from $r$ by a path in $\mathcal{G}_y$. 
In Fig.~3D of the main text, we present a 2D histogram of $\rho(r,y)$ (controlled for the weighted in-degree of the registrant $\text{w-in-deg}(r,y)$) combined for all years $y$ and all $\text{K-Street Registrants}(y)$ (c.f.~eq.~\eqref{eq:registrants_kstreet_def}). 
This plot is reproduced in Fig.~\ref{fig:sm:reach}C. 
In Fig.~\ref{fig:sm:reach}A, we present an analogous plot but only count the number of government entity associations, and in Fig.~\ref{fig:sm:reach}B, we consider only the number of legislator associations.
The combined number of associations (Fig.~\ref{fig:sm:reach}C) is the sum of the two scores. 
Nevertheless, we find that each of these scores in isolation (number of government associations and number of legislator associations) is a sufficient condition for high registrant income in and of itself. 


\section{Portfolio analysis}
\label{sec:sm:target_portfolio}
In this section, we describe the methodology of \textit{lobbying issue portfolio} and \textit{lobbying target portfolio} analysis presented in Fig.~3 of the main text, respectively. 

\subsection{Portfolio matrices}

\begin{figure}[h]
    \centering
    \includegraphics[width = \textwidth]{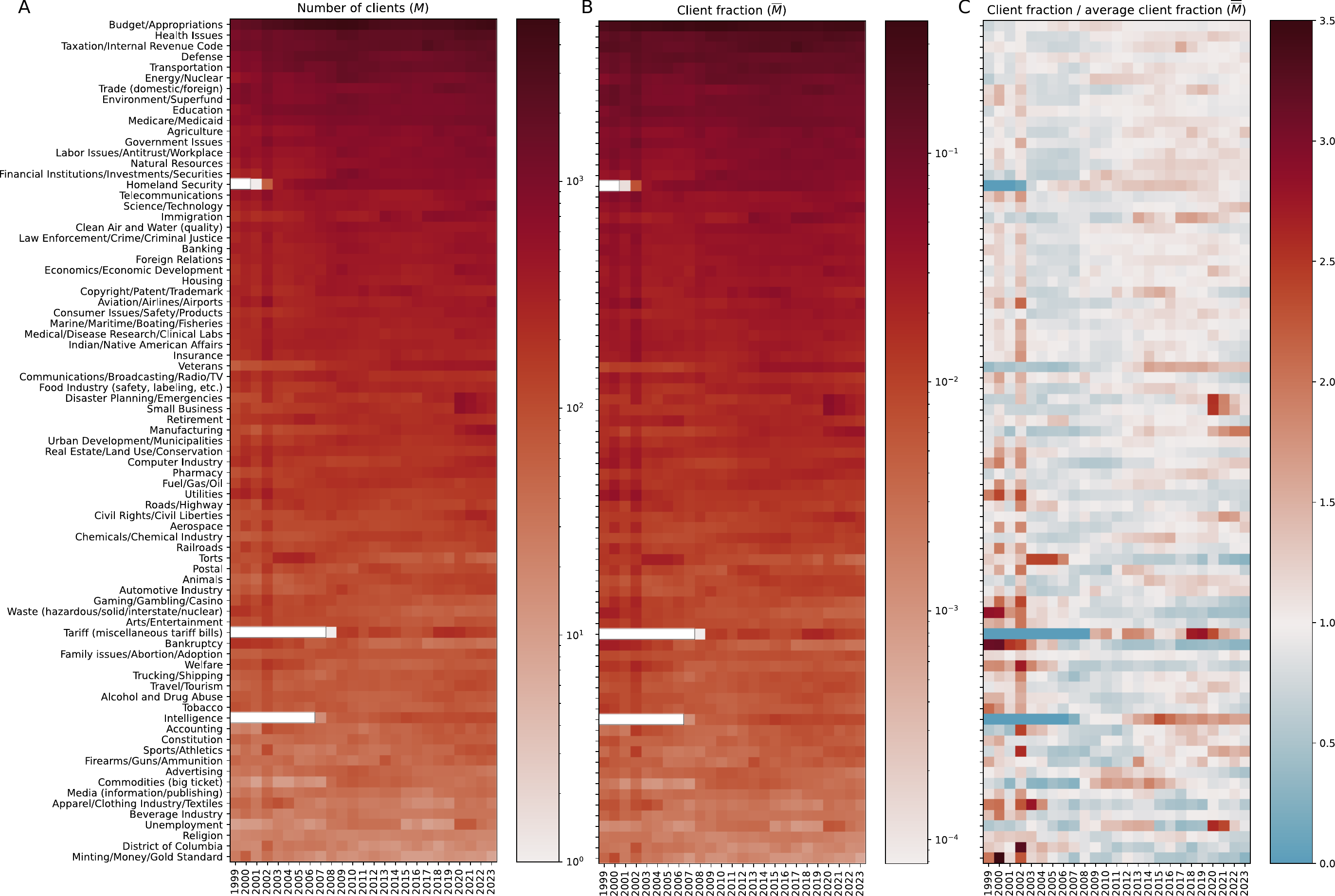}
    \caption{Issue portfolio analysis. (A) The number of clients lobbying on issue area $a$ in year $y$ ($M_{ay}$).  (B) The fraction of clients lobbying on issue area $a$ in year $y$ ($\overline{M}_{ay}$).  (B) Row-normalized fraction of clients lobbying on issue area $a$ in year $y$ ($\overline{\overline{M}}_{ay}$). }
    \label{fig:sm:client_group_mat_method}
\end{figure}

We start the lobbying issue portfolio analysis by counting the number of clients who in year $y$ lobbied on issue area~$a$ and combine these scores into a matrix $M$, such that 
\begin{equation}
    M_{ay} = | \left\{  c \in \text{Clients}(y): \exists_f~ c(f) = c \land a \in A(f) \land y(f)  = y \right\} |.
\end{equation}
This data is presented in Fig.~\ref{fig:sm:client_group_mat_method}A.
The number of clients lobbying on a given issue can also reflect the total number of clients lobbying in a given year (c.f.~Fig.~\ref{fig:sm:order_per_year}A). 
To discount potential spurious temporal trend, instead of looking at the number of clients, we compute the fraction of clients (Fig.~\ref{fig:sm:client_group_mat_method}B)
\begin{equation}
    \overline{M}_{gy} = \frac{M_{gy}}{|\text{Clients}(y)|}.
        \label{eq:mgy}
\end{equation}
Finally, to account for disparities in `popularity' of different areas, we compute the issue portfolio matrix $\overline{\overline{M}}$ by normalizing the matrix $\overline{M}$ row by row such that
\begin{equation}
    \overline{\overline{M}}_{gy} =\frac{\overline{M}_{gy}  }{  \frac{1}{N_y} \sum_y \overline{M}_{gy} },
    \label{eq:sm:m_barbar}
\end{equation}
where $N_y = 25 $ is the number of years considered. 
If $ \overline{\overline{M}}_{gy}>1$, it means that the fraction of reports mentioning entity $g$ in year $y$ is above the average for years 1999--2023.

The target portfolio analysis follows a very similar methodology, with the target portfolio matrix defined as
\begin{equation}
    \overline{\overline{M}}_{gy} =\frac{\overline{M}_{gy}  }{  \frac{1}{N_y} \sum_y \overline{M}_{gy} },
\end{equation}
where
\begin{equation}
    \overline{M}^I_{ay}  = \frac{  | \left\{  c \in \text{Clients}(y): \exists_f~ c(f) = c \land g \in G(f) \land y(f)  = y \right\} |   }{|\text{Clients}(y)|}.
\end{equation}

\begin{figure}[h]
    \centering
    \includegraphics[width = .75\textwidth]{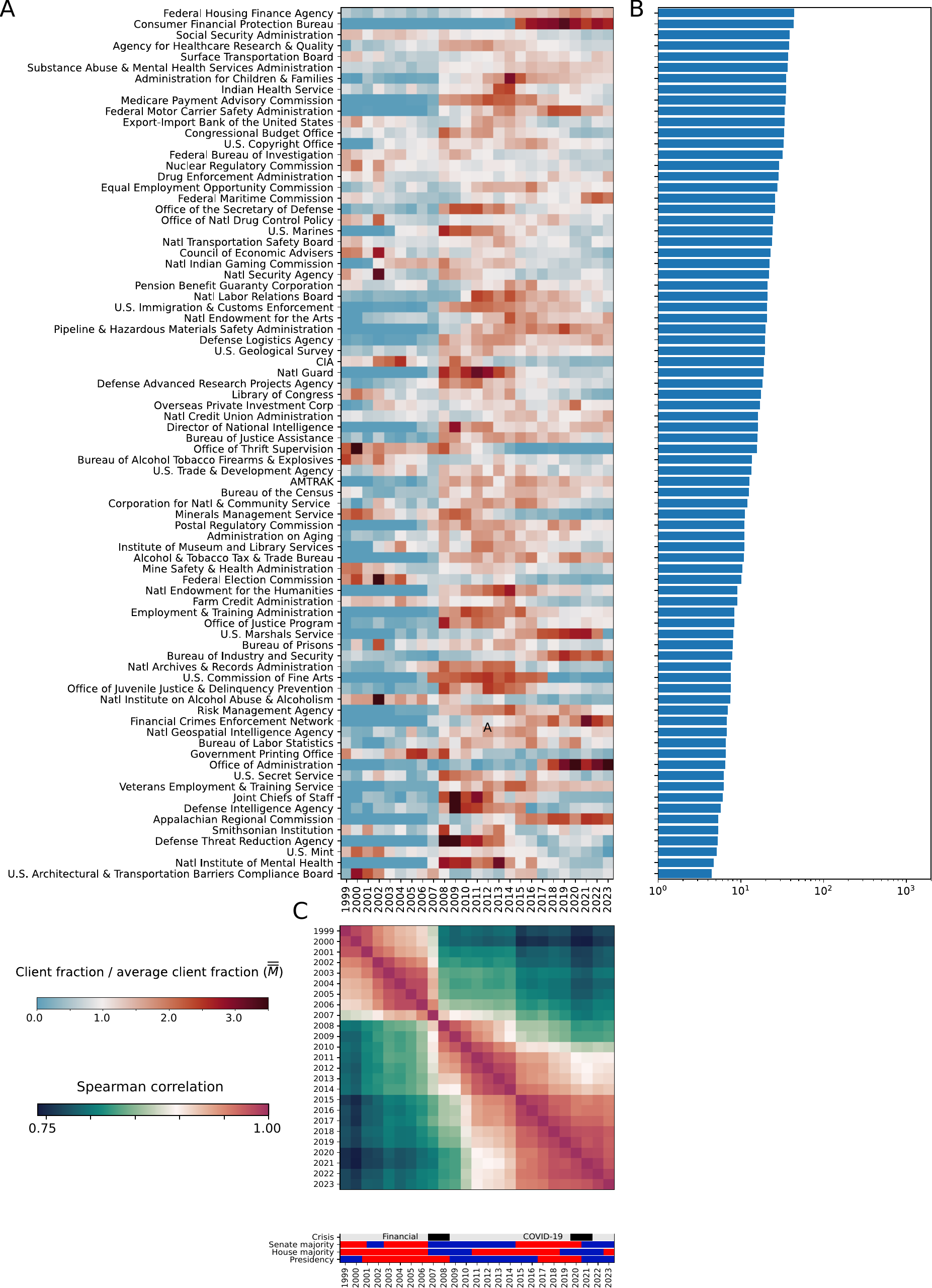}
    \caption{Target portfolio for less often mentioned government entities (80 -- 159 in the ranking). (A) Portfolio matrix ($\overline{\overline{M}}$). (B) Number of clients lobbying with different government entities ($M$).
    (C) Spearman correlation matrix for these government entities. }
    \label{fig:sm:gov_entity_mat_ctd}
\end{figure}

Figure~3 of the main text presents the target portfolios for the top 79 government entities, i.e., the entities with the largest average number of clients. 
In Figures~\ref{fig:sm:gov_entity_mat_ctd}, we show such scores for the next 79 entities in this ranking, as well as the corresponding year-to-year Spearman correlation matrix, which is somewhat different to the correlation matrix for the top 79 issues. 
Nevertheless, we note that many government entities in this range are approached by only a handful of clients, so the ranking can be sensitive to small variations. 
For this reason, we do not pursue the target portfolio analysis for all 250 government entities mentioned in the reports, but we focus on the most popular ones.

\subsection{Budget-based analysis}
\begin{figure*}
    \centering
    \includegraphics[width=.95\textwidth]{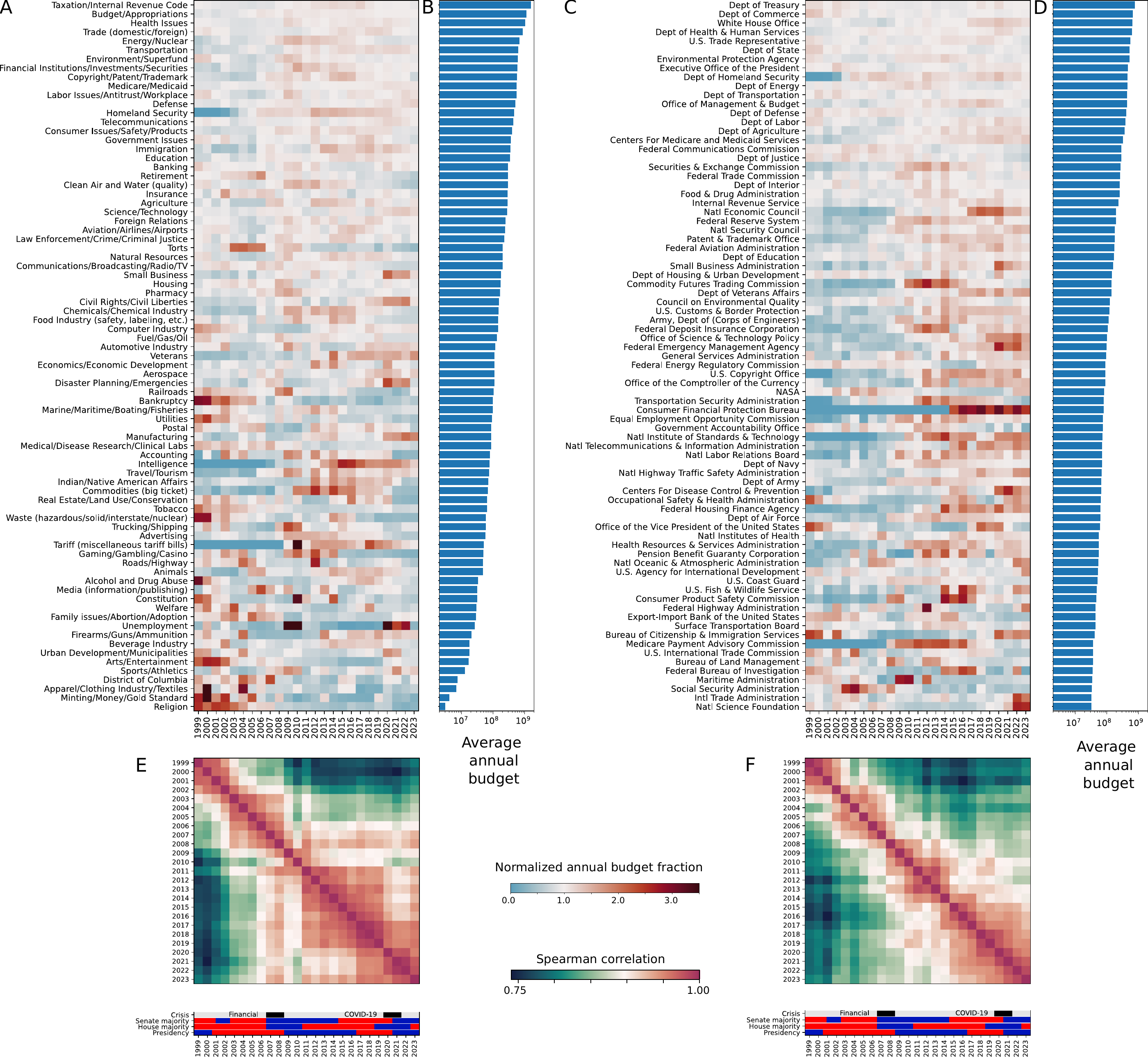}
    \caption{\textbf{Lobbying portfolio analysis based on budget.}     (A)~Fraction of the lobbying budget dedicated to a given issue, scaled by its average value in the period 1999--2023.
    (B)~Average annual lobbying budget associated with a given issue. 
    (C) Fraction of the lobbying budget that can be associated with lobbying a given government entity in a given year, scaled by the average value in the period 1999--2023.
    The government entities are listed according to the average annual budget. 
    In this Figure, we present only the top approached government entities; note that the ranking may be different from that in Fig.~3.
    (D)~Average annual lobbying budget dedicated to lobbying different government entities. 
    (E)~Year-to-year correlation of the \textit{issue portfolio vectors} (columns of the matrix in panel (A)).
    (F)~Year-to-year correlation of the \textit{government approach portfolios} presented as columns in panel (C). 
    }
    \label{fig:sm:portfolio_analysis_budget}
\end{figure*}

Instead of client group size, the lobbying portfolios can also be defined based on the lobbying budget, e.g., we could choose 
\begin{equation}
    \overline{M'}_{ay} =\frac{ \sum_f \mathbbm{1}[a \in A(f) ~\land~y(f)  = y ~\land~r(f) ] } {\sum_f \mathbbm{1}[ y(f)  = y]}.
\end{equation}
This idea is pursued in Fig.~\ref{fig:sm:portfolio_analysis_budget}, which can be directly compared with Fig.~3 of the main text.
We find that the correlation structure of the budget-based portfolios is less clear than for the client-group-based portfolios, e.g., the block-diagonal structure of Fig.~\ref{fig:sm:portfolio_analysis_budget}F appears to be less apparent than the block-diagonal structure of Fig.~3F of the main text. 

\subsection{Principal component analysis}
We can further analyze the similarity of annual portfolios by performing principal component analysis (PCA). 
For the issue portfolio, we normalize $\overline{\overline{M}}_{ay}$ defined in eq.~\eqref{eq:sm:m_barbar} one more time by column-wise subtraction of mean and division by standard deviation. 
Thus, we obtain the matrix $\overline{\overline{\overline{M}}}_{ay}$ that we decompose as
\begin{equation}
    \overline{\overline{\overline{M}}}_{ay} = \sum_i \sigma_i \vec{e}_i \vec{u}_i,  
\end{equation}
where $\sigma_1 \geq \sigma_2 \geq \dots $ are singular values, $\vec{e}_i \in \mathbb{R}^{N_{g} \times 1}$ are principal components, $N_a = 79$ is the number of issue areas, $\vec{u}_i \in \mathbb{R}^{1 \times N_y}$ are score vectors, and $N_y = 25$ is the number of analyzed years.
Each entry of a principal component corresponds to one government entity, while each entry of a score vector corresponds to one year. 

\begin{figure}
    \centering
    \includegraphics[width=.5\textwidth]{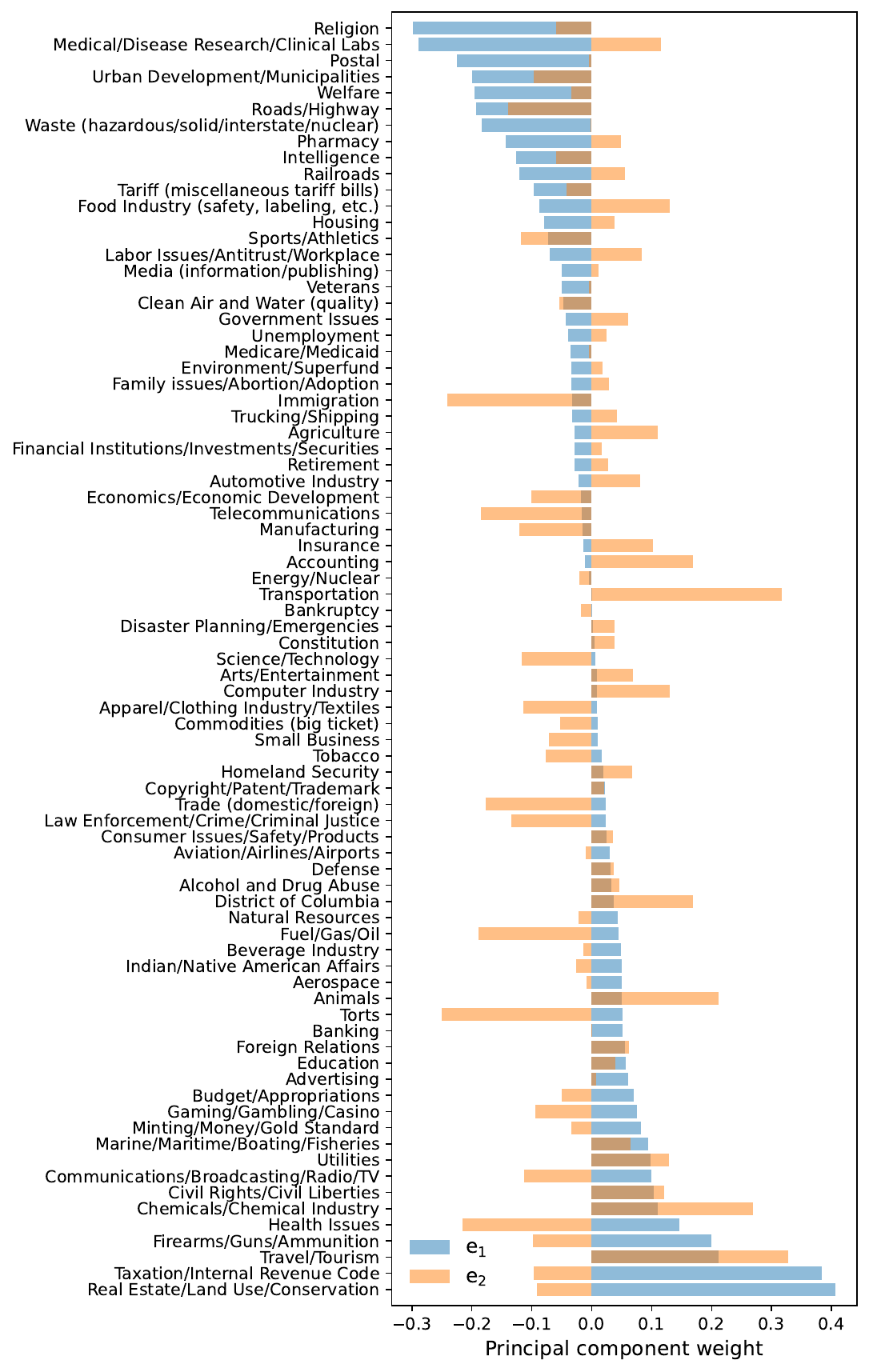}
    \caption{Weights of the first two principal components $\vec{e}_1$ and $\vec{e}_2$ of the issue portfolio matrix $\overline{\overline{\overline{M}}}_{ay}$. Components
    $\vec{e}_1$ and $\vec{e}_2$ explain more than $38\%$ and $13\%$ of variance respectively.
    Issue areas are sorted by increasing $\vec{e}_1$ weight.
    }
    \label{fig:sm:pca_issue_components}
\end{figure}

\begin{figure}
    \centering
    \includegraphics[width=.75\textwidth]{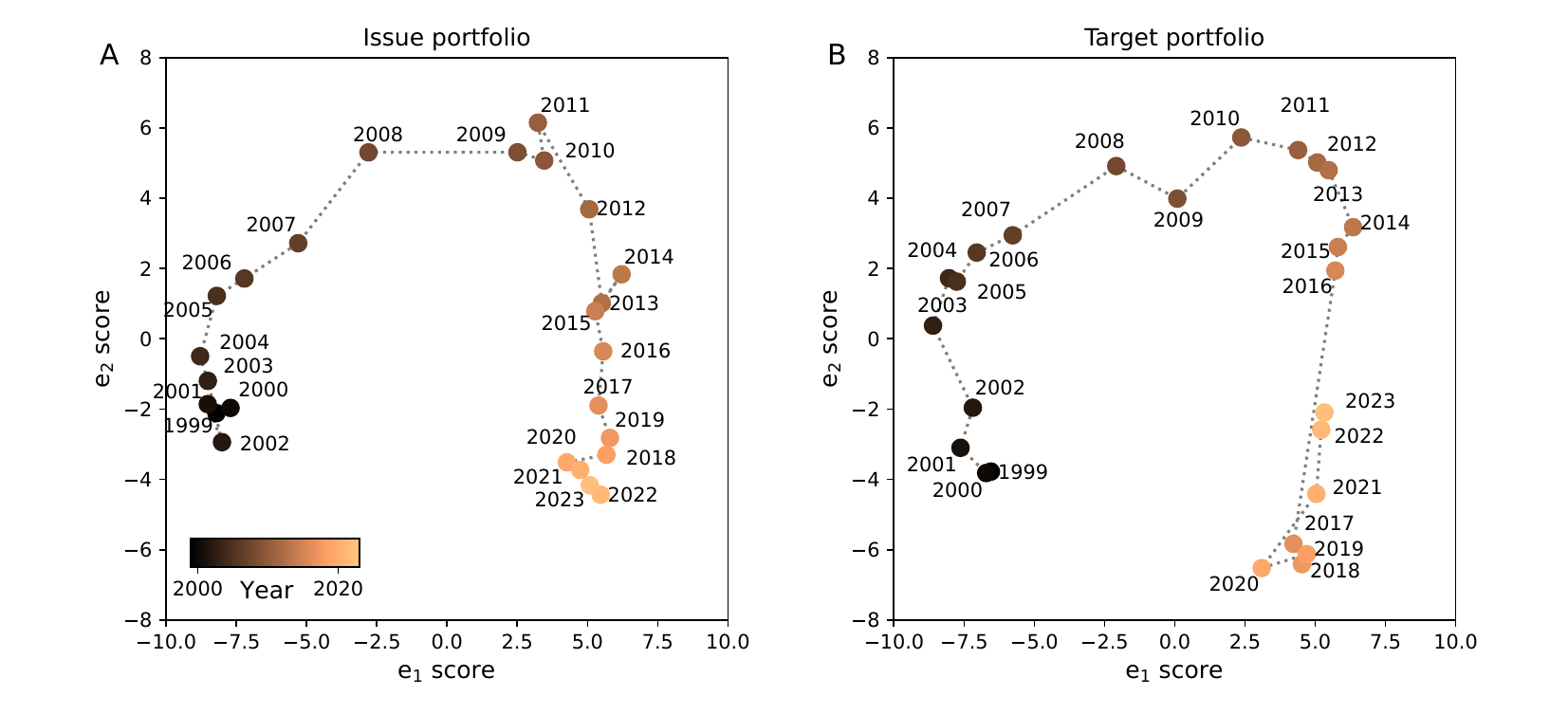}
    \caption{Projection of the annual portfolios onto 2D space spanned by the principal components $\vec{e}_1$ and $\vec{e}_2$.
    (A)~Issue portfolios.
    (B)~Target portfolios (top 79 entities).
    }
    \label{fig:sm:pca_years_paths}
\end{figure}

Figure~\ref{fig:sm:pca_issue_components} shows the weights of the two highest-variance principal components $\vec{e}_1$ and $\vec{e}_2$ that explain more than $48\%$ and $13\%$ of variance, respectively.
The projection of the target portfolio onto the 2D space $(\vec{e}_1, \vec{e}_2)$ is shown in Fig.~\ref{fig:sm:pca_years_paths}A.
We can observe a steady evolution along the circle-like path from early years (top right) to recent years (top left).
Also, note that the pairs of consecutive years (2007, 2008) and (2008, 2009) are considerably more dissimilar than others, suggesting significant changes in the target portfolio between the two years. These changes are coincident to two major external events: first, the global financial crisis; second, the adoption of reforms in the Honest Leadership and Open Government Act of 2007.

\begin{figure}
    \centering
    \includegraphics[width=.65\textwidth]{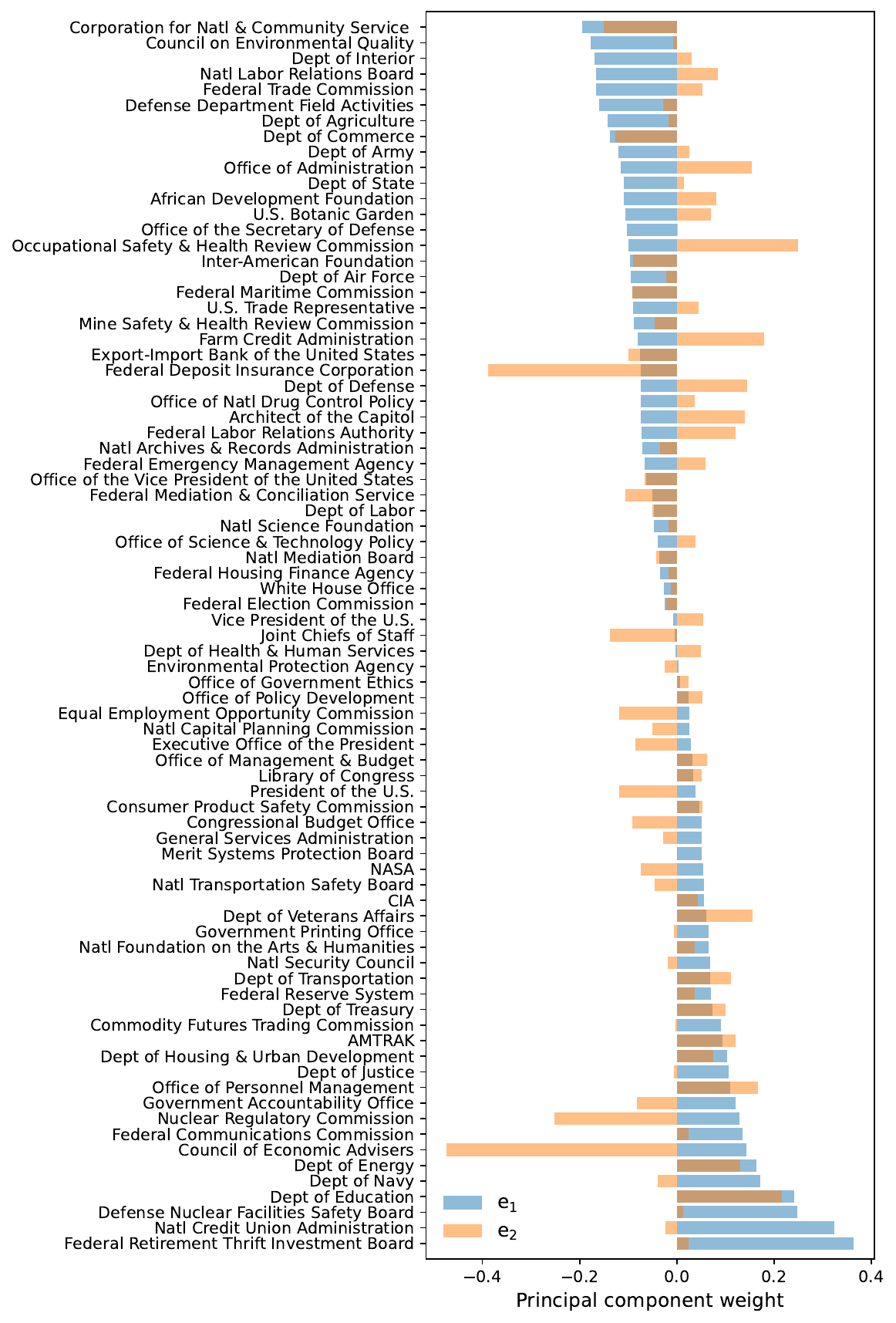}
    \caption{Weights of the first two principal components $\vec{e}_1$ and $\vec{e}_2$ of the target portfolio matrix $\overline{\overline{\overline{M}}}_{gy}$. Components
    $\vec{e}_1$ and $\vec{e}_2$ explain more than $42\%$ and $21\%$ of variance respectively.
    Government entities are sorted by increasing $\vec{e}_1$ weight.
    }
    \label{fig:sm:pca_gov_components}
\end{figure}

Analogously, we normalize the target portfolio matrices $\overline{\overline{M}}_{gy}$ and apply PCA to the obtained normalized matrices $\overline{\overline{\overline{M}}}_{gy}$.
Figure~\ref{fig:sm:pca_years_paths}B shows that, similarly to the issue portfolio, $\vec{e}_1$ also co-evolves with time.
For completeness, the weights of the first two principal components $\vec{e}_1$ and $\vec{e}_2$ are shown in Fig.~\ref{fig:sm:pca_gov_components}, but the interpretation of this result is beyond the scope of this study.

\clearpage

\section{Probabilistic analysis of bill-related lobby}
\label{sec:sm:probability}

In Fig.~4 of the article, we present a probabilistic analysis of lobbying associated with the Stop Online Piracy Act (SOPA).
Here, we provide additional methodological details of this analysis, and we include two additional examples of bill lobby analysis.

\subsection{Methodology}
\begin{figure}
    \centering
    \includegraphics[width=.65\textwidth]{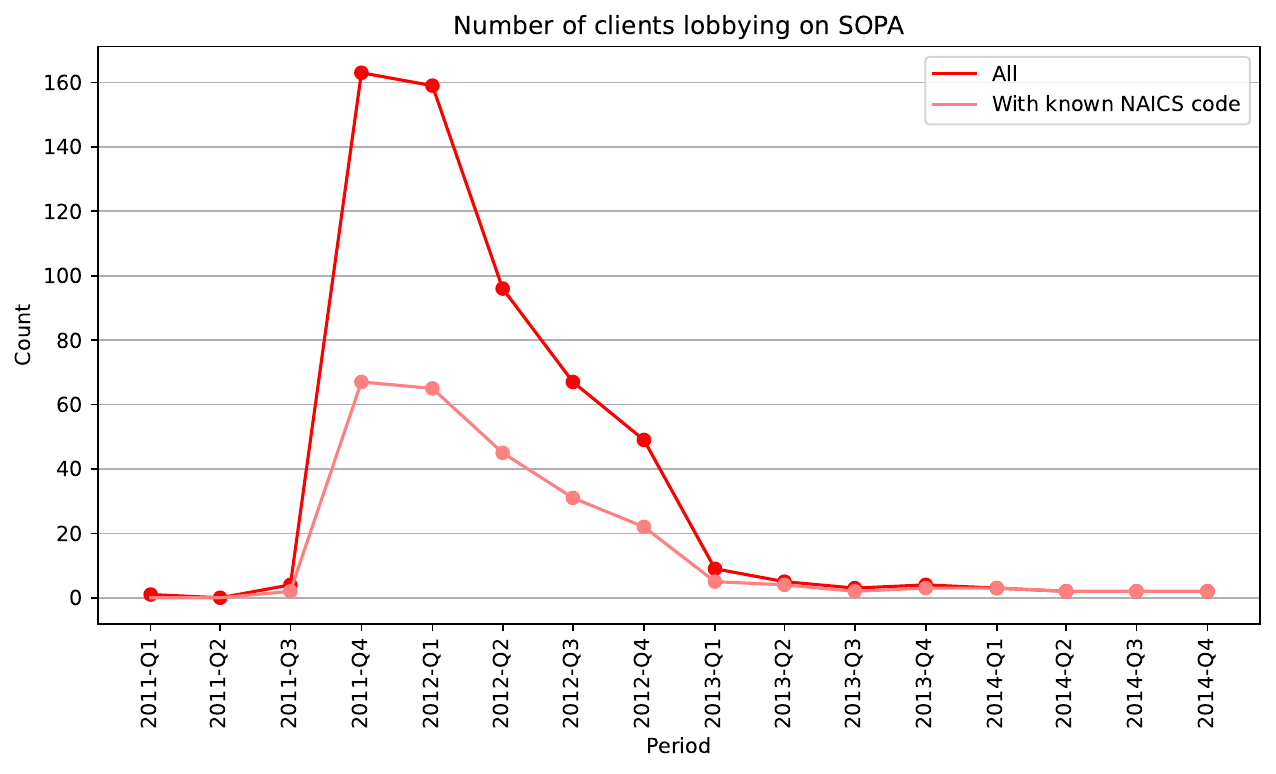}
    \caption{Number of clients lobbying on the Stop Online Piracy Act (SOPA) in a given quarter.}
    \label{fig:sm:sopa_numbers}
\end{figure}

Conceptually, the first step of our analysis is to draw one of the clients $C$ who lobbied in the last quarter of 2011, uniformly at random. 
The probability of choosing a specific client is simply $\frac{1}{|\text{Clients}(\text{Q4 2011})|}$ (c.f.~Sec.~\ref{sec:sm:network_construction:nodes}). 
As the database contains every LDA LD-2 filing, we can compute various empirical (frequentist) probabilities by filtering the database and counting.  
Thus, the probability that a randomly chosen client lobbied on a given bill is simply the fraction of clients that mentioned this bill in at least one filing. 
The number of clients lobbying on SOPA peaked in the fourth quarter of 2011 (Fig.~\ref{fig:sm:sopa_numbers}), which is the focus period of our analysis. 
For about $40 \%$ of them, we can map the client to an industry by linking the client to the S\&P Compustat database~\cite{compustat} and from there to the client's North American Industry Classification System (NAICS) code. The remainder of our analysis is restricted to these clients. Compustat tracks all publicly traded North American headquartered firms; clients lacking Compustat entries (and thus lacking NAICS codes we can access) are either non-North American multinationals, private firms, or non-firm entities like interest groups or non-governmental organizations.

\begin{figure}
    \centering
    \includegraphics[width=.95\textwidth]{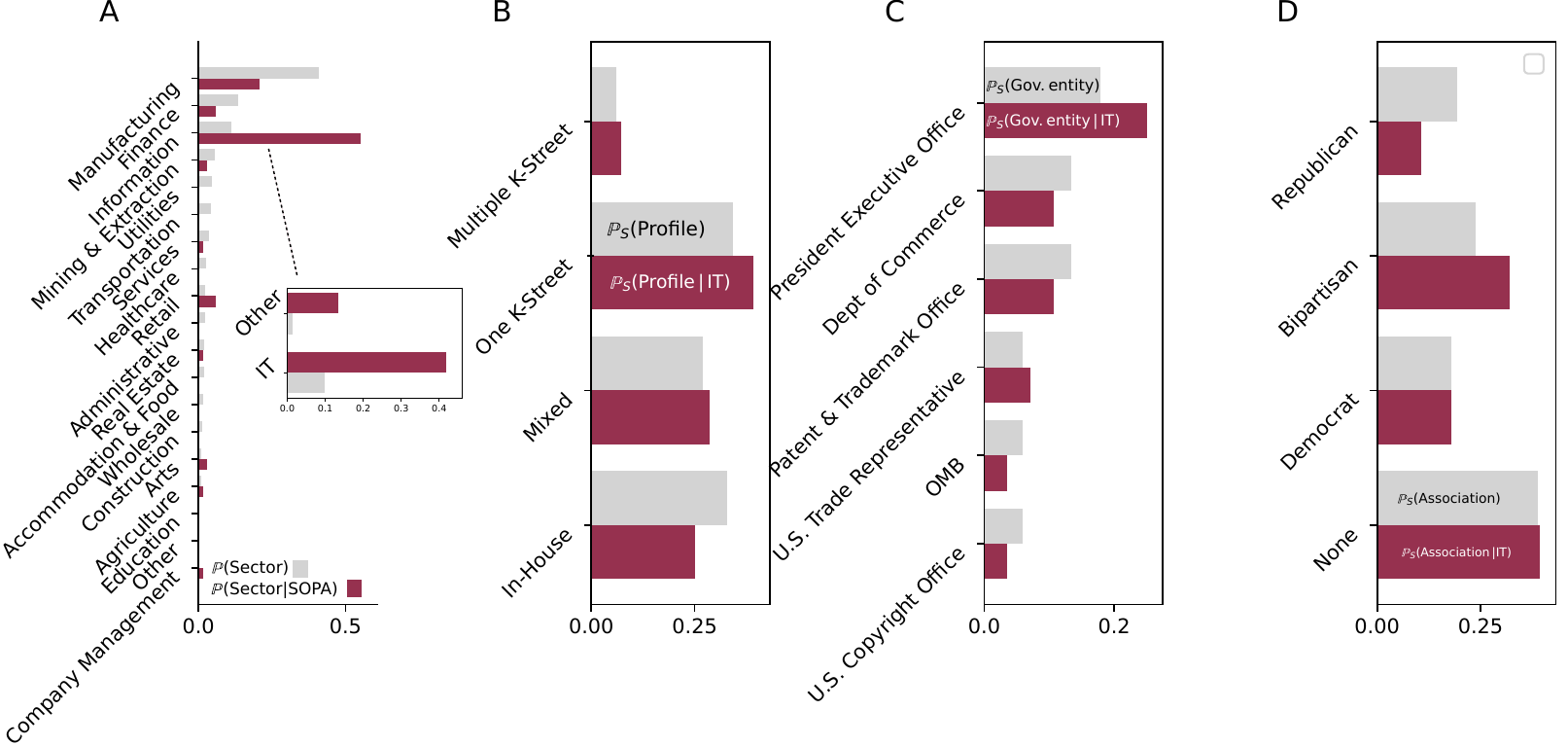}
    \caption{Unconditional and conditional probabilities for SOPA lobbying in the fourth quarter of 2011. 
    (A) Probability that a random client belongs to a given NAICS sector $\mathbb{P}(\text{Sector})$, compared to the conditional probability $\mathbb{P}(\text{Sector}|\text{SOPA})$.
    The sectors are sorted by the value of $\mathbb{P}(\text{Sector})$.
    The inset shows the Information sector subdivided into IT and non-IT companies.
    (B) Probability that a random client hires registrant(s) of a particular profile: for all clients lobbying on SOPA $\mathbb{P}_S(\text{Profile})$, and for those of them who belong to the Information industry $\mathbb{P}_S(\text{Profile}\, | \, \text{Inf})$.
    (C)  Probability that a random client lobbied with a particular government entity: for all clients lobbying on SOPA $\mathbb{P}_S(\text{Gov. Agency})$, and for those of them who belong to the Information industry $\mathbb{P}_S(\text{Gov. Agency}\, | \, \text{Inf})$.
    Note that multiple government entities can be approached at the same time. 
    For brevity, in this layer, we only show the 6 government entities with the largest $\mathbb{P}_S(\text{Gov. Agency})$.     
    (D) Probability that a random client hired a lobbyist with a documented association with a current legislator: for all clients lobbying on SOPA $\mathbb{P}_S(\text{Association})$, and for those of them who belong to the Information industry $P\mathbb{P}_S(\text{Association} \, | \, \text{Inf})$.
    }
    \label{fig:sm:sopa_marginals}
\end{figure}

NAICS codes are hierarchical six-digit numbers. Successive digits are increasingly more granular measures of industry. We take the first two digits of the NAICS code, which allows us to assign clients to one of the 20 broad client sectors (Tab.~\ref{tab:naics_codes}).
For the purposes of the probabilistic analysis, the sector assignment is treated as a random variable $C$, which can take one of the 20 values. 
The probability that a randomly selected client (with NAICS code) belongs to a given sector $\mathbb{P}(C \in \text{Sector})$ measures the relative frequency of different industry sectors participating in lobbying in the given filing quarter.
The conditional probability $\mathbb{P}(\text{Sector} |\text{SOPA})$ is the relative frequency of different sectors lobbying on SOPA (Fig.~\ref{fig:sm:sopa_marginals}A).  
The \textit{probability shift}
\[
\Delta \mathbb{P}(\text{Sector}\,|\, \text{SOPA}) = \mathbb{P}(C \in \text{Sector}\,|\, \text{SOPA}) - \mathbb{P}(C \in \text{Sector})
\]
is presented in Fig.~4B of the main manuscript.  

To investigate more specifically the lobbying strategies of the IT sector, we subdivide the Information industry sector (NAICS 2-digit code `51') into `IT' and `Other' companies. 
To this end, we use the more fine-grained 4-digit NAICS code. 
Specifically, in our analysis, the `Other' companies comprise: `Motion Picture and Sound Recording Industries' sector (code 512), `Newspaper, Periodical, Book, and Directory Publishers' (code 5131), and `Broadcasting and Content Providers' (code 516). 
All other companies with a NAICS code starting with `51' are classified as `IT'.
In other words, in our analysis, the `IT' sector is a subset of the `Information' sector comprised of companies that are not explicitly related to content creation.    

\begin{table}[h!]
\centering
\caption{NAICS Two-Digit Sector Codes, Full Descriptions, and Abbreviated Names.}
\begin{tabular}{cll}
\toprule
\textbf{NAICS Code} & \textbf{Client Sector} $C$ & \textbf{Abbreviated Name} \\
\midrule
11 & Agriculture, Forestry, Fishing and Hunting & Agriculture \\
21 & Mining, Quarrying, and Oil and Gas Extraction & Mining \& Extraction \\
22 & Utilities & Utilities \\
23 & Construction & Construction \\
31-33 & Manufacturing & Manufacturing \\
42 & Wholesale Trade & Wholesale \\
44-45 & Retail Trade & Retail \\
48-49 & Transportation and Warehousing & Transportation / Warehousing \\
51 & Information & Information \\
52 & Finance and Insurance & Finance \\
53 & Real Estate and Rental and Leasing & Real Estate \\
54 & Professional, Scientific, and Technical Services & Services \\
55 & Management of Companies and Enterprises & Company Management \\
56 & Administrative and Support and Waste Management and Remediation Services & Administrative \\
61 & Educational Services & Education \\
62 & Health Care and Social Assistance & Healthcare \\
71 & Arts, Entertainment, and Recreation & Arts \\
72 & Accommodation and Food Services & Accommodation \& Food \\
81 & Other Services (except Public Administration) & Other \\
92 & Public Administration & Public Administration \\
\bottomrule
\end{tabular}
\label{tab:naics_codes}
\end{table}

As described in the final section of the main manuscript, we can also use the probabilistic framework to analyze/infer lobbying strategies.
First, we analyze the random variable that describes the \textit{registrant profile} hired by the randomly drawn client. 
The registrant profile, or 'Profile' for short, is a random variable that can take one of four values: In-House (if the client lobbied on SOPA only through their own lobbying department), Mixed (if the client lobbied on SOPA both as a self-filer, and through at least one K-Street registrant), One K-Street (if the client lobbied on SOPA through one K-Street firm only), and Multiple K-Street (if the client lobbied on SOPA through more than one K-Street registrant, but not through an In-House lobbying department).
In Fig.~\ref{fig:sm:sopa_marginals}B, we compare the registrant profile probability for information industry clients $\mathbb{P}_S(\text{Profile}\, | \, \text{Inf}) $, where $\mathbb{P}_S$ indicates a restriction to the clients that lobbied on SOPA, to the same distribution for all the clients lobbying on SOPA $\mathbb{P}_S(\text{Profile})$.
The probability shift, which is the difference of these two quantities, is presented in Fig.~4C of the main manuscript. 

Secondly, we look at the event of approaching different government entities (c.f.~Sec.~\ref{sec:sm:target_portfolio}). 
The probability $\mathbb{P}_S(\text{Gov. entity})$ denotes the probability that a given government entity was mentioned in the same section as SOPA in one of the lobbying reports of the client (Fig.~\ref{fig:sm:sopa_marginals}C). 
As each client can approach multiple government entities, 
$\sum_{\text{Entities} }\mathbb{P}_S(\text{Gov. entity}) \neq 1$. 
For brevity, we present only the 6 most lobbied agencies (other than the Senate and House of Representatives, which are included in virtually all lobbying filings involving legislation). 
We also incorporate the entries that pointed to the White House Office into the Executive Office of the President. 

Finally, in Fig.~\ref{fig:sm:sopa_marginals}D, we look at the probability that one of the lobbyists hired by the randomly selected client had a past professional association with a current legislator from one of the main parties (c.f.~Fig.~\ref{fig:sm:legislator_connections}). 
The coarse-grained random variable that we name `Association', can take one of the four mutually exclusive values: Republican (only), Democratic (only), Bipartisan (connections to both parties, possibly via different lobbyists), and None (no documented past affiliation).
The significance of bipartisan lobby is further investigated in Sec.~\ref{sec:polarization}. 

\subsection{Additional examples}

\subsubsection{American Clean Energy and Security Act of 2009 (ACES) }

\begin{figure}
    \centering
    \includegraphics[width=.65\textwidth]{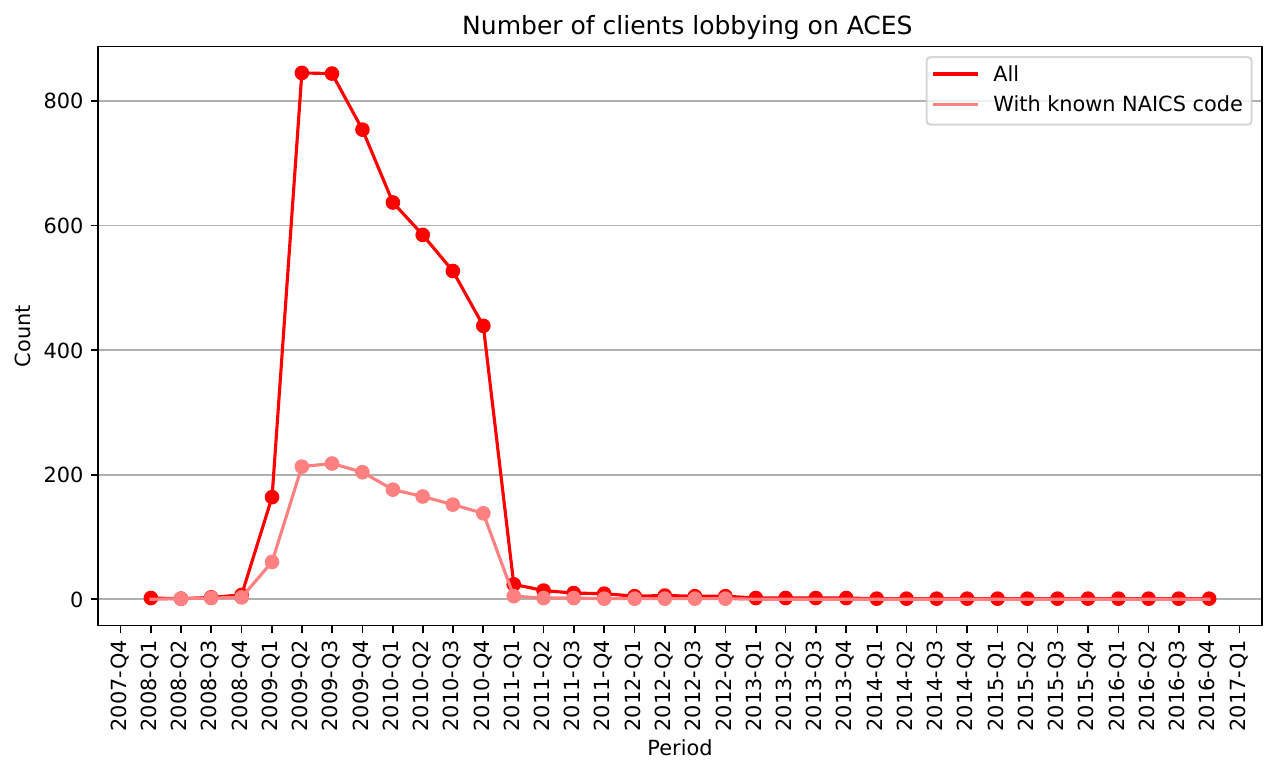}
    \caption{Number of clients lobbying on the American Clean Energy and Security Act of 2009 (ACES) in a given quarter.
    }
    \label{fig:sm:aces_numbers}
\end{figure}

\begin{figure}
    \centering
    \includegraphics[width=.95\textwidth]{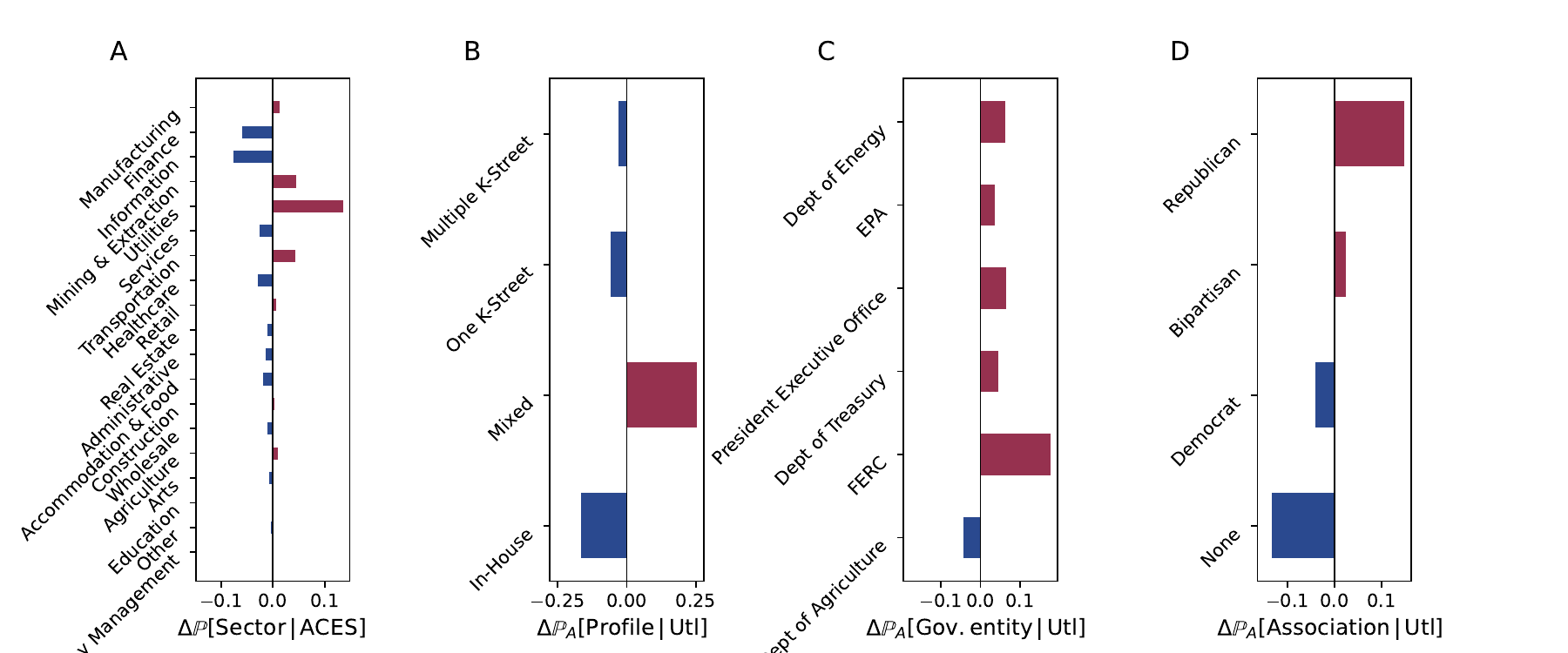}
    \caption{Probability shifts for ACES  lobbying (analogous to Fig.~4A-D of the manuscript) in the first half of 2009.
    (A) Probability that a random client belongs to a given NAICS sector $\mathbb{P}(\text{Sector})$, compared to the conditional probability $P\mathbb{P}(\text{Sector} \, | \, ACES)$.
    The sectors are sorted by $\mathbb{P}(\text{Sector})$.
(B) Probability shift comparing the probability that a client from the Utilities sector~('Utl') hired a registrant of a particular type, compared to the same quantity for all the clients lobbying on ACES with known NAICS sector.  
    (C)  Analogous quantity associated with the probability of approaching a particular government entity. 
    (D) Analogous quantity associated with the probability of hiring a lobbyist with a documented association with a current legislator.
    }
    \label{fig:sm:aces_marginals}
\end{figure}

The American Clean Energy and Security Act of 2009 (ACES), also known as the Waxman‑Markey bill, is another example of a bill that attracted significant lobby, with more than 800 clients in the first half of 2009 (Fig.~\ref{fig:sm:aces_numbers}). 
In Fig.~\ref{fig:sm:aces_marginals}A, we show the probability gain associated with the ACES lobbying, and in Fig.~\ref{fig:sm:aces_marginals}B-D we show the probability shifts that characterize the lobby of the Utilities sector (the most prominent ACES lobby based on $\Delta \mathbb{P} ( \text{Sector} \, | \, \text{ACES} ) $ ).
One interesting feature is the Republican leaning of their lobby, which correlates with the fact that the ACES bill never passed the Senate due to expected Republican filibuster.

\subsubsection{Coronavirus Aid, Relief, and Economic Security (CARES) Act of 2020}

\begin{figure}
    \centering
    \includegraphics[width=.65\textwidth]{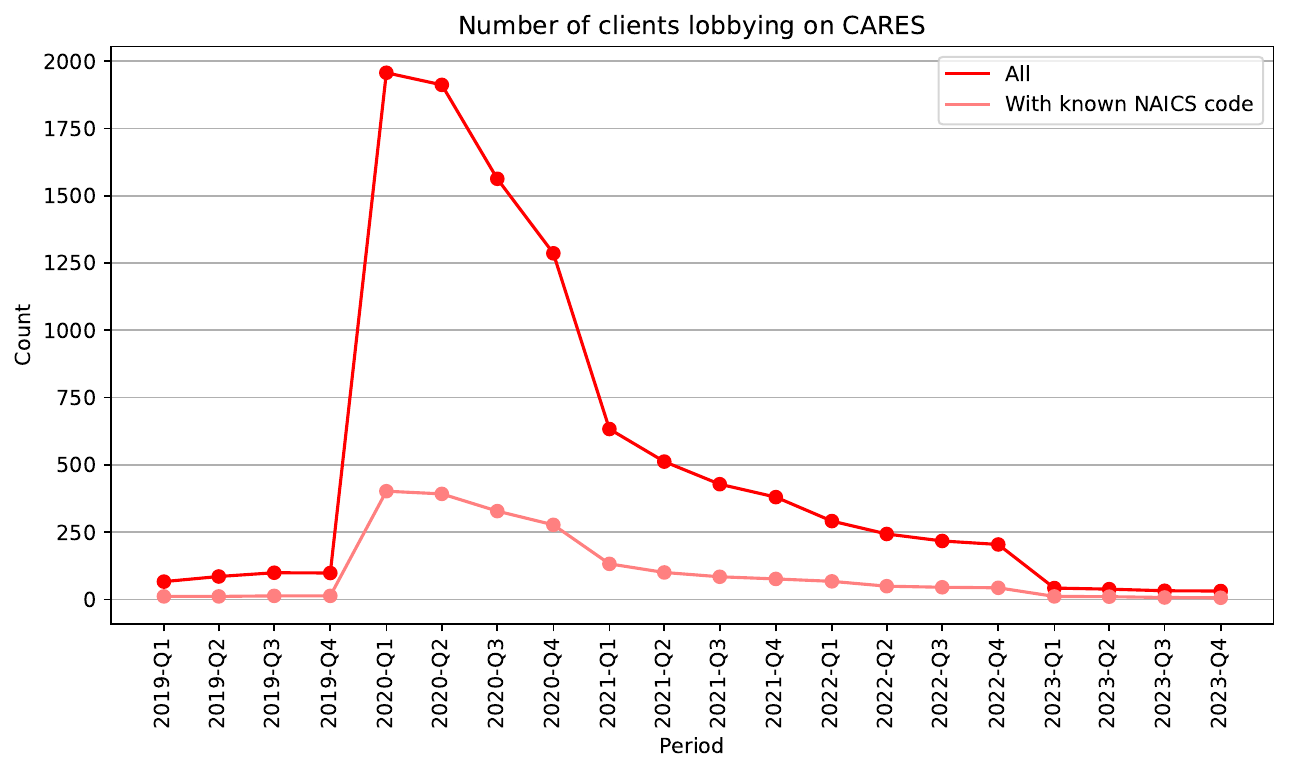}
    \caption{Number of clients lobbying on Coronavirus Aid, Relief, and Economic Security Act of 2020 (CARES) in a given quarter.
    }
    \label{fig:sm:care_numbers}
\end{figure}

\begin{figure}
    \centering
    \includegraphics[width=.95\textwidth]{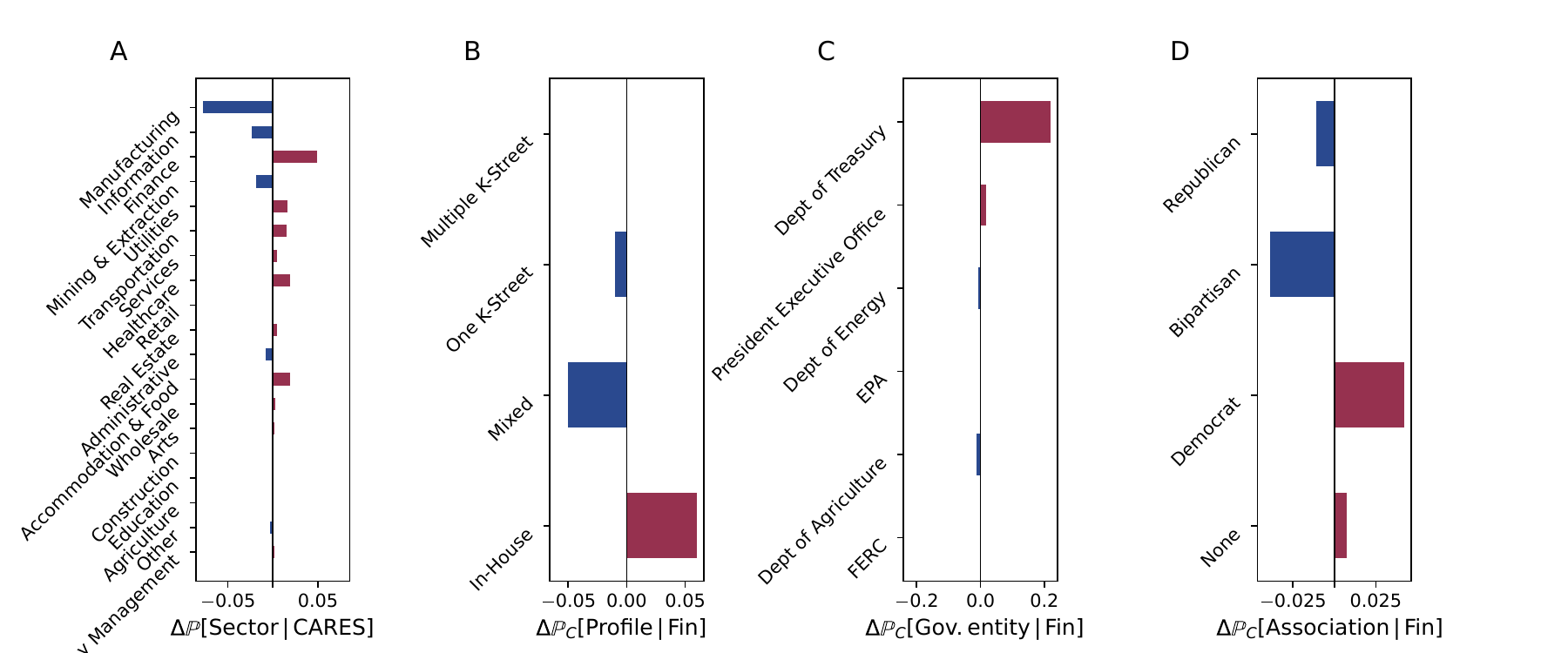}
    \caption{Probability shifts for CARES  lobbying (analogous to Fig.~4A-D of the manuscript, and Fig.~\ref{fig:sm:aces_marginals}) for the first quarter of 2020.
    (A) Probability that a random client belongs to a given NAICS sector $\mathbb{P}(\text{Sector})$, compared to the conditional probability $P\mathbb{P}(\text{Sector} \, | \, \text{CARES})$. 
    The sectors are sorted by $\mathbb{P}(\text{Sector})$.
    (B) Probability gain comparing the probability that a client from the Finance sector (`Fin') hired a registrant of a particular type, compared to the same quantity for all the clients lobbying on CARES with known NAICS sector.  
    (C)  Analogous quantity associated with the probability of approaching a particular government entity. 
    (D) Analogous quantity associated with the probability of hiring a lobbyist with a documented association with a current legislator.
    }
    \label{fig:sm:cares_marginals}
\end{figure}

In contrast to the bills analyzed in the previous sections, the CARES Act has been signed into law. 
The number of interested clients was also larger by an order of magnitude (Fig.~\ref{fig:sm:care_numbers}), reflecting the fact that the bill involved extremely broad-based government spending and income support that had implications for virtually all American firms. Clients from the financial sector were among those interested in the bill (Fig.~\ref{fig:sm:cares_marginals}A).
They were somewhat more likely to lobby through an In-House registrant than an average client (Fig.~\ref{fig:sm:cares_marginals}B), and more likely to lobby the Department of Treasury (Fig.~\ref{fig:sm:cares_marginals}C). 
Some differences were observed between them and other clients in terms of known partisan leaning of the lobbyists (Fig.~\ref{fig:sm:cares_marginals}D), but they are smaller in magnitude than the ones we found in Fig.~4, or Fig.~\ref{fig:sm:aces_marginals}.

\clearpage

\section{Polarization analysis}
\label{sec:polarization}

To get more insights into political trends in lobbying, we can use our data to quantify political polarization in more depth. 
Our analysis is exploratory in nature. Here, we explain its methodology. 

\subsection{Relation-based measures}

In order to compare the polarization of lobbyists (trend unknown) to the polarization of lawmakers (polarization known to progress), we need a methodology that could be applied to both. 
In Fig.~4D, we used a simple network-based measure that can be applied to any set of entities $\mathcal{S}$ subdivided into two subsets $\mathcal{S_1}, \mathcal{S_2}$, which are joined by some kind of relational ties.

In case of legislators, $\mathcal{S}_1, \mathcal{S}_2$ correspond to representatives/senators who belong to one of the two main political parties, e.g., we can take 
\[
\mathcal{S}_1(y) : = \text{Democratic Senators}(y) = \{p: \text{Party}(p,y) = \text{Democratic},~\text{Chamber}(p,y) = \text{Senate}  \}, 
\]
\[
\mathcal{S}_2(y) : = \text{Republican Senators}(y) = \{p: \text{Party}(p,y) = \text{Republican},~\text{Chamber}(p,y) = \text{Senate}  \} ,
\] 
and the superset is simply 
\[
\mathcal{S} = \mathcal{S}_1(y) \cup \mathcal{S}_2(y). 
\]
Thus, we neglect here the independent legislators, but their number is small, and it is not always straightforward to define how they might contribute to polarization.

The relations between legislators can be defined in all sorts of ways. 
Here, we focus on bill cosponsoring. 
We can envisage the legislators as nodes in the multigraph, and we add one edge ($p_1,p_2$) between any pair of politicians for each bill they co-sponsored together. 
We treat each chamber separately, neglect resolutions, and treat sponsors and cosponsors in the same way.

The edges can connect members of the same party or members of two parties. 
To evaluate the degree of polarization, we look at the number of bipartisan cosponsorships. 
Specifically, we compute 
\begin{equation}
\text{Legislator Bipartisan Index} (y) = \frac{\left| (p_1,p_2) : p_1 \in \mathcal{S}_1, p_2 \in \mathcal{S}_2  \right | }{  \left|  (p_1,p_2) : p_1, p_2 \in \mathcal{S} \right|  }.
\label{eq:sm:bipartisan}
\end{equation}
We compute the bipartisan index for House and Senate separately, and we report it in Fig.~4D of the main paper. 
The fact that it has been consistently decreasing throughout the 21st century can be interpreted as a symptom of Congress polarization. \\

Crucially, we can use a very similar measure to assess polarization among the lobbyists. 
Now, we define 
\[
\mathcal{S}_1(y) : = \{l \in \text{Lobbyists}(y) :  p(l,y) \cap   \text{Democratic Legislators}(y) \neq \emptyset , p(l,y) \cap   \text{Republican Legislators}(y) = \emptyset \}, 
\]
\[
\mathcal{S}_2(y) : = \{l \in \text{Lobbyists}(y) : p(l,y) \cap   \text{Republican Legislators}(y) \neq \emptyset , p(l,y) \cap   \text{Democratic Legislators}(y) = \emptyset  \}, 
\]
where 
\[
\text{Democratic Legislators}(y) = \{p: \text{Party}(p,y) = \text{Democratic} \},
\]
\[
\text{Republican Legislators}(y) = \{p: \text{Party}(p,y) = \text{Republican} \}.
\]
Again, we neglect lobbyists associated with independent legislators, and lobbyists associated with legislators from both of the major parties, but such cases are rare  (c.f.~Sec.~\ref{sec:sm:connections_data}). 

As before, $\mathcal{S} = \mathcal{S}_1(y) \cup \mathcal{S}_2(y)$ and we want to treat the lobbyists as nodes of a graph, with links representing some kind of ties. 
In our analysis, we considered two different kinds of links. 
First, we constructed the graph based on professional filings, i.e., we add a link $(l_1, l_2) $ for every filing where the two lobbyists featured together. 
This led to the collaboration-based bipartisan index. 
As an alternative, we consider a graph where an edge is added for every bill that the two lobbyists lobbied on together (potentially for different clients). 
This lead to the bipartisan index based on legislative interest. 
Both of these indices are computed with the same formula
\begin{equation}
\text{Lobbyist Bipartisan Index} (y) = \frac{\left| (l_1,l_2) : l_1 \in \mathcal{S}_1, l_2 \in \mathcal{S}_2  \right | }{  \left|  (l_1,l_2) : l_1, l_2 \in \mathcal{S} \right|  }, 
\end{equation}
which is exactly analogous to~\eqref{eq:sm:bipartisan}, except the edges are constructed based on a different notion of a relationship. 
The lobbyists bipartisan indices do not seem to decrease in the same way they do for legislators. 
Thus, as far as this simple measure is concerned, we do not find symptoms of increasing polarization in the lobbying industry. 

\subsection{Reach-based measures}

\begin{figure}
    \centering
    \includegraphics[width=.99\textwidth]{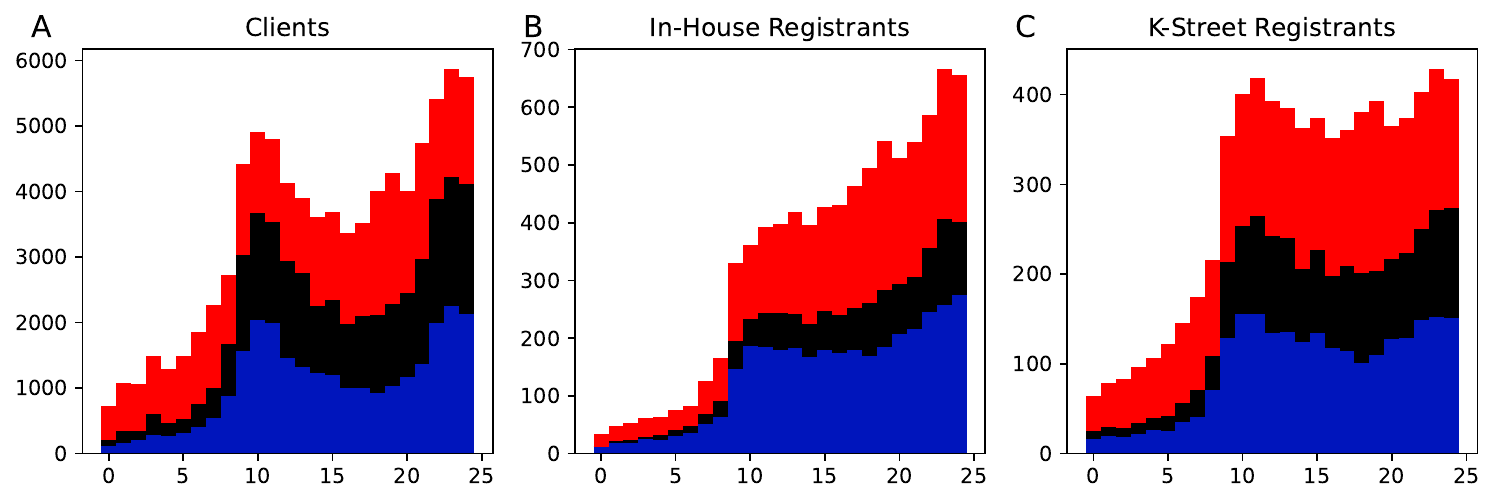}
    \caption{Reach-based polarization analysis. (A) The number of clients with a democratic profile (blue), a bipartisan profile (black), and a republican profile (red). (B) Analogous plot for In-House registrants. (C) Analogous plot for K-Street registrants. In all cases, the relative number of the bipartisan profile agents increases in time, which could be indicative of depolarization.  }
    \label{fig:sm:polarization_reach}
\end{figure}

We can also introduce a complementary measure of polarization in lobbying, focused on registrants and clients 
We employ here the notion of \textit{reach} introduced in Sec.~\ref{sec:reach_centrality}.
For example, we say that an agent (registrant or client) can `reach' the Democratic Party, if it employs a lobbyist with a historical association to a Democratic legislator. 
Based on this idea, we divide clients and registrants into four categories: those who are not connected to any political party, agents with documented connections to the Democratic Party only (democratic profile), agents with documented connections to the Republican Party only (republican profile), and agents connected to both Democrats and Republicans (bipartisan profile). 
We neglect the first group, and the sizes of the other three groups across the years are shown in Fig~\ref{fig:sm:polarization_reach}. 

The relative size of the bipartisan set can be used as a measure of polarization, e.g., for clients, we compute  
\[
\frac{|\text{Bipartisan profile clients} |}{|\text{Bipartisan profile clients} | + |\text{Democratic profile clients} | + |\text{Republican profile clients|}}.
\]
We note that these measures increase for all the agent types, which indicates that polarization in lobbying might even be decreasing. 

\begin{figure}
    \centering
    \includegraphics[width=.99\textwidth]{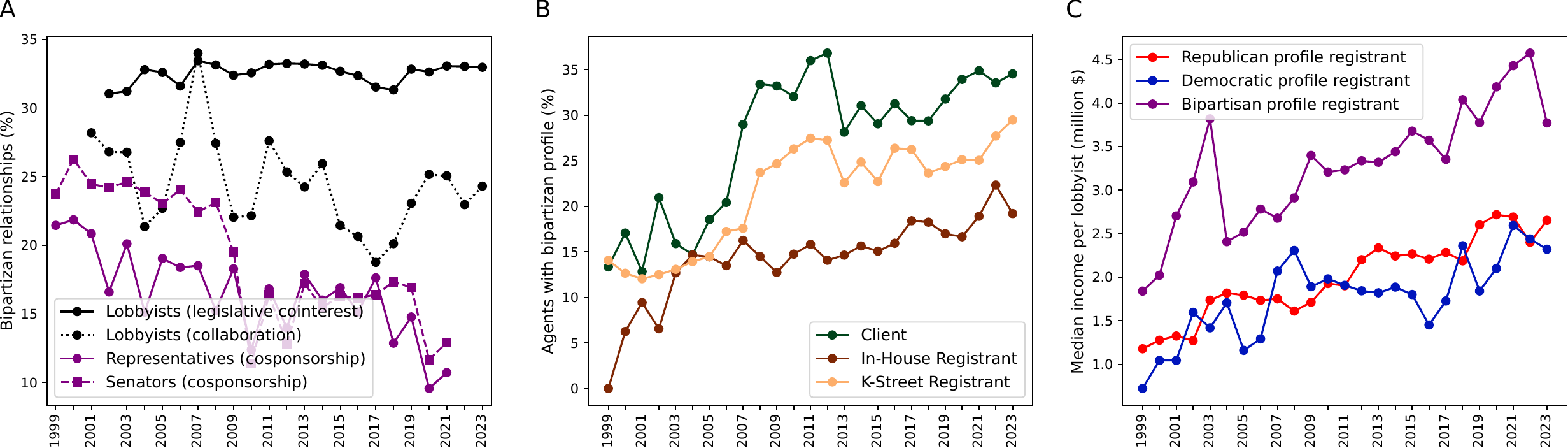}
    \caption{(A) The share of bipartisan professional relationships can be used to evaluate the degree of polarization. 
    Our analysis suggests that, unlike the polarization of the lawmakers, the polarization of the lobbying industry is not increasing. 
    (B) Our data indicates that the share of lobbying agents (clients and registrants) with bipartisan profiles increases.  
    (C) Registrants employing lobbyists with known historical connections to both of the major political parties, on average, generate higher income per lobbyist.   }
    \label{fig:sm:polarization}
\end{figure}

In Fig.~\ref{fig:sm:polarization}, we compare the polarization of lobbyists and legislators. 
Perhaps unsurprisingly, our data analysis reveals strong evidence for an increasing polarization among the legislators, highlighted by the fact that the rate of bipartisan co-sponsorship of bills in the U.S. Senate has dropped by about 40\% over the last 25 years. 
Nevertheless, a similar trend towards polarization has not occurred within the lobbying industry. 
The number of instances when the lobbyists of opposite political leaning work together or lobby on the same bill has remained relatively stable (Fig.~\ref{fig:sm:polarization}A). Moreover, the proportion of clients and registrants that employ a bipartisan team of lobbyists is weakly increasing, suggesting that both clients and registrants recognize a need to build connections with both parties to maximize impact~(Fig.~\ref{fig:sm:polarization}B). 
The incentive on the registrant side is clear: bipartisan teams of lobbyists generate more income ~(Fig.~\ref{fig:sm:polarization}C). Thus, our data support the hypothesis that lobbying is a strategic, rather than ideological activity.

\subsubsection*{Modularity}
As an alternative to the bipartisan index, we could also take the \textit{modularity} of the relationship graph~\cite{Newman_book}, partitioned into $\mathcal{S}_1(y)$ and $ \mathcal{S}_2(y)$. 
Let $A_{ij}$ be the number of bills co-sponsored by legislators $p_i$ and $p_j$ (edge weight), $k_i$ be the total number of co-sponsorship ties of legislator $i$ (node degree), and $m$ be the total number of edges. The modularity is defined as 
\[
Q(y) = \frac{1}{2m} \sum_{i,j} [A_{ij} - \frac{k_ik_j}{2m}] \mathbbm{1} [ \text{Party}(p_1,y) =   \text{Party}(p_2,y) ], 
\]
and analogously for the lobbyists. 

\begin{figure}
    \centering
    \includegraphics[width=.45\textwidth]{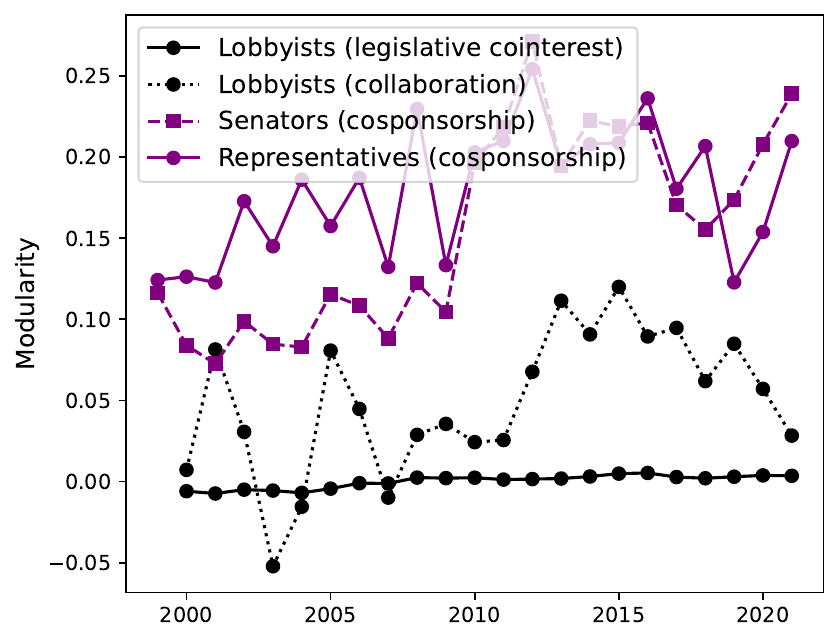}
    \caption{Modularity of the cross-partisan collaboration graphs - an analogue of Fig.~\ref{fig:sm:polarization}A with a different measure. }
    \label{fig:sm:modularity}
\end{figure}

The results are shown in Fig.~\ref{fig:sm:modularity}.
For lobbyists, the modularity is close to zero, which indicates a lack of polarization. 
For legislators, we find consistently higher values of modularity, and in the case of senators, we observe a marked increase in modularity over time. 

\section{Data and code availability }
\label{sec:data_availability}

Anonymized relational data, sufficient to reproduce the results of this paper, as well as the code used to analyze the data and produce the figures, can be accessed at \url{https://osf.io/y2hnm/?view_only=9a14b7896f834c6c9290eee85dde5bb7}.
Researchers who are interested in interfacing with the broader \textsf{LobbyView} database can do so by visiting our website, \href{https://lobbyview.org}{lobbyview.org}. 
Registration is free and instant, and allows access to the CSV data downloads and query API for report, client, bill, issue, and lobbyist-level data, as well as issue text and client-politician connection dyad data.
Researchers who require access to more detailed or granular data or fields not exposed in the CSVs or API are invited to contact the authors for access as necessary.

\bibliographystyle{Science}
\bibliography{scibib}